\documentclass[11pt]{article}

\usepackage{amsmath}
\usepackage{amssymb}
\usepackage{enumerate}
\usepackage{cases}
\usepackage{multirow}
\usepackage{subfigure}
\usepackage{graphicx}
\usepackage{appendix}
\usepackage{pgf,tikz}
\usetikzlibrary{arrows}
\usepackage{mathrsfs} %

\graphicspath{{./}{./Figures/}}
\DeclareMathOperator{\sign}{sgn}

\begin{document}
\newcommand{\Ai}{{\rm Ai}}
\newcommand{\tf}[1]{{\begin{scriptsize} \it #1 \end{scriptsize}}}
\newcommand{\Bi}{{\rm Bi}}
\newcommand{\F}{{Fig.~}}
\newcommand{\Fs}{{Figs~}}
\newcommand{\cF}{\mathcal{F}}
\newcommand{\includegraphicsDave}[2][]{\includegraphics[#1]{#2_LOWRES_xy}}		%
\newcommand{\done}[2]{\dfrac{d {#1}}{d {#2}}}
\newcommand{\donet}[2]{\frac{d {#1}}{d {#2}}}
\newcommand{\pdone}[2]{\dfrac{\partial {#1}}{\partial {#2}}}
\newcommand{\pdonet}[2]{\frac{\partial {#1}}{\partial {#2}}}
\newcommand{\pdonetext}[2]{\partial {#1}/\partial {#2}}
\newcommand{\pdtwo}[2]{\dfrac{\partial^2 {#1}}{\partial {#2}^2}}
\newcommand{\pdtwot}[2]{\frac{\partial^2 {#1}}{\partial {#2}^2}}
\newcommand{\pdtwomix}[3]{\dfrac{\partial^2 {#1}}{\partial {#2}\partial {#3}}}
\newcommand{\pdtwomixt}[3]{\frac{\partial^2 {#1}}{\partial {#2}\partial {#3}}}
\newcommand{\bs}[1]{\mathbf{#1}}
\newcommand{\bx}{\mathbf{x}}
\newcommand{\by}{\mathbf{y}}
\newcommand{\bd}{\mathbf{d}} 
\newcommand{\bn}{\mathbf{n}} 
\newcommand{\bP}{\mathbf{P}} 
\newcommand{\bp}{\mathbf{p}} 
\newcommand{\ol}[1]{\overline{#1}}
\newcommand{\rf}[1]{(\ref{#1})}
\newcommand{\xt}{\mathbf{x},t}
\newcommand{\hs}[1]{\hspace{#1mm}}
\newcommand{\vs}[1]{\vspace{#1mm}}
\newcommand{\eps}{\varepsilon}
\newcommand{\ord}[1]{\mathcal{O}\left(#1\right)} 
\newcommand{\oord}[1]{o\left(#1\right)}
\newcommand{\Ord}[1]{\Theta\left(#1\right)}
\newcommand{\PhiF}{\Phi_{\rm freq}}
\newcommand{\real}[1]{{\rm Re}\left[#1\right]} 
\newcommand{\im}[1]{{\rm Im}\left[#1\right]}
\newcommand{\hsnorm}[1]{||#1||_{H^{s}(\bs{R})}}
\newcommand{\hnorm}[1]{||#1||_{\tilde{H}^{-1/2}((0,1))}}
\newcommand{\norm}[2]{\left\|#1\right\|_{#2}}
\newcommand{\normt}[2]{\|#1\|_{#2}}
\newcommand{\on}[1]{\Vert{#1} \Vert_{1}}
\newcommand{\tn}[1]{\Vert{#1} \Vert_{2}}
\newcommand{\ts}{\tilde{s}}
\newcommand{\darg}[1]{\left|{\rm arg}\left[ #1 \right]\right|}
\newcommand{\bnabla}{\boldsymbol{\nabla}}
\newcommand{\dive}{\boldsymbol{\nabla}\cdot}
\newcommand{\curl}{\boldsymbol{\nabla}\times}
\newcommand{\Phixy}{\Phi(\bx,\by)}
\newcommand{\PhiOxy}{\Phi_0(\bx,\by)}
\newcommand{\dxPhixy}{\pdone{\Phi}{n(\bx)}(\bx,\by)}
\newcommand{\dyPhixy}{\pdone{\Phi}{n(\by)}(\bx,\by)}
\newcommand{\dxPhiOxy}{\pdone{\Phi_0}{n(\bx)}(\bx,\by)}
\newcommand{\dyPhiOxy}{\pdone{\Phi_0}{n(\by)}(\bx,\by)}

\newcommand{\rd}{\mathrm{d}}
\newcommand{\R}{\mathbb{R}}
\newcommand{\N}{\mathbb{N}}
\newcommand{\Z}{\mathbb{Z}}
\newcommand{\C}{\mathbb{C}}
\newcommand{\K}{{\mathbb{K}}}
\newcommand{\ri}{{\mathrm{i}}}
\newcommand{\re}{{\mathrm{e}}} 

\newcommand{\cA}{\mathcal{A}}
\newcommand{\cC}{\mathcal{C}}
\newcommand{\cS}{\mathcal{S}}
\newcommand{\cD}{\mathcal{D}}
\newcommand{\cone}{{c_{j}^\pm}}
\newcommand{\ctwo}{{c_{2,j}^\pm}}
\newcommand{\cthree}{{c_{3,j}^\pm}}

\newtheorem{thm}{Theorem}[section]
\newtheorem{lem}[thm]{Lemma}
\newtheorem{defn}[thm]{Definition}
\newtheorem{prop}[thm]{Proposition}
\newtheorem{cor}[thm]{Corollary}
\newtheorem{rem}[thm]{Remark}
\newtheorem{conj}[thm]{Conjecture}
\newtheorem{ass}[thm]{Assumption}
\newtheorem{example}[thm]{Example} 
\newcommand{\X}{X}
\newcommand{\Y}{Y}
\newcommand{\x}{x}
\newcommand{\y}{y}
\title{Contour integral solutions of the parabolic wave equation}

\author{D.\ P.\ Hewett\footnotemark[1] \footnotemark[3], J.\ R.\ Ockendon\footnotemark[2], V.\ P.\ Smyshlyaev\footnotemark[1]}

\maketitle

\renewcommand{\thefootnote}{\fnsymbol{footnote}}
\footnotetext[1]{Department of Mathematics, University College London, London, UK}
\footnotetext[2]{Oxford Centre for Industrial and Applied Mathematics, Mathematical Institute, Oxford, UK}
\footnotetext[3]{Corresponding author, email {\tt d.hewett\char'100ucl.ac.uk}}
\renewcommand{\thefootnote}{\arabic{footnote}}
\begin{abstract}
We present a simple systematic construction and analysis of solutions of the two-dimensional parabolic wave equation that exhibit far-field localisation near a given algebraic plane curve. 
Our solutions are complex contour integral superpositions of elementary plane wave solutions with polynomial phase, the desired localisation being associated with the coalescence of saddle points. %
Our solutions provide a unified framework in which to describe some classical phenomena in two-dimensional high frequency wave propagation, including smooth and cusped caustics, whispering gallery and creeping waves, and tangent ray diffraction by a smooth boundary. 
We also study a subclass of solutions exhibiting localisation near a cubic parabola, and discuss their possible relevance to the study of the canonical inflection point problem governing the transition from whispering gallery waves to creeping waves.

\end{abstract}
\section{Introduction}
\label{Intro}

In the study of short wavelength linear two-dimensional wave propagation described by the dimensionless Helmholtz equation 
\begin{align}
\label{eqn:HE}
\pdtwo{\varphi}{\x }+\pdtwo{\varphi}{\y} + k^2 \varphi =0
\end{align}
in the limit as $k\to\infty$, 
one often seeks a wave field which is localised in the neighbourhood of a certain plane curve $\mathscr{C}$, for example a caustic or shadow boundary associated with a WKB (geometrical optics) approximation, or the boundary of a scattering obstacle. 
By ``localised'' we mean that the field is concentrated near $\mathscr{C}$, decaying rapidly in the normal direction to $\mathscr{C}$.  

As is well known (see e.g. \cite{BaKi:79,BaBu:91,OckTew:12}), many such `thin-layer' solutions of \rf{eqn:HE} can be locally described using the \textit{parabolic} or \textit{paraxial} approximation. 
This approximation pertains when the modulation length of a plane wave in the direction of propagation, $\x$, is short compared to that in the transverse direction, $\y$. In such a situation we may write %
\begin{align}
\label{eqn:scaling}
\x =k^{-\lambda} \X, \qquad \y=k^{-(1+\lambda)/2}\Y,
\end{align}
for some $0\leq\lambda<1$, %
and set the complex wave function to be 
$\varphi = A \re^{\ri k\x }$. 
We then find that, to lowest order as $k\to\infty$,  
$A(\X,\Y)$ satisfies the so-called parabolic 
wave equation (PWE)
\begin{align}
\label{eqn:PWE}
2\ri \pdone{A}{\X}+\pdtwo{A}{\Y} = 0.
\end{align}
The PWE \rf{eqn:PWE} is only a local approximation to the Helmholtz equation, and in order for 
$A(\X,\Y)$ to match with a global WKB approximation $\varphi(\x ,\y)$ of \rf{eqn:HE} that is localised near the curve $\mathscr{C}$ when $k$ is large, we require the far-field expansion of $A$ as $\X,\Y\to\infty$ to give rise to at least two waves that are asymptotically coincident on $\mathscr{C}$ (e.g.\ the incoming and outgoing rays associated with a smooth caustic). 

In this paper we present a simple methodology for the systematic construction and analysis of PWE solutions exhibiting far-field localisation near algebraic curves 
\begin{align}
\label{eqn:Bdy}
(-\Y)^l =C_{l,m}\kappa^{m} \X^{l+m}, 
\end{align}
where $l,m$ are positive integers, 
\begin{align}
\label{eqn:ClmDef}
C_{l,m}=\frac{l^{2l}}{\left(m(l+m)\right)^l}
\end{align}
is a normalisation constant, 
and $\kappa>0$ is a shape parameter, related to curvature. Precisely, the form of \rf{eqn:Bdy}-\rf{eqn:ClmDef} means that the curvature of \rf{eqn:Bdy} as $\X$ approaches $0$ is asymptotically $\kappa^{m/l}|\X|^{m/l-1}$. %
In order that \rf{eqn:Bdy} corresponds to a curve in the original $(\x,\y)$-plane which is independent of the wavenumber $k$, we require that the scaling parameter $\lambda$ in \rf{eqn:scaling} satisfies
\begin{align}
\label{eqn:lambda}
\lambda = \frac{l}{l+2m}.
\end{align}

The existence of such localised solutions is well known. Two familiar examples, described in detail in \cite{OckTew:12}, are 
\begin{enumerate}[(i)]
\item \label{i}
$(l,m)=(1,1)$, $\lambda=1/3$, corresponding to localisation near a parabola $\Y=-\kappa \X^2/2$, which arises in the study of smooth caustics, creeping waves, whispering gallery waves, and the Fock-Leontovich tangent ray diffraction problem;
\item \label{ii}
$(l,m)=(2,1)$, $\lambda=1/2$, corresponding to localisation near the curve $\Y^2=16\kappa \X^3/9$, which arises in the study of cusped caustics. 
\newcounter{count}
\setcounter{count}{\value{enumi}}
\end{enumerate}
Other important, but less well-understood examples include
\begin{enumerate}[(i)]
\setcounter{enumi}{\value{count}}
\item \label{iii}
$(l,m)=(1,2)$, $\lambda=1/5$, corresponding to localisation near a cubic parabola $\Y=- \kappa^2 \X^3/6$. This arises in the (as yet not fully solved) canonical inflection point problem governing the concave-convex transition from whispering gallery waves to creeping waves on a smooth boundary \cite{Pop:79,Pop:79a,Pop:82,Pop:82a,Pop:86,PopKra:86,BabSmy:86,BabSmy:87,Kaz:03};
\item $(l,m)=(1,m)$ with $m\geq 3$, generalising \rf{iii} to model the scattering of whispering gallery modes by a higher order zero of the boundary curvature \cite{PopPsh:83,NakShiTan:89,Kaz:05}.
\end{enumerate}

The class of solutions we consider take the form of contour integral superpositions of elementary plane wave solutions of \rf{eqn:PWE}. %
Specifically, we study solutions of the form
\begin{align}
\label{eqn:AInt}
A(\X,\Y) &= \int_\Gamma F(t) \,\re^{\ri p(\X,\Y,t)} \,\rd t,
\end{align}
where
\begin{align}
\label{eqn:POriginal}
p(\X,\Y,t)&=- \Y t^m -\frac{\X t^{2m}}{2}+\frac{\alpha mt^{2m+l}}{(2m+l)},
\end{align}
$\alpha>0$ is a parameter, $F$ is an analytic prefactor, and $\Gamma$ is a contour in the complex $t$-plane beginning and ending at infinity, such that the integral \rf{eqn:AInt} (and the derivatives necessary for \rf{eqn:PWE} to hold) exist and are continuous for real $\X,\Y$. 
Given $l,m$ and $\kappa$, we will see that, 
for 
appropriate choices of $\alpha$, $\Gamma$ and $F$, \rf{eqn:AInt}-\rf{eqn:POriginal} can describe solutions localised near \rf{eqn:Bdy} when 
$|\Y|^l\sim |\X|^{l+m}\to\infty$, with the localisation being associated with the coalescence of saddle points in the method of steepest descent.\footnote{Shadow boundary phenomena give rise to the Fresnel integral
\[{\rm Fr}\left(\frac{\Y}{\sqrt{2\X}}\right) = \frac{1}{2\pi \ri}\int_\Gamma \frac{\re^{\ri(-\X t^2/2 - \Y t)}}{t}\,\rd t, \qquad \X>0,\,\Y\in\R,\]
with $\Gamma$ going from $\re^{-\ri\pi/4}\infty$ to $\re^{3\ri\pi/4}\infty$ and passing below the pole at $t=0$. This is of the form \rf{eqn:AInt} with $l=0$ and $m=1$ (i.e.\ $\lambda = 0$), but the localisation near $Y=0$ arises from coalescence of the saddle with the pole.}
 
PWE solutions of the form \rf{eqn:AInt} are already well known in the study of wave fields near smooth and cusped caustics, where one encounters the `canonical diffraction integrals' of catastrophe theory (see e.g.\ \cite{Ber:80,arnol2003catastrophe,kryukovskii2009construction} and the final chapter of the NIST handbook \cite[\S36]{DLMF}, edited by Berry and Howls), which have 
$\Gamma$ the real axis and $F(t)\equiv 1$. 
The existence of a representation of the form \rf{eqn:AInt}, with $\Gamma$ the real axis, is guaranteed for these caustic fields by the fact that the functions $A(\X,\Y)$ in question solve PWE globally on the whole of $\R^2$ and are bounded at infinity.\footnote{More generally, let 
$A(\X,\Y)$ be any tempered distribution solving \rf{eqn:PWE} on the whole of $\R^2$ in a distributional sense. Taking a Fourier transform of \rf{eqn:PWE} gives 
$(s - t^2/2)\hat{A}(s,t)=0$, 
which implies that 
$\hat{A}(s,t) = f(t)\delta(s-t^2/2)$ 
for some distribution $f(t)$ (note that no derivatives of the delta function arise). Hence by Fourier inversion
\begin{align*}
A(\X,\Y) = \int_{-\infty}^\infty \int_{-\infty}^\infty f(t)\delta(s-t^2/2) \re^{-\ri (\X s + \Y t)}\,\rd s\rd t
=\int_{-\infty}^\infty f(t) \re^{-\ri (\X t^2/2 + t \Y)}\,\rd t.
\end{align*}
}
Of course, many important PWE solutions, including the canonical models for whispering gallery waves, creeping waves, and tangent ray diffraction, do not enjoy this boundedness property.  Nonetheless, we shall show that the solutions listed above can all be represented by \rf{eqn:AInt}, although in general the contour $\Gamma$ must be complex, and the prefactor $F(t)$ must differ from $1$. When $F(t)$ has exponential behaviour at infinity, this has the effect of modifying the saddle point structure and shifting the localisation curve. Poles of $F(t)$ allow the switching on/off of different fields components as the contour $\Gamma$ is deformed across them (see e.g.\ \cite{He:14} and \S\ref{sec:BVPs} below).

One objective of the paper is to present all the solutions mentioned above in a single unified framework, and outline a simple methodology for their far-field analysis. A second objective is to promote 
the ansatz \rf{eqn:AInt} as a possible framework in which to seek solutions of as yet unsolved canonical problems such as the inflection point problem and its higher order analogues. Limited investigations in this direction have already been made by Kazakov in \cite{Kaz:03,Kaz:05}, where it was shown that with $F(t)\equiv 1$ a certain choice of contour $\Gamma$ leads to a localised solution with the character of incoming whispering gallery waves. But the idea has not been fully explored. Our contribution is to provide a systematic qualitative analysis of the far-field behaviour of all the other possible contour choices relating to Kazakov's integral. In particular, we identify a contour that leads to asymptotic localization capturing outgoing creeping waves. We also explain how choosing $F(t)\neq 1$ appropriately can force these solutions to asymptotically satisfy homogeneous boundary conditions on \rf{eqn:Bdy}, and discuss some of the issues involved in trying to use integrals of the form \rf{eqn:AInt} to solve the full inflection point boundary value problem. 

The structure of the paper is as follows. In \S\ref{sec:general} we outline our basic methodology for determining the qualitative far-field behaviour of \rf{eqn:AInt}, and identifying when localisation occurs. 
In \S\ref{sec:poly} we consider the case where the prefactor $F(t)\equiv 1$. In this case it is possible to provide a complete classification of the possible choices of contour $\Gamma$ and the symmetry relations satisfied by the resulting solutions. 
By dividing the $(\X,\Y)$-plane into appropriate regions based on an analysis of the different saddle point configurations that can occur, and the locations of the relevant Stokes and anti-Stokes lines, we are able to deduce the kinds of far-field localisation that can occur for each possible choice of integration contour. Our methodology applies to general $(l,m)$ but we shall focus in particular on cases \rf{i}-\rf{iii} above. We shall show that this framework is sufficient to describe many of the known PWE solutions appearing in wave propagation. But we also describe many apparently new solutions exhibiting exotic behaviour. We will also give asymptotic formulae near the localisation and, for comparison, plots of $|A(\X,\Y)|$ and $\real{A\re^{\ri k\x}}$ in cases where the computation of the integral in \rf{eqn:AInt} is not too difficult. 

In \S\ref{sec:BVPs} and \S\ref{sec:Discussion} we turn to the question of whether PWE solutions of the form \rf{eqn:AInt} can satisfy homogeneous Dirichlet (soft) or Neumann (hard) boundary conditions on \rf{eqn:Bdy}. We first revisit the well-studied parabolic case $(l,m)=(1,1)$, showing that one can retrieve paradigm models for whispering gallery and creeping waves by taking $F(t)$ in \rf{eqn:AInt} to be a certain exponential in $t$, and that, as was shown recently in \cite{He:14}, \rf{eqn:AInt} can also describe the solution of the classical Fock-Leontovich tangent ray diffraction problem, with $F(t)$ the Pekeris caret function (or Fock integral). 
In \S\ref{sec:Discussion} we then discuss the possibility of using \rf{eqn:AInt} to solve the inflection point problem.

As a postscript to this introductory section, we note that in much of the literature on thin layer wave phenomena
(e.g.\ \cite{Pop:79,Pop:79a,Pop:82,Pop:82a,PopPsh:83,Pop:86,PopKra:86,BabSmy:86,BabSmy:87,Kaz:03,Kaz:05}) fields are expressed not in Cartesian coordinates but rather in curvilinear coordinates adapted to the curve $\mathscr{C}$. This simplifies the statement of boundary conditions in boundary value problems, but complicates the governing equation. To allow easy comparison with this literature, we now provide an explicit connection between the curvilinear viewpoint and our Cartesian framework, generalising the connections made for certain special cases in \cite{OckTew:12}. 
Our localisation curve $\mathscr{C}$ is described in Cartesian coordinates by
\[ Y\pm C_{l,m}^{1/l}\kappa^{m/l}X^{1+m/l}=0,\] 
where the $\pm$ symbol, describing different branches of the curve, is required only when $l$ is even (when $l$ is odd we always take $+$). 
Changing to local curvilinear coordinates  
\[t=X, \qquad z = Y\pm C_{l,m}^{1/l}\kappa^{m/l}X^{1+m/l}, \]
the curve $\mathscr{C}$ is now described by $z=0$, and setting
\[ A(X,Y) = \exp\left[-\ri \left(\pm\frac{l}{m}\kappa^{m/l}X^{m/l}Y + \frac{l^3(3m+l)}{2m^2(m+l)(2m+l)}\kappa^{2m/l}X^{2m/l+1} \right)\right]u(t,z),\]
one finds that%
\[2\ri u_t + u_{zz} \pm 2\kappa^{m/l}zt^{m/l-1}u = 0, \] 
which reduces to the governing equations appearing in \cite{Pop:79,Pop:79a,Pop:82,Pop:82a,PopPsh:83,Pop:86,PopKra:86,BabSmy:86,BabSmy:87,Kaz:03,Kaz:05} under the appropriate choices of $m,l$ and $\kappa$.

\section{\label{sec:general}Far-field analysis}

Our general approach to analysing the far-field behaviour of \rf{eqn:AInt} is as follows. We first rewrite \rf{eqn:AInt} in outer variables $(\x,y)$ and rescale the integration variable, setting (recall \rf{eqn:scaling} and \rf{eqn:lambda})
\begin{align}
\label{eqn:scaling2}
\X=k^{\lambda} \x = k^{l/(l+2m)}\x, \qquad \Y=k^{(1+\lambda)/2}\y = k^{(l+m)/(l+2m)}\y, 
\qquad t = k^{\lambda/l} \tau = k^{1/(l+2m)} \tau,
\end{align}
to obtain 
\begin{align}
\label{eqn:AInt2}
A = k^{\frac{1}{l+2m}}\int_{\tilde{\Gamma}} F\left(k^{\frac{1}{l+2m}}\tau\right) \re^{\ri k \phi(\tau)}\,\rd \tau,
\end{align}
where $\tilde{\Gamma}$ is the rescaled integration contour and 
\begin{align}
\label{eqn:Phase}
\phi(\tau) = - \y\tau^m -\frac{\x\tau^{2m}}{2} + \frac{\alpha m\tau^{2m+l}}{(2m+l)}.
\end{align}
We then compute the behaviour of \rf{eqn:AInt2} as $k\to\infty$ using the method of steepest descent, which involves deforming $\tilde\Gamma$ onto the appropriate steepest descent contour and evaluating the local contribution from the relevant saddle points. If $F(t)$ possesses poles and/or branch points then, depending on the required contour deformation, their contribution may also need to be computed. 
We note that if the integration contour $\tilde\Gamma$ is the real axis and the contributing saddles are also real, the leading order behaviour may be obtained by the method of stationary phase, without the need for contour deformation. This was the approach taken in \cite{OckTew:12}, and our own explicit calculations will also exploit this.

Assuming that $F(t)$ behaves sub-exponentially at infinity (for example if $F(t)$ is a polynomial or a rational function), the saddle point structure of \rf{eqn:AInt2} is governed entirely by the function $\phi(\tau)$.
The saddle points are the roots of $\phi'(\tau)=0$, and hence solve
\begin{align}
\label{eqn:StatPointsPrelim}
\tau^{m-1}\left(- \y  -\x\tau^{m} + \alpha\tau^{m+l}\right)=0.
\end{align}
Asymptotic localisation is associated with the existence of saddle points of order greater than one, i.e.\ points where both $\phi'(\tau)=0$ and $\phi''(\tau)=0$, the latter condition being equivalent to
\begin{align}
\label{eqn:CoalescePrelim}
\begin{cases}
 -\x + \alpha(1+l)\tau^{l}=0, & m=1,\\
\tau^{m-2}\left(-(m-1)\y  -(2m-1)\x\tau^m + \alpha(2m+l-1)\tau^{m+l}\right)=0, & m\geq 2.
\end{cases}
\end{align}
We observe that when $\x=\y=0$, $\tau=0$ is a saddle point of order $2m+l-1$. When $\x\neq 0$ and $\y=0$, $\tau=0$ is a saddle point of order $2m-1$. When 
$\y\neq 0$, $\tau=0$ is a saddle point of order $m-1$. 
Non-zero saddle points are non-zero solutions of
\begin{align}
\label{eqn:StatPoints}
- \y  -\x\tau^{m} + \alpha\tau^{m+l}=0,
\end{align}
and, eliminating $\y$ between \rf{eqn:CoalescePrelim} and \rf{eqn:StatPoints}, we see that, for $m\geq 1$, localisation occurs where, in addition to \rf{eqn:StatPoints}, we have that 
\begin{align}
\label{eqn:Coalesce}
-m\x + \alpha(m+l)\tau^{l}=0.
\end{align}
Eliminating $\tau$ between \rf{eqn:StatPoints} and \rf{eqn:Coalesce} it follows that, if we take
\begin{align}
\label{eqn:alphaDef}
\alpha = \frac{1}{\kappa}\frac{m}{l+m}\left(\frac{m}{l}\right)^{l/m},
\end{align}
then, where localisation occurs, it does so on the curve \rf{eqn:Bdy}, with $\tau^l =(l/m)^{l/m}\kappa \x$. 

Whether or not localisation actually occurs on \rf{eqn:Bdy} depends on whether the steepest descent contour passes through the relevant saddle points, which, in turn, depends on the choice of initial contour $\Gamma$. 
When localisation does occur, and corresponds to a second order saddle point, according to the pioneering paper \cite{ChFrUr:57} the far-field behaviour should be described by an Airy function whose argument is proportional to distance normal to the localisation curve \rf{eqn:Bdy}.  
When $m\geq 2$ other kinds of localisation are also possible because of the existence of the saddle point of order $2m-1$ at $\tau=0$ when $\y=0$. 
In the next section we investigate these issues in the special case where $F(t)\equiv 1$. Then in \S\ref{sec:BVPs} and \S\ref{sec:Discussion} we consider the more general case where $F(t)\not\equiv 1$, where the possible asymptotic behaviour can be more complicated. In particular we note that if $F(t)$ behaves exponentially at infinity, the localisation curve is shifted away from \rf{eqn:Bdy}. Furthermore, if $F(t)$ has singularities then these may interact with the saddle points to generate further localisation phenomena. 

\section{\label{sec:poly}The case $F(t)\equiv 1$}
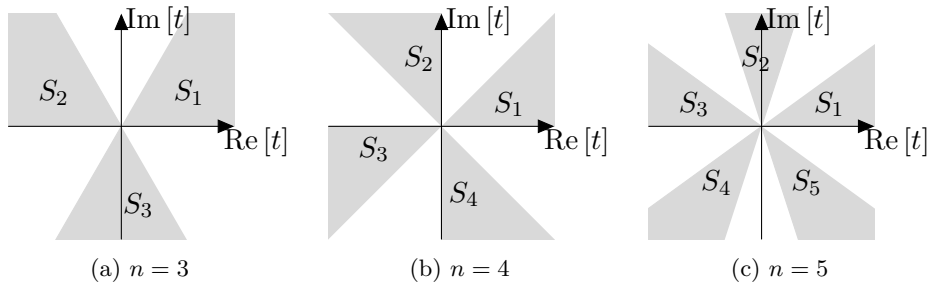
\begin{figure}[t]
\colorlet{lightgray}{black!15}
\begin{center}
\subfigure[\hs{-5}(a) $n=3$]{
\begin{tikzpicture}[line cap=round,line join=round,>=triangle 45,x=1.0cm,y=1.0cm, scale=2]
\def\xm{0.75};
\def\ym{0.75};
\pgfmathsetmacro{\rad}{2*sqrt(\xm^2+\ym^2)}
\def\theta{60};
\pgfmathsetmacro{\costheta}{cos(\theta)};
\pgfmathsetmacro{\sintheta}{sin(\theta)};
\pgfmathsetmacro{\costwotheta}{cos(2*\theta)};
\pgfmathsetmacro{\sintwotheta}{sin(2*\theta)};
\pgfmathsetmacro{\costhreetheta}{cos(3*\theta)};
\pgfmathsetmacro{\sinthreetheta}{sin(3*\theta)};
\pgfmathsetmacro{\cosfourtheta}{cos(4*\theta)};
\pgfmathsetmacro{\sinfourtheta}{sin(4*\theta)};
\pgfmathsetmacro{\cosfivetheta}{cos(5*\theta)};
\pgfmathsetmacro{\sinfivetheta}{sin(5*\theta)};
\pgfmathsetmacro{\cossixtheta}{cos(6*\theta)};
\pgfmathsetmacro{\sinsixtheta}{sin(6*\theta)};
\begin{scope}
\clip (-\xm,-\ym) rectangle (\xm,\ym);
	\filldraw [lightgray] (0,0) -- (\rad,0) -- (\rad*\costheta,\rad*\sintheta) ;
	\filldraw [lightgray] (0,0) -- (\rad*\costwotheta,\rad*\sintwotheta) -- (\rad*\costhreetheta,\rad*\sinthreetheta) ;
	\filldraw [lightgray] (0,0) -- (\rad*\cosfourtheta,\rad*\sinfourtheta) -- (\rad*\cosfivetheta,\rad*\sinfivetheta) ;
\end{scope}
\draw (\xm+0.15,-0.1) node {$\real{ t}$};
\draw [->] (-\xm,0) -- (\xm,0);
\draw [->] (0,-\ym) -- (0,\ym);
\draw (0.25,\ym-0.05) node {$\im{ t}$};
\draw (0.6*\xm,0.3*\ym) node {$S_1$};
\draw (-0.6*\xm,0.3*\ym) node {$S_2$};
\draw (0.15*\xm,-0.7*\ym) node {$S_3$};
\end{tikzpicture} 
}
\subfigure[\hs{-5}(b) $n=4$]{
\begin{tikzpicture}[line cap=round,line join=round,>=triangle 45,x=1.0cm,y=1.0cm, scale=2]
\def\xm{0.75};
\def\ym{0.75};
\pgfmathsetmacro{\rad}{2*sqrt(\xm^2+\ym^2)}
\def\theta{45};
\pgfmathsetmacro{\costheta}{cos(\theta)};
\pgfmathsetmacro{\sintheta}{sin(\theta)};
\pgfmathsetmacro{\costwotheta}{cos(2*\theta)};
\pgfmathsetmacro{\sintwotheta}{sin(2*\theta)};
\pgfmathsetmacro{\costhreetheta}{cos(3*\theta)};
\pgfmathsetmacro{\sinthreetheta}{sin(3*\theta)};
\pgfmathsetmacro{\cosfourtheta}{cos(4*\theta)};
\pgfmathsetmacro{\sinfourtheta}{sin(4*\theta)};
\pgfmathsetmacro{\cosfivetheta}{cos(5*\theta)};
\pgfmathsetmacro{\sinfivetheta}{sin(5*\theta)};
\pgfmathsetmacro{\cossixtheta}{cos(6*\theta)};
\pgfmathsetmacro{\sinsixtheta}{sin(6*\theta)};
\pgfmathsetmacro{\cosseventheta}{cos(7*\theta)};
\pgfmathsetmacro{\sinseventheta}{sin(7*\theta)};
\pgfmathsetmacro{\coseighttheta}{cos(8*\theta)};
\pgfmathsetmacro{\sineighttheta}{sin(8*\theta)};
\begin{scope}
\clip (-\xm,-\ym) rectangle (\xm,\ym);
		\filldraw [lightgray] (0,0) -- (\rad,0) -- (\rad*\costheta,\rad*\sintheta) ;
	\filldraw [lightgray] (0,0) -- (\rad*\costwotheta,\rad*\sintwotheta) -- (\rad*\costhreetheta,\rad*\sinthreetheta) ;
	\filldraw [lightgray] (0,0) -- (\rad*\cosfourtheta,\rad*\sinfourtheta) -- (\rad*\cosfivetheta,\rad*\sinfivetheta) ;
		\filldraw [lightgray] (0,0) -- (\rad*\cossixtheta,\rad*\sinsixtheta) -- (\rad*\cosseventheta,\rad*\sinseventheta) ;

\end{scope}
\draw (\xm+0.15,-0.1) node {$\real{ t}$};
\draw [->] (-\xm,0) -- (\xm,0);
\draw [->] (0,-\ym) -- (0,\ym);
\draw (0.25,\ym-0.05) node {$\im{ t}$};
\draw (0.6*\xm,0.2*\ym) node {$S_1$};
\draw (-0.2*\xm,0.6*\ym) node {$S_2$};
\draw (-0.6*\xm,-0.2*\ym) node {$S_3$};
\draw (0.2*\xm,-0.6*\ym) node {$S_4$};
\end{tikzpicture} 
}
\subfigure[\hs{-5}(c) $n=5$]{
\begin{tikzpicture}[line cap=round,line join=round,>=triangle 45,x=1.0cm,y=1.0cm, scale=2]
\def\xm{0.75};
\def\ym{0.75};
\pgfmathsetmacro{\rad}{2*sqrt(\xm^2+\ym^2)}
\def\theta{36};
\pgfmathsetmacro{\costheta}{cos(\theta)};
\pgfmathsetmacro{\sintheta}{sin(\theta)};
\pgfmathsetmacro{\costwotheta}{cos(2*\theta)};
\pgfmathsetmacro{\sintwotheta}{sin(2*\theta)};
\pgfmathsetmacro{\costhreetheta}{cos(3*\theta)};
\pgfmathsetmacro{\sinthreetheta}{sin(3*\theta)};
\pgfmathsetmacro{\cosfourtheta}{cos(4*\theta)};
\pgfmathsetmacro{\sinfourtheta}{sin(4*\theta)};
\pgfmathsetmacro{\cosfivetheta}{cos(5*\theta)};
\pgfmathsetmacro{\sinfivetheta}{sin(5*\theta)};
\pgfmathsetmacro{\cossixtheta}{cos(6*\theta)};
\pgfmathsetmacro{\sinsixtheta}{sin(6*\theta)};
\pgfmathsetmacro{\cosseventheta}{cos(7*\theta)};
\pgfmathsetmacro{\sinseventheta}{sin(7*\theta)};
\pgfmathsetmacro{\coseighttheta}{cos(8*\theta)};
\pgfmathsetmacro{\sineighttheta}{sin(8*\theta)};
\pgfmathsetmacro{\cosninetheta}{cos(9*\theta)};
\pgfmathsetmacro{\sinninetheta}{sin(9*\theta)};
\pgfmathsetmacro{\costentheta}{cos(10*\theta)};
\pgfmathsetmacro{\sintentheta}{sin(10*\theta)};
\begin{scope}
\clip (-\xm,-\ym) rectangle (\xm,\ym);
	\filldraw [lightgray] (0,0) -- (\rad,0) -- (\rad*\costheta,\rad*\sintheta) ;
	\filldraw [lightgray] (0,0) -- (\rad*\costwotheta,\rad*\sintwotheta) -- (\rad*\costhreetheta,\rad*\sinthreetheta) ;
	\filldraw [lightgray] (0,0) -- (\rad*\cosfourtheta,\rad*\sinfourtheta) -- (\rad*\cosfivetheta,\rad*\sinfivetheta) ;
	\filldraw [lightgray] (0,0) -- (\rad*\cossixtheta,\rad*\sinsixtheta) -- (\rad*\cosseventheta,\rad*\sinseventheta) ;
		\filldraw [lightgray] (0,0) -- (\rad*\coseighttheta,\rad*\sineighttheta) -- (\rad*\cosninetheta,\rad*\sinninetheta) ;
\end{scope}
\draw (\xm+0.15,-0.1) node {$\real{ t}$};
\draw [->] (-\xm,0) -- (\xm,0);
\draw [->] (0,-\ym) -- (0,\ym);
\draw (0.25,\ym-0.05) node {$\im{ t}$};
\draw (0.6*\xm,0.2*\ym) node {$S_1$};
\draw (-0.05*\xm,0.6*\ym) node {$S_2$};
\draw (-0.6*\xm,0.2*\ym) node {$S_3$};
\draw (-0.4*\xm,-0.5*\ym) node {$S_4$};
\draw (0.4*\xm,-0.5*\ym) node {$S_5$};
\end{tikzpicture} 
}
\caption{Sectors (shaded) in the complex $t$-plane in which $\im{t^n}$ is positive (so that $\re^{\ri t^n}$ decays exponentially as $|t|\to\infty$ with $\arg{t}$ fixed) in the cases $n=3,4,5$. %
}
\label{Sectors}
\end{center}
\end{figure}

When $F(t)\equiv 1$, 
the set of permissible integration contours $\Gamma$ in \rf{eqn:AInt} can be described explicitly. Since $\alpha>0$, a permissible integration contour $\Gamma$ must begin and end at infinity in one of the sectors $S_j:=\{t:(2j-2)\pi/(2m+l)< \arg{t} < (2j-1)\pi/(2m+l)\}$, $j\in\{1,\ldots,(2m+l)\}$  
(see \F\ref{Sectors} for an illustration for $n:=2m+l=3,4,5$). 
Let $A_{ij}$ denote \rf{eqn:AInt} with integration contour $\Gamma_{ij}$ beginning at infinity in $S_i$ and ending at infinity in $S_j$. Clearly $A_{ij}=-A_{ji}$, so there are, a priori, $\binom{2m+l}{2}$ different solutions. But also $A_{ij}+ A_{jk} = A_{ik}$, $i,j,k\in\{1,2,\ldots,2m+l\}$,
so that in particular %
$A_{21}+A_{32}+\ldots+A_{(2m+l)(2m+l-1)}+A_{1(2m+l)}=0$, and hence only $2m+l-1$ of the $A_{ij}$ are linearly independent. 
Furthermore, the assumption that $F(t)\equiv 1$ implies further useful symmetry relations satisfied by $A_{ij}$.  
Specifically, the fact that 
\begin{align}
\label{eqn:pRelnGeneral}
p(\X,\Y,t)=(-1)^l p\big((-1)^l \X,(-1)^{m+l}\Y,-t\big)
\end{align}
means that
\begin{align}
\label{eqn:SymmetryGeneral}
A_{ij}(\X,\Y)=
\begin{cases}
-\overline{A_{i'j'}(-\X,-(-1)^{m} \Y)}, & l\textrm{ odd},\\
-A_{i''j''}(\X,(-1)^m \Y), & l\textrm{ even},
\end{cases}
\end{align}
where for $l$ odd $\Gamma_{i'j'}$ is the reflection of $\Gamma_{ij}$ in the imaginary $t$-axis, and for $l$ even $\Gamma_{i''j''}$ is the image of $\Gamma_{ij}$ under the map $t\to -t$. %

In \S\ref{sec:i} and \S\ref{sec:ii} we review the well-studied cases $(l,m)=(1,1)$ (parabolic caustic) and $(l,m)=(2,1)$ (cusped caustic) in a format that is helpful for understanding the case $(l,m)=(1,2)$ (the cubic parabola), which will be considered in \S\ref{sec:Inflection}.

\subsection{Parabola ($(l,m)=(1,1)$, $2m+l=3$, $\lambda = 1/3$, $\alpha=1/(2\kappa)$)}
\label{sec:i}

\begin{figure}[t!]
\begin{center}
\subfigure[(a)]{\includegraphics[width=32mm]{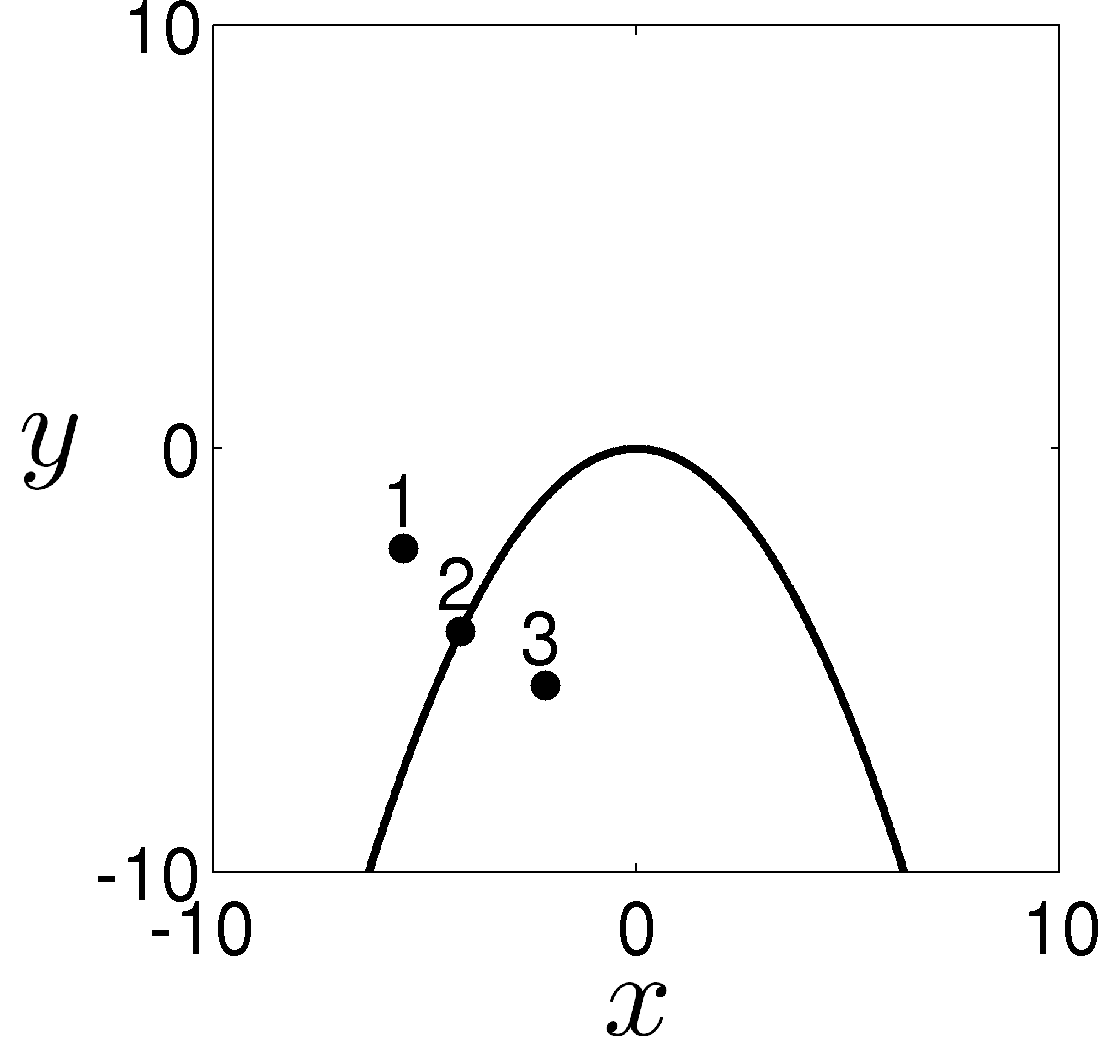}
}
\subfigure[(b) Point 1]{\includegraphics[width=30mm]{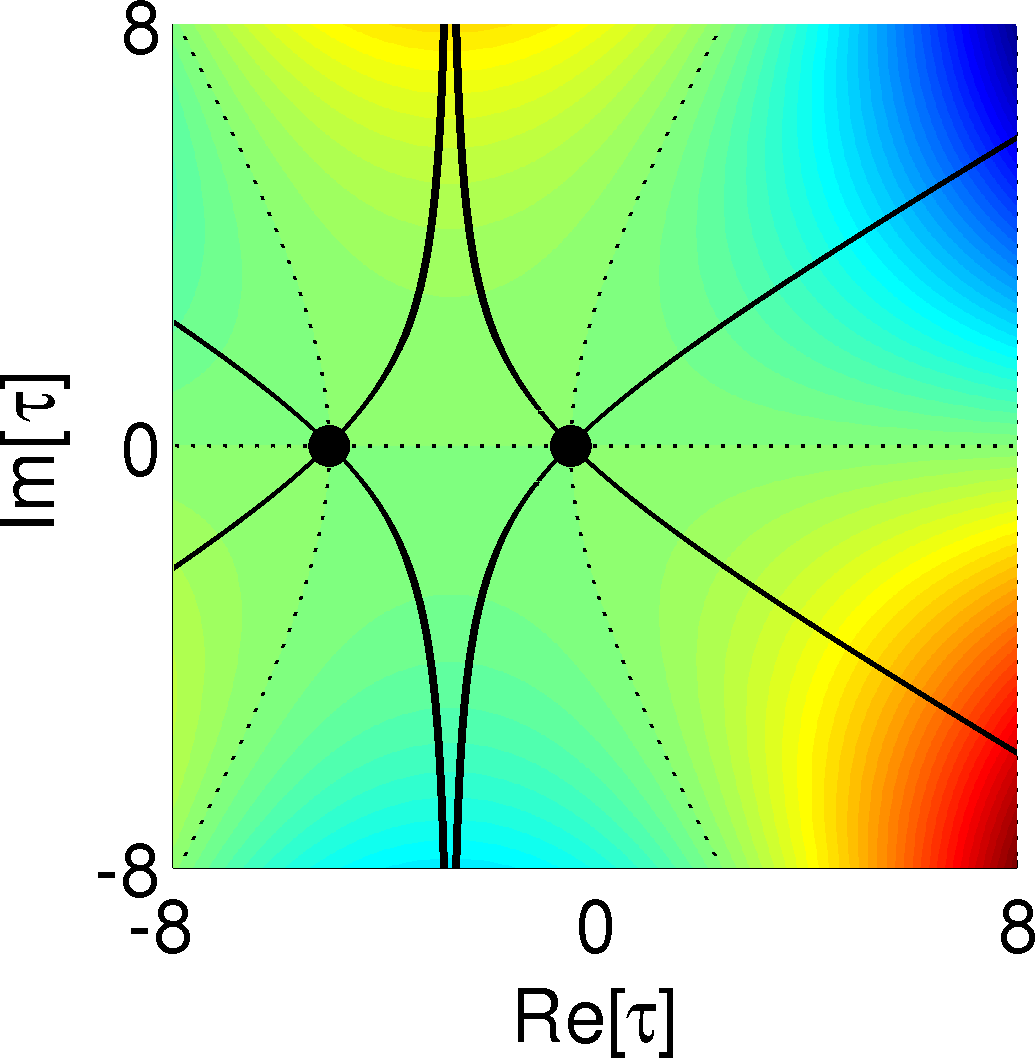}
}
\subfigure[(c) Point 2]{\includegraphics[width=30mm]{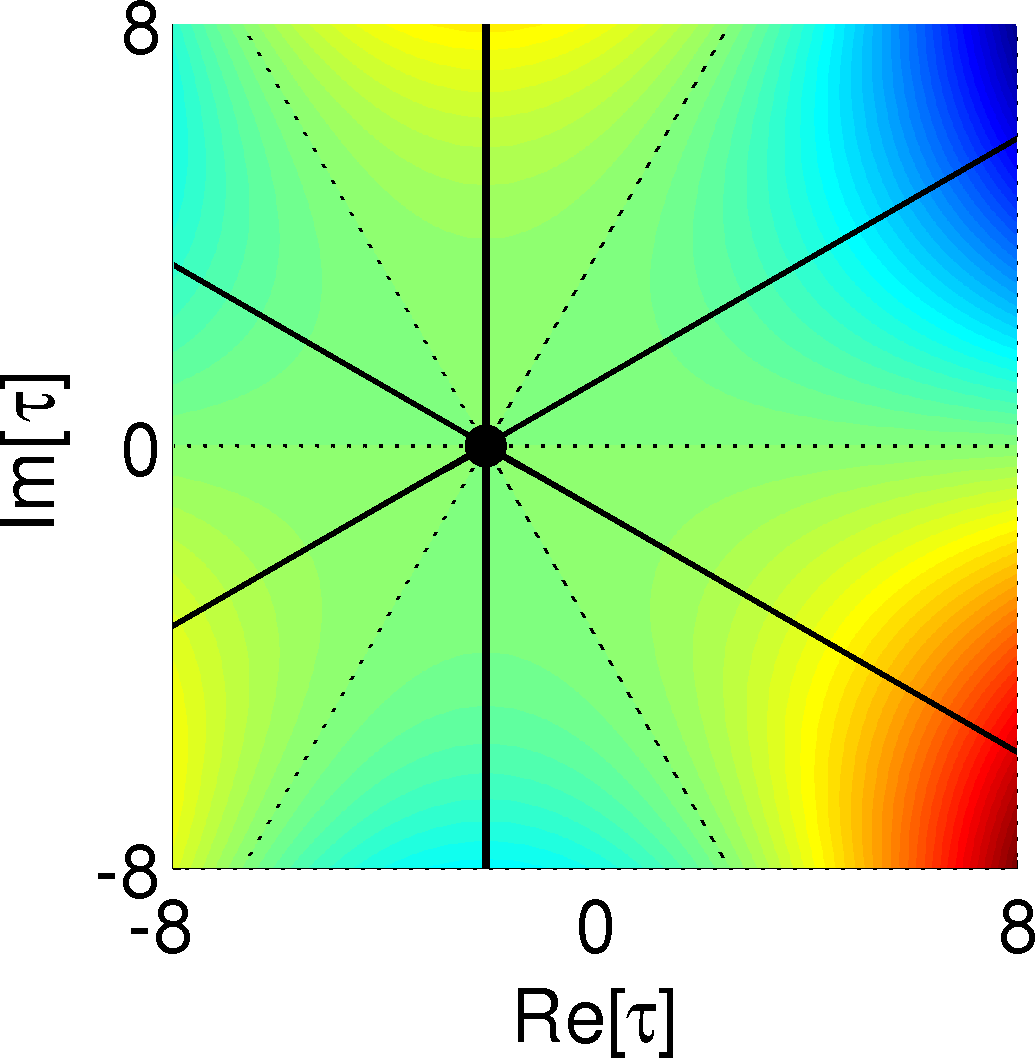}
}
\subfigure[(d) Point 3]{\includegraphics[width=30mm]{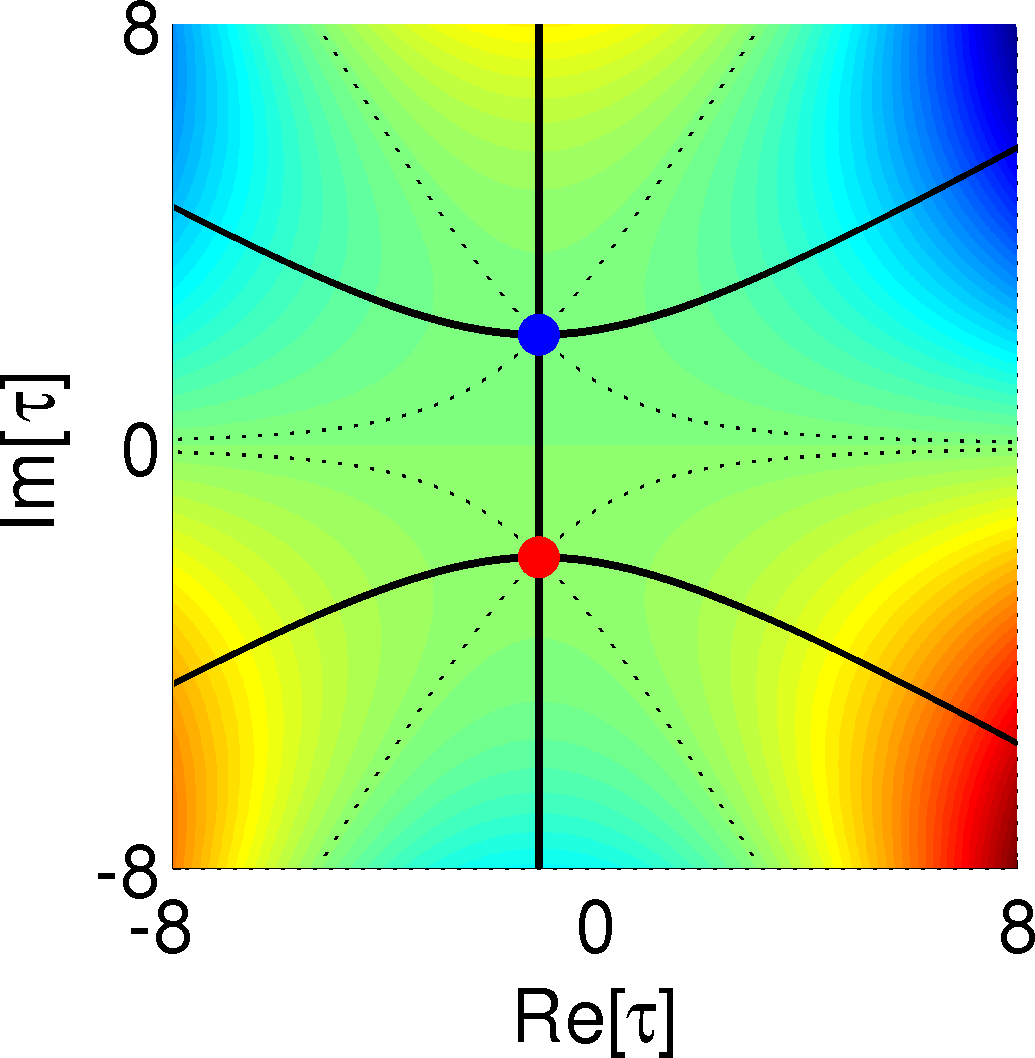}
}
\caption{Saddle point configurations for $(l,m)=(1,1)$ 
with $\alpha=1$. (a) The localisation curve $\y=-\x^2/4$. (b)-(d) Plots of $\im{\phi(\tau)}$ at the points 1-3 in (a), with red corresponding to $\im{\phi}<0$ (i.e. a `hill' in the real part of the integrand) and blue to $\im{\phi}>0$ (i.e. a `valley'). The saddle points are coloured as follows: red $\Rightarrow$ $\im{\phi}<0$ (giving an exponentially large contribution in the far-field), blue $\Rightarrow$ $\im{\phi}>0$ (exponentially small) and black $\Rightarrow$ $\im{\phi}=0$. The curves of constant $\real{\phi}$ (solid line) and constant $\im{\phi}$ (dotted line) through each of the saddle points are also plotted.}
\label{fig:AirySaddlesFull}
\end{center}
\end{figure}

This well-studied case, in which localisation is near $\Y+\kappa \X^2/2=0$, is the prototype for all our examples. 
With $F(t)\equiv1$, we have the solutions 
\begin{align}
\label{eqn:AIntParabolic}
A_{ij} = \int_{\Gamma_{ij}} \re^{\ri (-\Y t-\X t^2/2 + \alpha t^3/3)} \,\rd t = \int_{\Gamma_{ij}} \re^{\ri (-\Y t-\X t^2/2 + t^3/(6\kappa))} \,\rd t,
\end{align}
where $\Gamma_{ij}$ goes from $S_i$ to $S_j$ in \F\ref{Sectors}(a). 
The scalings \rf{eqn:scaling2} are $\X=k^{1/3} \x$, $\Y=k^{2/3}\y$, $t = k^{1/3} \tau$,
and the saddle points (the solutions of the quadratic equation \rf{eqn:StatPoints}) are at %
\begin{align}
\label{eqn:ParabSaddles}
\tau_\pm=\kappa\left(\x\pm\sqrt{\x^2+2\y/\kappa}\right),%
\end{align}
where, for definiteness, we assume the principal branch of the square root. There are then three possibilities, illustrated in \F\ref{fig:AirySaddlesFull} for the case $\alpha=1$ ($\kappa=1/2$):
\begin{itemize}
\item 
If $\x^2+2\y/\kappa>0$ (point 1 in \F\ref{fig:AirySaddlesFull}(a)), 
there are two distinct real saddle points (see \F\ref{fig:AirySaddlesFull}(b)). 
\item
If $\x^2+2\y/\kappa=0$ (point 2 in \F\ref{fig:AirySaddlesFull}(a)), 
there is a double real saddle point (see \F\ref{fig:AirySaddlesFull}(c)). 
\item
If $\x^2+2\y/\kappa<0$ (point 3 in \F\ref{fig:AirySaddlesFull}(a)), 
there is a pair of complex conjugate saddle points (see \F\ref{fig:AirySaddlesFull}(d)). 
\end{itemize}

\begin{figure}[t!]
\begin{center}
\hs{-5}
\subfigure[(a) $|A_{21}|$]{\includegraphicsDave[width=40mm]{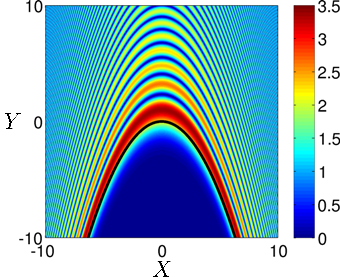}
}
\subfigure[(b) $\real{A_{21}\re^{\ri k\x}}$]{\includegraphicsDave[width=88mm]{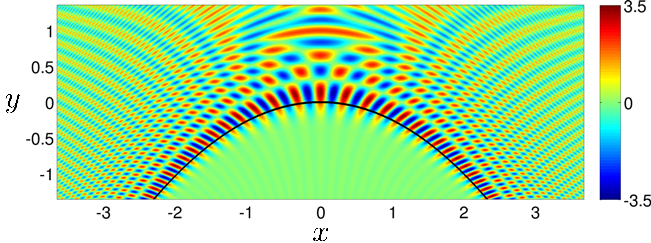}
}\\
\hs{-5}
\subfigure[(c) $|A_{32}|$]{\includegraphicsDave[width=40mm]{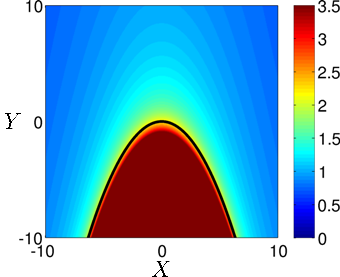}
}
\subfigure[(d) $\real{A_{32}\re^{\ri k\x}}$]{\includegraphicsDave[width=88mm]{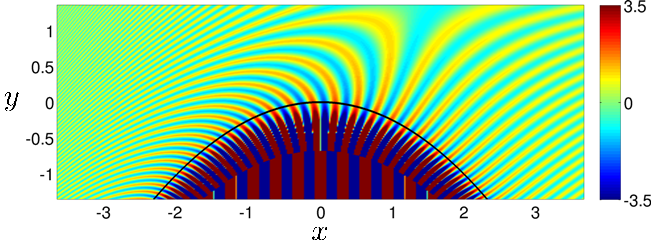}
}
\caption{Plots of $|A(\X,\Y)|$ and $\real{A\re^{\ri k \x}}$ for $(l,m)=(1,1)$, with $\lambda=l/(l+2m)=1/3$ and $k=20$. The localisation curve is superimposed in black. (The plotting artefacts in (d) in the field under the curve are caused by the fact that the field is exponentially large here.)}
\label{fig:AiryFieldPlots}
\end{center}
\end{figure}

The far-field behaviour of \rf{eqn:AIntParabolic} depends on the choice of integration contour. 
There are $\binom{3}{2}=3$ solutions $A_{ij}$, two of which are linearly independent, satisfying the symmetry relations (cf.\ \rf{eqn:SymmetryGeneral}) 
\begin{align}
A_{21}(\X,\Y) = \overline{A_{21}(-\X,\Y)}, \qquad
\label{eqn:AirySym}
A_{13}(\X,\Y) = \overline{A_{32}(-\X,\Y)}.
\end{align}
Furthermore, \rf{eqn:pRelnGeneral} implies that the steepest descent contour for $A_{21}(\X,\Y)$ can be obtained from that for $A_{21}(-\X,\Y)$ by reflection in the imaginary $\tau$-axis. A similar relationship exists between the steepest descent contours for $A_{13}(\X,\Y)$ and $A_{32}(-\X,\Y)$. %
As a result, there are essentially only two distinct far-field behaviours (modulo reflections in the $\Y$-axis), namely that of $A_{21}$ and $A_{32}$. Plots of $|A|$ and the real part of the associated approximate solutions $A\re^{\ri k\x}$ of the Helmholtz equation are given for both cases in \F\ref{fig:AiryFieldPlots}.  
Schematic illustrations, along with typical steepest descent contours, can be found in \Fs\ref{fig:AirySaddlesA21} and \ref{fig:AirySaddlesA32}. 

\def\psize{20mm}
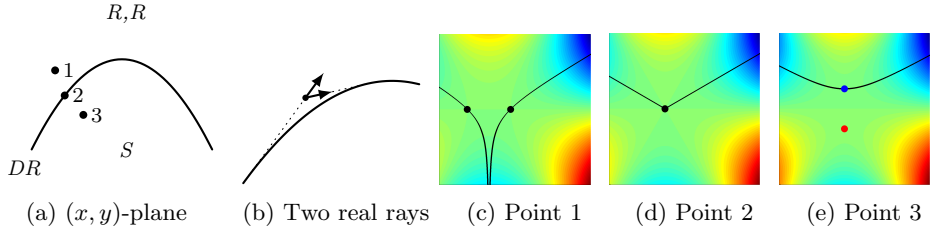
\begin{figure}[t!]
\begin{center}
\subfigure[(a) $(\x,\y)$-plane] {
\begin{tikzpicture}[line cap=round,line join=round,>=triangle 45,x=0.3cm,y=0.3cm]
\def\size{4};
\def\Ta{R,R};\def\Tb{DR\quad};\def\Tc{S};
\draw (0,\size/2) node {\tf\Ta};
\draw[thick, smooth,domain=-\size:\size,variable=\t] plot({-\t},{-1*(\t)^2/4}) node[below] {\tf\Tb};
\draw (0,-\size) node {\tf\Tc};
\begin{scriptsize}
\fill (-2.9594,-0.49197) circle (1.5pt);
\draw (-2.3594,-0.49197) node {1};
\fill (-2.5342,-1.6056) circle (1.5pt);
\draw (-1.9342,-1.6056) node {2};
\fill (-1.7089,-2.4657) circle (1.5pt);
\draw (-1.1089,-2.4657) node {3};
\end{scriptsize}
\end{tikzpicture}
}
\subfigure[(b) Two real rays] {
\begin{tikzpicture}[line cap=round,line join=round,>=latex,x=0.3cm,y=0.3cm,scale=0.6]
\draw[thick,smooth,domain=-11:2,variable=\t] plot({\t},{-1*((\t)^2/16)});
\draw [thick,->] (-6.4,-1.25) -- (-4.4,-0.8);
\draw [thick,->] (-6.4,-1.25) -- (-5.05,0.6);
\draw [dotted] (-6.4,-1.25)-- (-10.97,-7.52);
\draw [dotted] (-4.4,-0.8)-- (-1.83,-0.21);
\fill (-6.4,-1.25) circle (2pt);
\end{tikzpicture}
}
\subfigure[(c) Point 1]{\includegraphics[width=\psize]{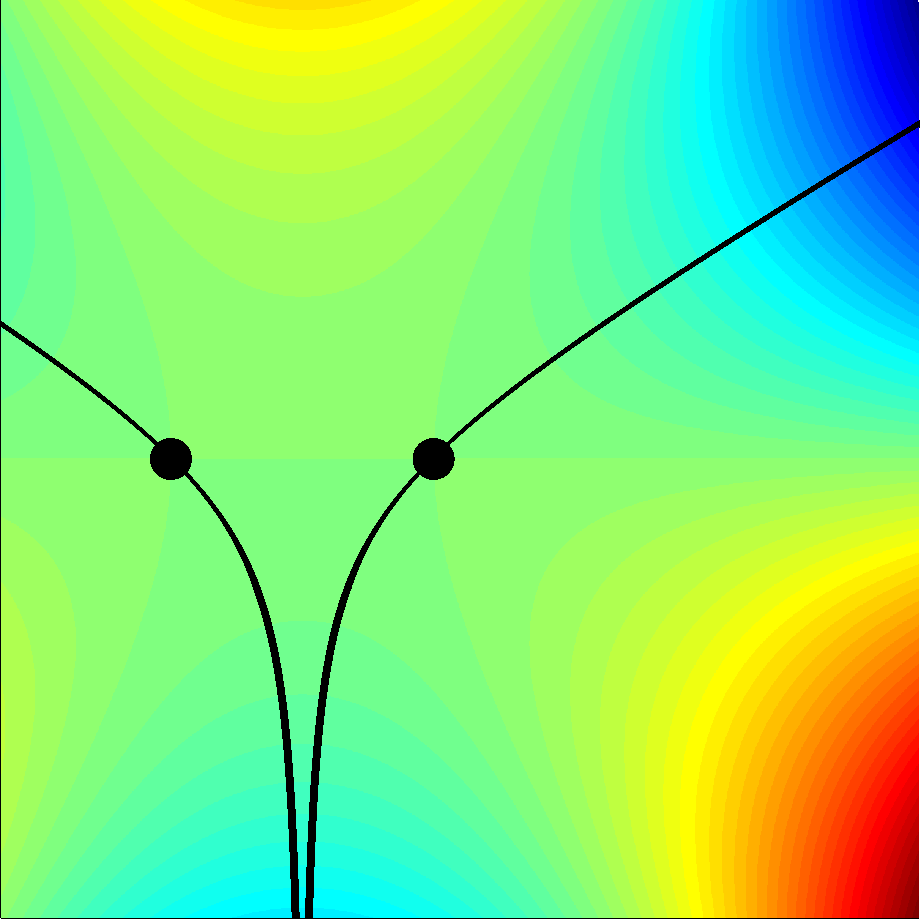}
}
\subfigure[(d) Point 2]{\includegraphics[width=\psize]{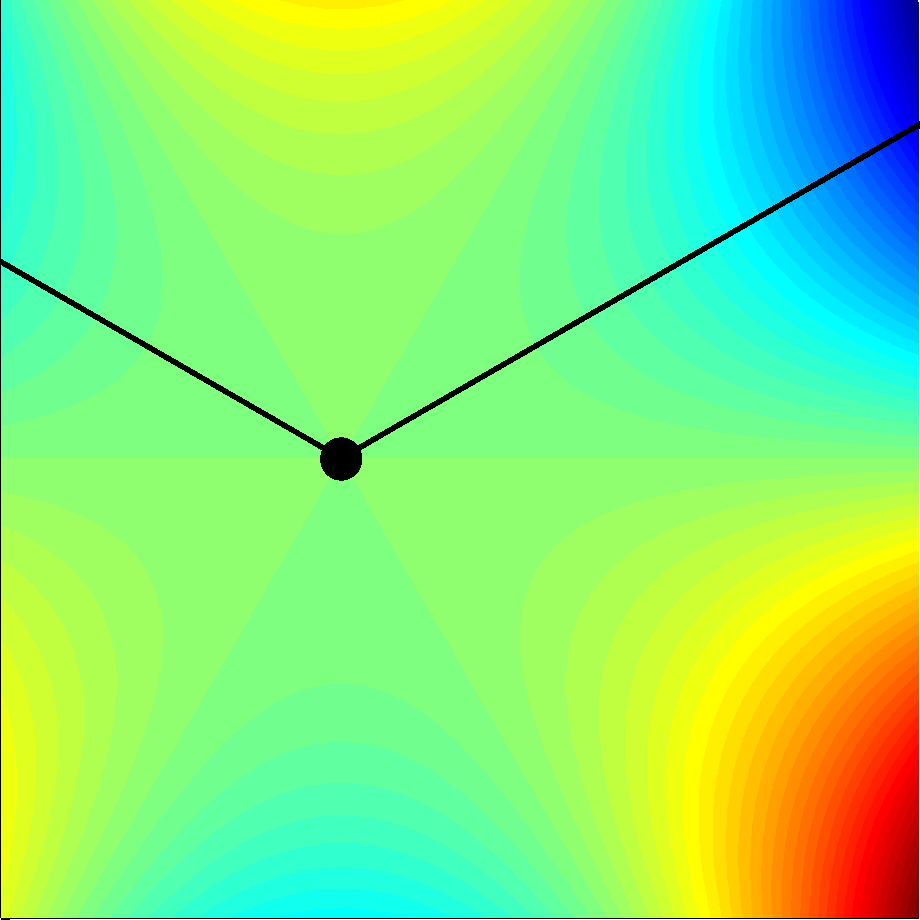}
}
\subfigure[(e) Point 3]{\includegraphics[width=\psize]{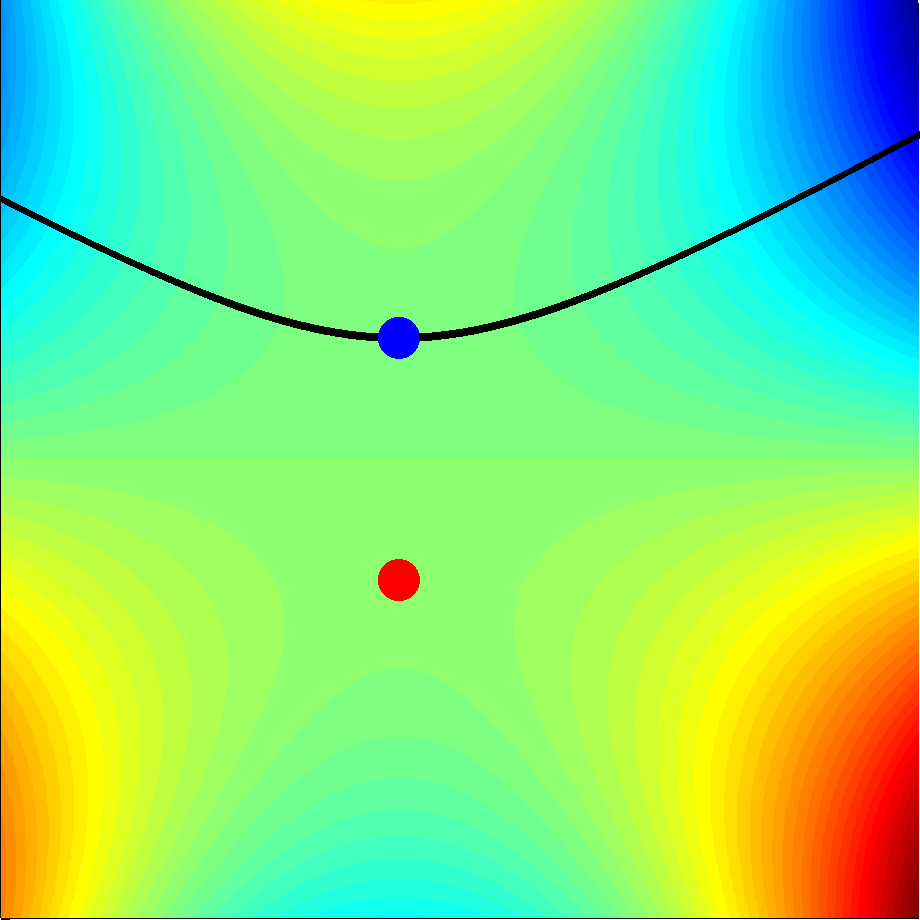}
}
\caption{Steepest descent contours for $A_{21}$ in the case $(l,m)=(1,1)$, $\alpha=1$, for points 1-3 from \F\ref{fig:AirySaddlesFull}(a). Key to (a): R=real, DR=double real, S=exponentially small.
}
\label{fig:AirySaddlesA21}
\end{center}
\end{figure}

The solution $A_{21}$ (\F\ref{fig:AiryFieldPlots}(a)-(b)) describes the field near a smooth (parabolic) caustic. 
Indeed, deforming the contour $\Gamma_{21}$ onto the real axis and making a change of variable reveals that
\begin{align}
\label{eqn:AAiry}
A_{21}=\int_{-\infty}^\infty \re^{\ri(-\Y t-\X t^2/2+t^3/(6\kappa))} \,\rd t
= 2\pi (2\kappa)^{1/3} \re^{-\ri\kappa(\X\Y+\kappa \X^3/3)}\Ai\left[-(2\kappa)^{1/3}\left(\Y+\frac{\kappa \X^2}{2}\right)\right],
\end{align}
where $\Ai$ is the Airy function 
\begin{align}
\label{AiryRealDef}
\Ai(z)=\frac{1}{2\pi}\int_{-\infty}^\infty \re^{\ri(tz+t^3/3)}\,\rd t, \qquad z\in\R. 
\end{align}
Formula \rf{eqn:AAiry} is valid for all $(\X,\Y)$, even on the localisation curve $\Y+\kappa \X^2/2=0$, and it permits evaluation of the far-field behaviour using the well-known asymptotics of the Airy function. However, we want to emphasize that, even without 
\rf{eqn:AAiry},
one can determine the far-field behaviour 
by a more general direct asymptotic analysis of the integral $A_{21}$. 
We first note that:
\begin{itemize}
\item 
Above the caustic (cf.\ \F\ref{fig:AirySaddlesA21}(c)) the steepest descent contour passes through the two real saddle points $\tau_\pm$ corresponding to the two families of real rays that envelop the caustic. For the solution of the Helmholtz equation to which the approximation $A_{21}\re^{\ri k\x}$ matches, one can check that $\tau_-$ corresponds to the ray outgoing from the caustic, and $\tau_+$ to the ray incoming towards the caustic, 
as illustrated in the familiar ray picture in \F\ref{fig:AirySaddlesA21}(b). %
\item On the caustic the steepest descent contour comprises the two segments in \F\ref{fig:AirySaddlesA21}(d), the degenerate saddle describing the localisation first mentioned in \cite{ChFrUr:57}. 
\item
Below the caustic (cf.\ \F\ref{fig:AirySaddlesA21}(e)) the only saddle point encountered by the steepest descent contour is 
$\tau_+ = \kappa(\x+\ri \sqrt{-(\x^2+2\y/\kappa)})$, 
where the imaginary part of the phase \rf{eqn:Phase} is equal to 
$(\kappa^2/3)(-(\x^2+2 \y/\kappa))^{3/2}>0$. 
This saddle point corresponds to the exponentially decaying wave generated by the complex rays that exists below the parabola. (This qualitatively resembles a whispering gallery wave - for an explicit connection see \S\ref{sec:BVPs}.)
\end{itemize}
Then, to investigate the localisation behaviour quantitatively we let
\begin{align*}
\label{}
\x = \x_0+\delta \x^*, \quad \y = -\kappa \x_0^2/2+\delta \y^*, \quad \tau = \kappa \x_0 + \tau',
\end{align*}
where $\delta,\tau'$ are small when $k\gg1$, in a way that will shortly be made precise. Deforming $\Gamma_{21}$ to the real axis, and assuming without loss of generality (cf.~\rf{eqn:AirySym}) that $\x_0>0$, the phase is then
\begin{align}
\label{eqn:AiryPhase}
\frac{\kappa^2 \x_0^3}{6}-\delta\left( \kappa \x_0 \y^*+\frac{\kappa^2 \x_0^2}{2}\x^*\right) - \tau'\left( \delta \y^*+\delta \kappa \x_0 \x^*\right) - \frac{(\tau')^2\delta \x^*}{2} +\frac{(\tau')^3}{6\kappa}. 
\end{align}
Taking $\delta=k^{-2/3}$ and $\tau'=k^{-1/3}(2\kappa)^{1/3}\zeta$ gives the lowest order approximation to $A_{21}$ as%
\begin{align}
(2\kappa)^{1/3}\exp\left[\ri\left(k\frac{\kappa^2 \x_0^3}{6}\right.\right. & \left.\left. - \,\, k^{1/3}\left( \kappa \x_0 \y^* + \frac{1}{2}\kappa^2 \x_0^2 \x^* \right)\right)\right]\notag \\
&\times\int_{-\infty}^\infty \exp{\left[ \ri\left(-(2\kappa)^{1/3}(\y^*+\kappa \x_0 \x^*)\zeta + \frac{\zeta^3}{3}\right) \right]}\,\rd \zeta \notag\\
 = 2\pi (2\kappa)^{1/3}\exp\left[\ri\left(k\frac{\kappa^2 \x_0^3}{6}\right.\right. & \left.\left. - \,\, k^{1/3}\left( \kappa \x_0 \y^* + \frac{1}{2}\kappa^2 \x_0^2 \x^* \right)\right)\right]\Ai\left[-(2\kappa)^{1/3}(\y^*+\kappa \x_0 \x^*)\right].
 \label{eqn:AiryAsympt}
\end{align}
Note that Airy functions describe both the exact PWE solution \rf{eqn:AAiry} and the far field behaviour \rf{eqn:AiryAsympt} in the vicinity of the parabola $\Y+\kappa \X^2/2=0$ on which the PWE solution is localised. Indeed, the term $\y^*+\kappa \x_0\x^*$ in \rf{eqn:AiryAsympt} is proportional to distance along the normal to the parabola, and is the local expansion of the normal distance as measured by $\Y+\kappa \X^2/2$ in \rf{eqn:AAiry}. This verifies that \rf{eqn:AiryAsympt} agrees with the local expansion of \rf{eqn:AAiry} near $\X=k^{1/3}\x_0$, $\Y=k^{2/3}\y_0$. As shown in \cite{ChFrUr:57}, it would have been possible to retrieve the full solution \rf{eqn:AAiry} by including all the terms in the stationary phase expansion leading to \rf{eqn:AiryAsympt}.
Only the curvature of the parabola at $\Y=0$ (given by $\kappa$) determines the far field wave structure.
In view of the scalings leading to \rf{eqn:PWE}, the PWE solution is only valid for \mbox{$\X\ll k^{-1/3}$} and $\Y\ll k^{-2/3}$, 
and to extend the solution beyond this regime we would have to use appropriate curvilinear coordinates, as in \cite{OckTew:12}.

\def\psize{20mm}
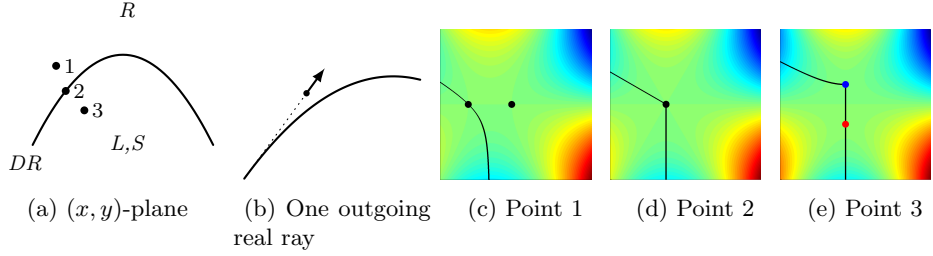
\begin{figure}[t!]
\begin{center}
\subfigure[(a) $(\x,\y)$-plane] {
\begin{tikzpicture}[line cap=round,line join=round,>=triangle 45,x=0.3cm,y=0.3cm]
\def\size{4};
\def\Ta{R};\def\Tb{DR\quad};\def\Tc{L,S};
\draw (0,\size/2) node {\tf\Ta};
\draw[thick, smooth,domain=-\size:\size,variable=\t] plot({-\t},{-1*(\t)^2/4}) node[below] {\tf\Tb};
\draw (0,-\size) node {\tf\Tc};
\begin{scriptsize}
\fill (-2.9594,-0.49197) circle (1.5pt);
\draw (-2.3594,-0.49197) node {1};
\fill (-2.5342,-1.6056) circle (1.5pt);
\draw (-1.9342,-1.6056) node {2};
\fill (-1.7089,-2.4657) circle (1.5pt);
\draw (-1.1089,-2.4657) node {3};
\end{scriptsize}
\end{tikzpicture}
}
\subfigure[(b) One outgoing real ray] {
\begin{tikzpicture}[line cap=round,line join=round,>=latex,x=0.3cm,y=0.3cm,scale=0.6]
\draw[thick,smooth,domain=-11:2,variable=\t] plot({\t},{-1*((\t)^2/16)});
\draw [thick,->] (-6.4,-1.25) -- (-5.05,0.6);
\draw [dotted] (-6.4,-1.25)-- (-10.97,-7.52);
\fill (-6.4,-1.25) circle (2pt);
\end{tikzpicture}
}
\subfigure[(c) Point 1]{\includegraphics[width=\psize]{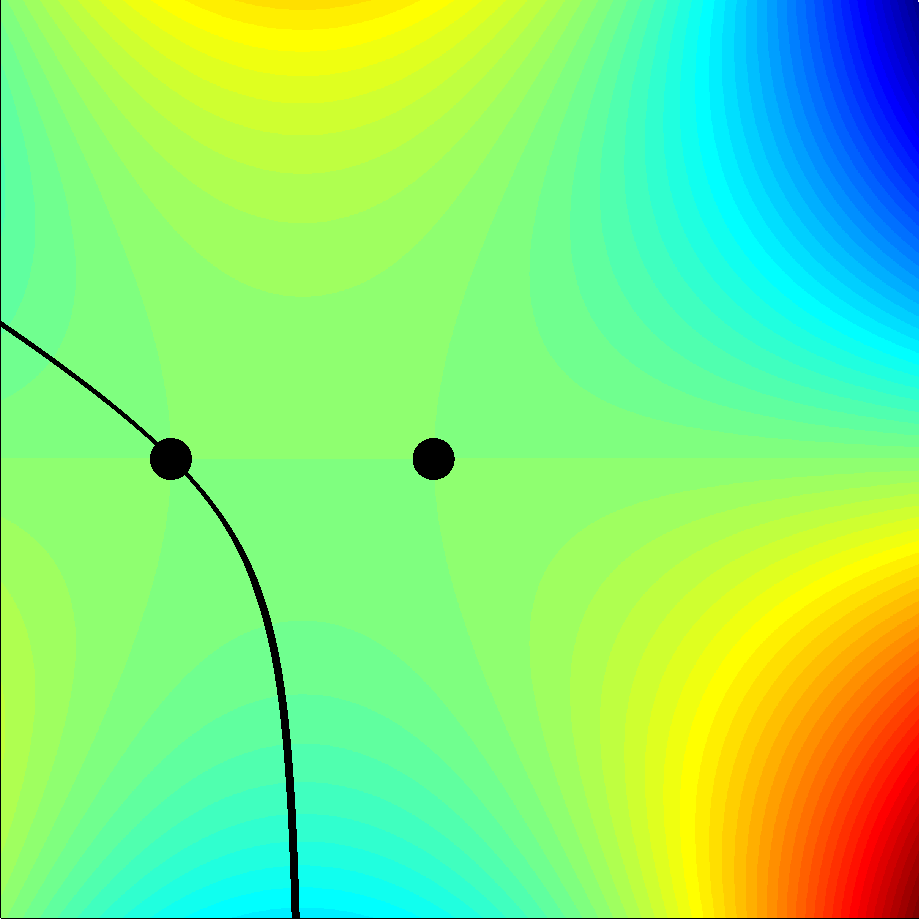}
}
\subfigure[(d) Point 2]{\includegraphics[width=\psize]{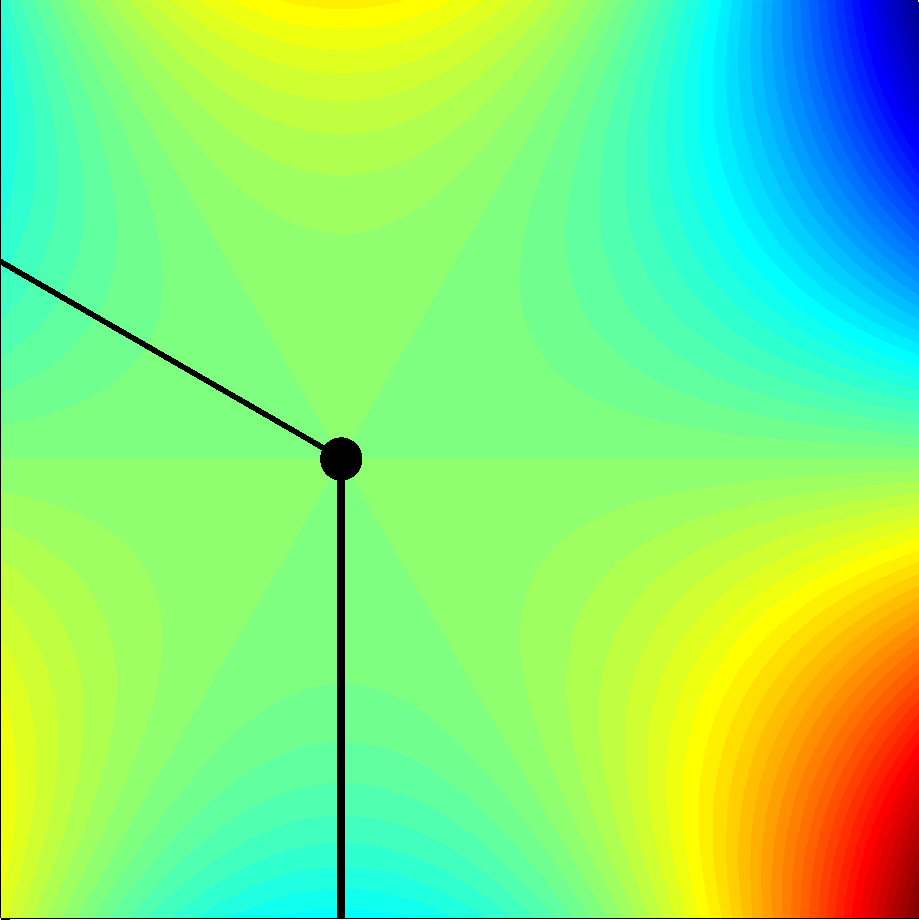}
}
\subfigure[(e) Point 3]{\includegraphics[width=\psize]{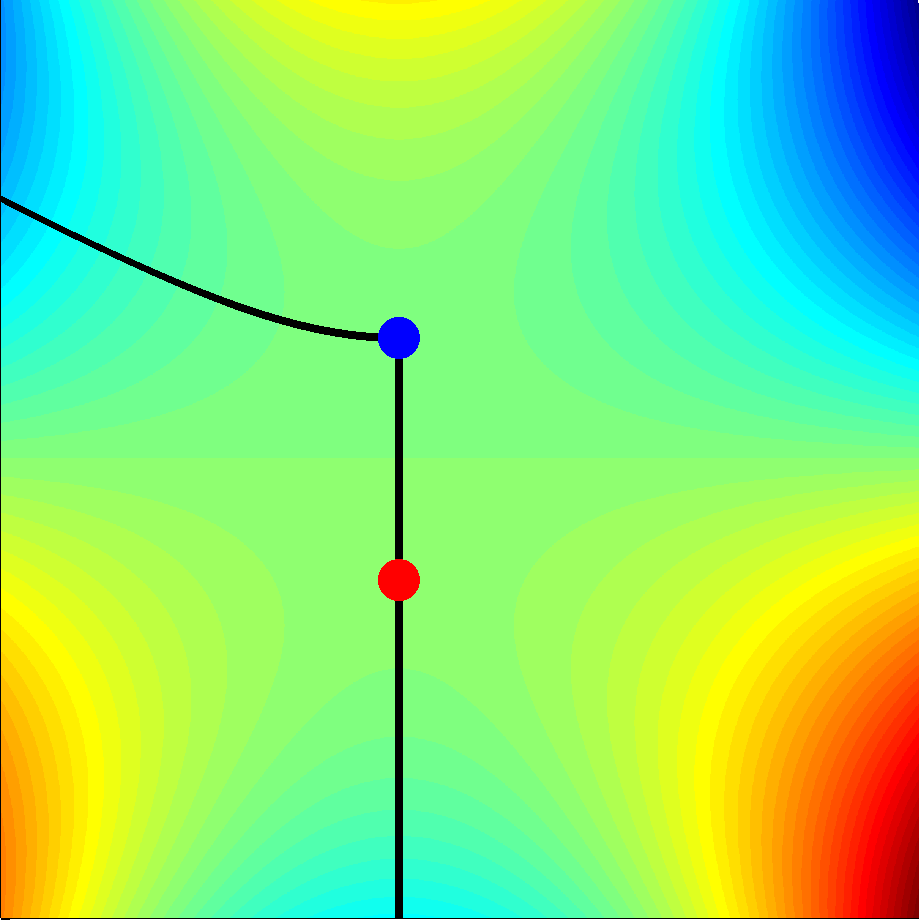}
}
\caption{Steepest descent contours for $A_{32}$ 
in the case $(l,m)=(1,1)$, $\alpha=1$, evaluated at points 1-3 from \F\ref{fig:AirySaddlesFull}(a). Key: R=real, DR=double real, L=exponentially large, S=exponentially small.
}
\label{fig:AirySaddlesA32}
\end{center}
\end{figure}

The solution $A_{32}$ (\F\ref{fig:AiryFieldPlots}(c)-(d)) can also be expressed in terms of the Airy function. Deforming the contour in \rf{AiryRealDef} onto $\Gamma_{21}$ gives an integral representation for $\Ai(z)$ valid for all $z\in\C$. A rotation in the complex $t$-plane then gives (cf.\ \rf{eqn:AAiry}-\rf{AiryRealDef})
\begin{align}
\label{eqn:AAiryA32}
A_{32}
&= \re^{2\ri \pi/3}(2\pi) (2\kappa)^{1/3} \re^{-\ri\kappa(\X \Y +\kappa \X^3/3)}\Ai\left[\re^{-\ri \pi/3}(2\kappa)^{1/3}\left(\Y+\frac{\kappa \X^2}{2}\right)\right].
\end{align}
As for $A_{21}$, the far-field behaviour can be obtained using the well-known asymptotic behaviour of the Airy function for complex argument. However, again it is possible to proceed without \rf{eqn:AAiryA32} and instead pursue a direct steepest descent analysis. We do not present all the details, but merely record the qualitative solution structure, which resembles that of creeping waves (see \S\ref{sec:BVPs}):
\begin{itemize}
\item 
Above the parabola (cf.\ \F\ref{fig:AirySaddlesA32}(c)) the steepest descent contour passes through just one of the two real saddle points, $\tau_-$. In the solution of the Helmholtz equation to which $A_{32}\re^{\ri k \x}$ matches, this corresponds to a single ray outgoing from the parabola (cf.\ \F\ref{fig:AirySaddlesA32}(b)). %
\item On the parabola the steepest descent contour comprises the two segments in \F\ref{fig:AirySaddlesA32}(d).%
\item
Below the parabola (cf.\ \F\ref{fig:AirySaddlesA32}(e)) the steepest descent contour passes through both complex saddle points, the lower one giving an exponentially large contribution as $k\to\infty$. Such a configuration could only be realistic if there was a source of energy in the outer wavefield described by the Helmholtz equation, as for example in the acoustic ring source (see \cite{Cha:94,CLOT:99}). 
\end{itemize}
As mentioned previously, the behaviour of $A_{13}$ can be easily deduced from that of $A_{32}$ using the symmetry relation \rf{eqn:AirySym}. But, for completeness, we present typical saddle point configurations for $A_{13}$ in \F\ref{fig:AirySaddlesA31}. The single family of tangent rays is now incoming rather than outgoing. %

\def\psize{20mm}
\begin{figure}[t!]
\begin{center}
\subfigure[(a) $(\x,\y)$-plane] {
\begin{tikzpicture}[line cap=round,line join=round,>=triangle 45,x=0.3cm,y=0.3cm]
\def\size{4};
\def\Ta{R};\def\Tb{DR\quad};\def\Tc{L,S};
\draw (0,\size/2) node {\tf\Ta};
\draw[thick, smooth,domain=-\size:\size,variable=\t] plot({-\t},{-1*(\t)^2/4}) node[below] {\tf\Tb};
\draw (0,-\size) node {\tf\Tc};
\begin{scriptsize}
\fill (-2.9594,-0.49197) circle (1.5pt);
\draw (-2.3594,-0.49197) node {1};
\fill (-2.5342,-1.6056) circle (1.5pt);
\draw (-1.9342,-1.6056) node {2};
\fill (-1.7089,-2.4657) circle (1.5pt);
\draw (-1.1089,-2.4657) node {3};
\end{scriptsize}
\end{tikzpicture}
}
\subfigure[(b) One ingoing real ray] {
\begin{tikzpicture}[line cap=round,line join=round,>=latex,x=0.3cm,y=0.3cm,scale=0.6]
\draw[thick,smooth,domain=-11:2,variable=\t] plot({\t},{-1*((\t)^2/16)});
\draw [thick,->] (-6.4,-1.25) -- (-4.4,-0.8);
\draw [dotted] (-4.4,-0.8)-- (-1.83,-0.21);
\fill (-6.4,-1.25) circle (2pt);
\end{tikzpicture}
}
\subfigure[(c) Point 1]{\includegraphics[width=\psize]{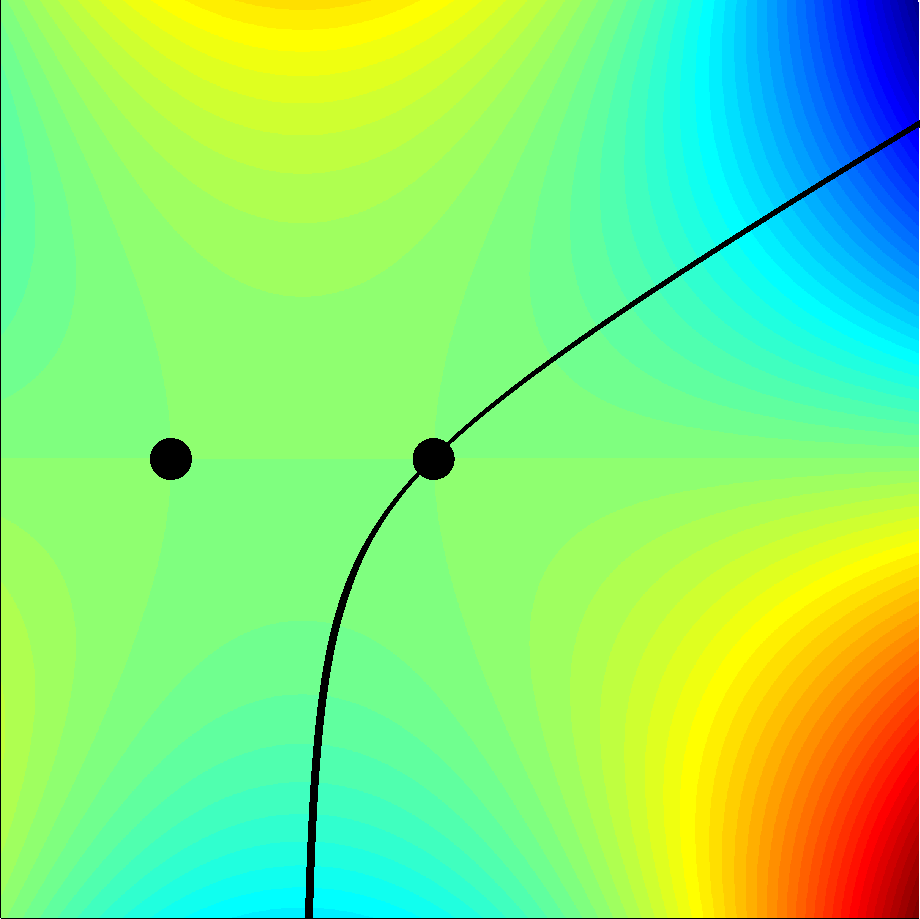}
}
\subfigure[(d) Point 2]{\includegraphics[width=\psize]{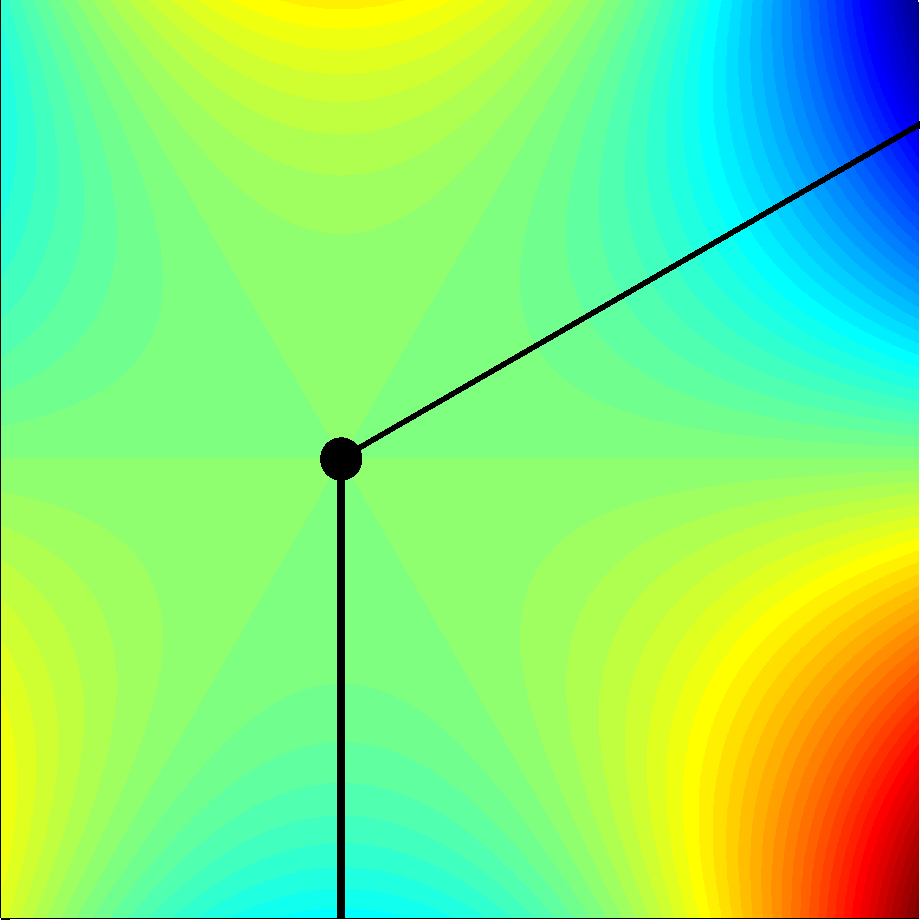}
}
\subfigure[(e) Point 3]{\includegraphics[width=\psize]{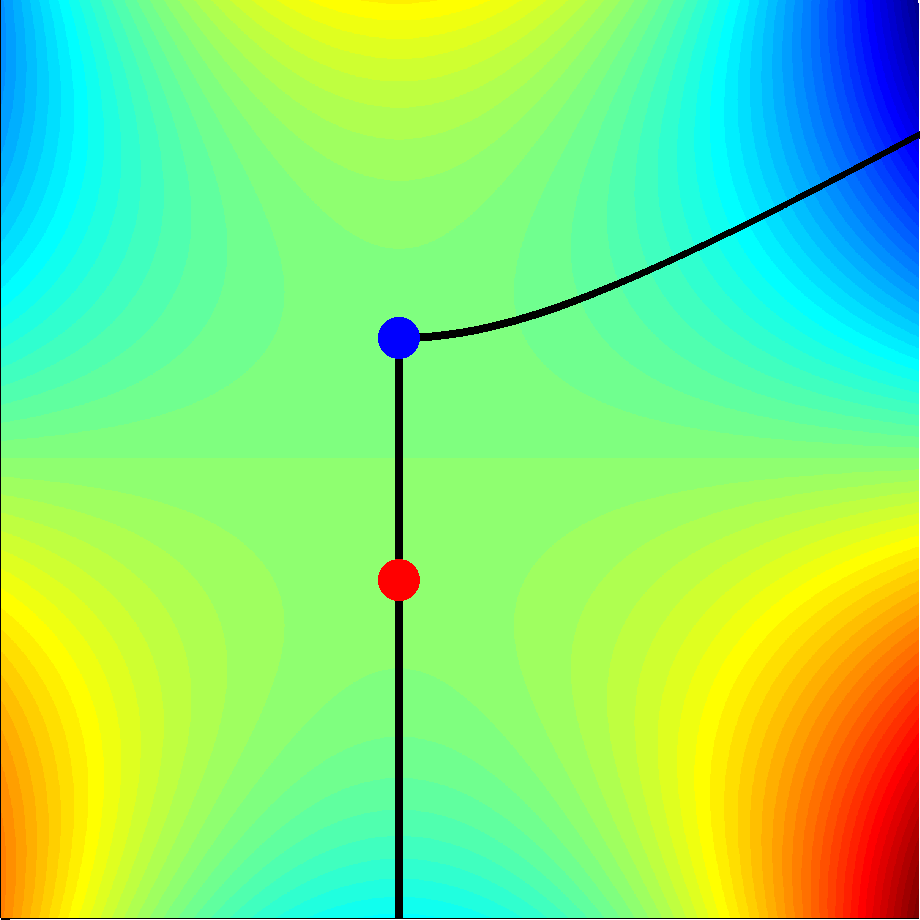}
}
\caption{Steepest descent contours for $A_{13}$ in the case $(l,m)=(1,1)$, $\alpha=1$, evaluated at points 1-3 from \F\ref{fig:AirySaddlesFull}(a). Key: R=real, DR=double real, L=exponentially large, S=exponentially small.
}
\label{fig:AirySaddlesA31}
\end{center}
\end{figure}
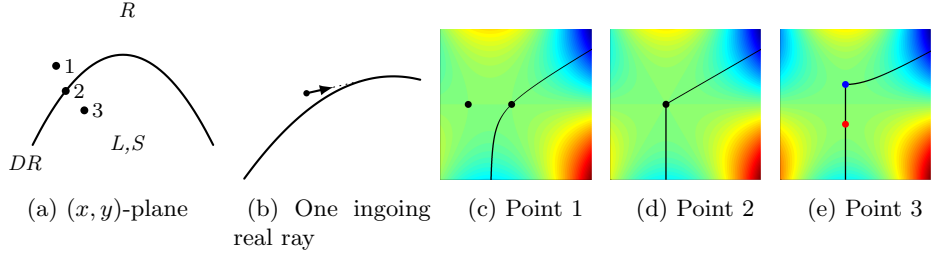

\subsection{\label{sec:ii} Cusp ($(l,m)=(2,1)$, $2m+l=4$, $\lambda=1/2$, $\alpha=1/(12\kappa)$)}

\begin{figure}[t]
\def\size{25mm}
\begin{center}
\subfigure[(a)]{\includegraphics[width=28mm]{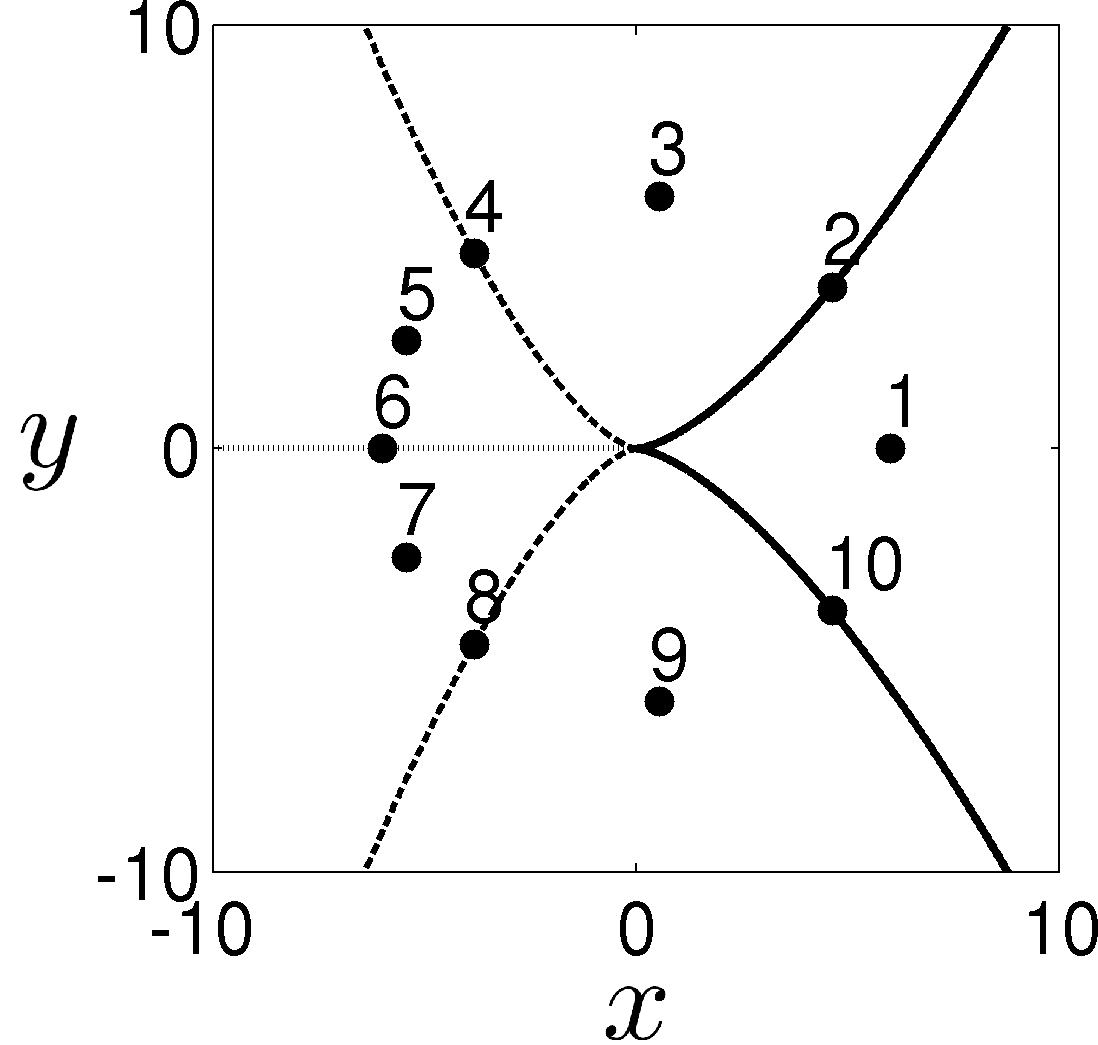}
}
\subfigure[(b) Point 1]{\includegraphics[width=\size]{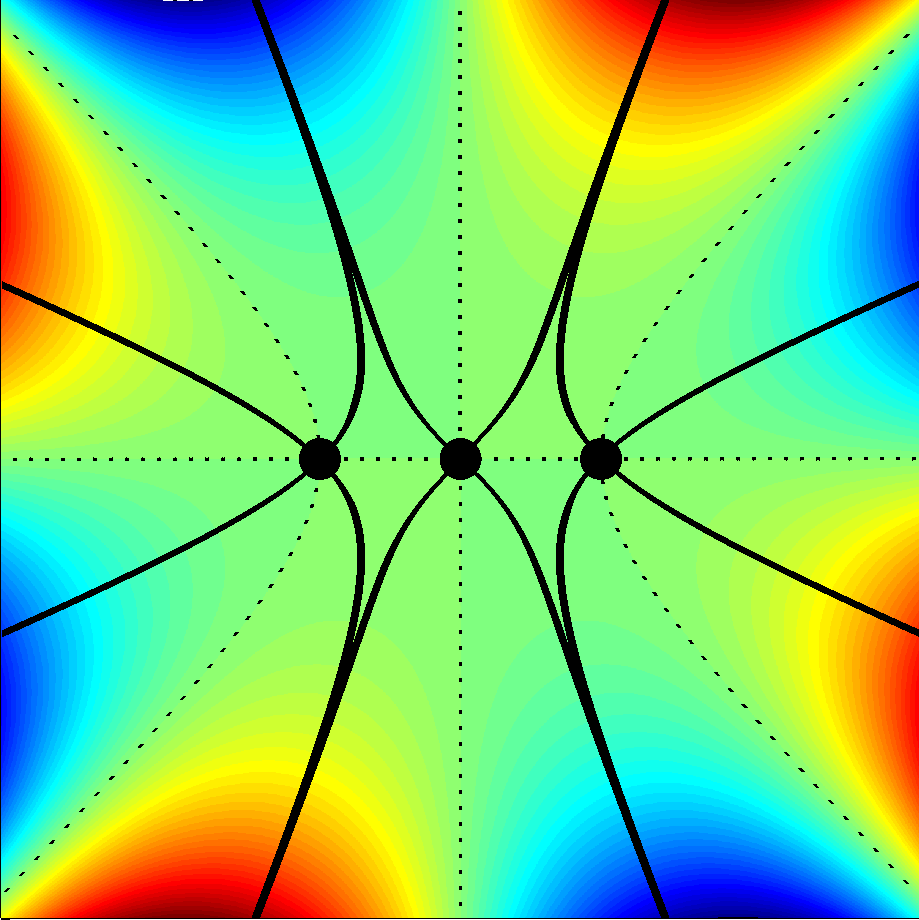}
}
\subfigure[(c) Point 2]{\includegraphics[width=\size]{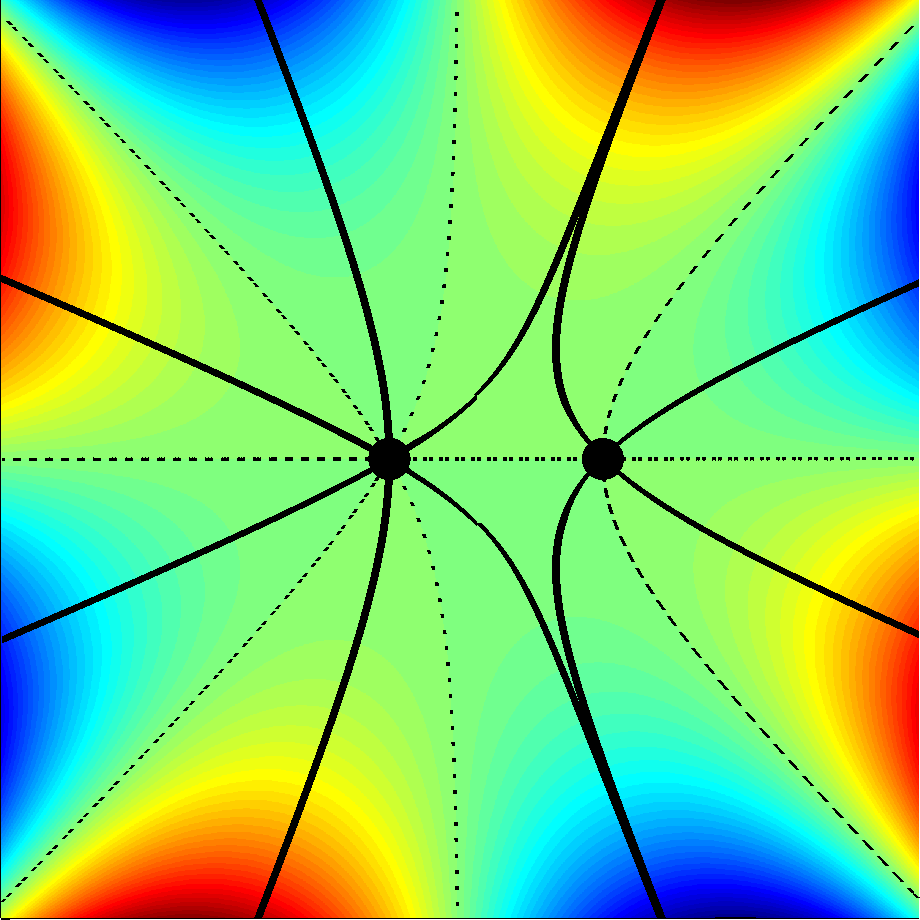}
}
\subfigure[(d) Point 3]{\includegraphics[width=\size]{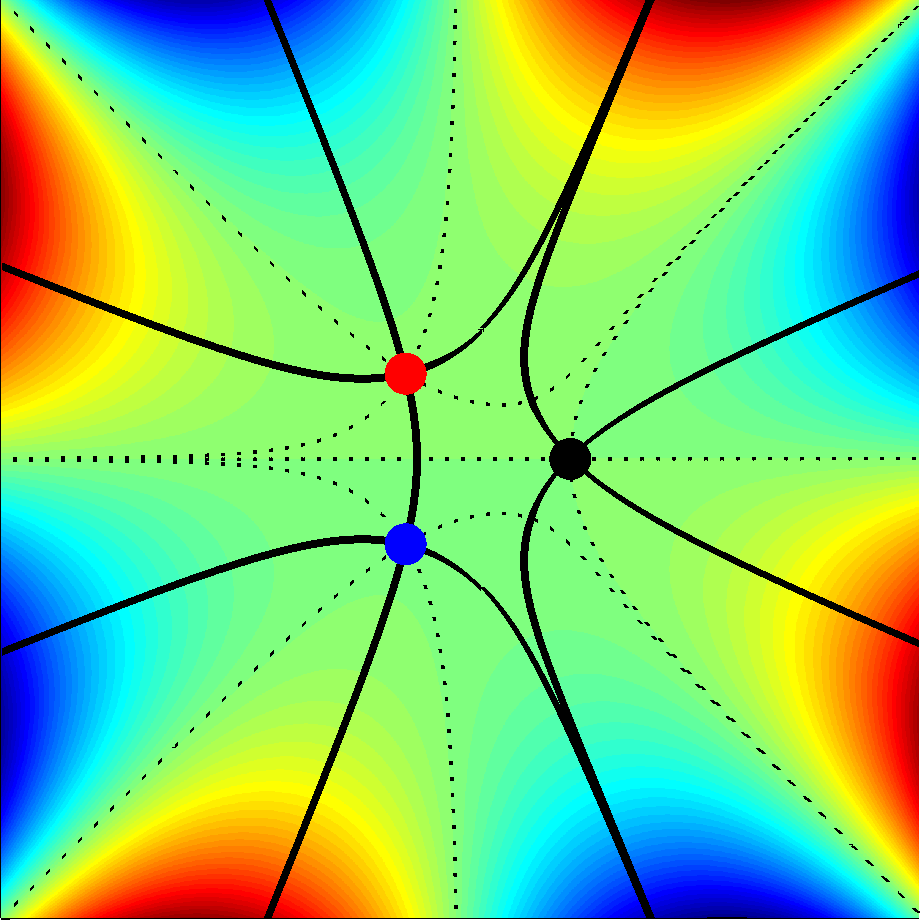}
}
\subfigure[(e) Point 4]{\includegraphics[width=\size]{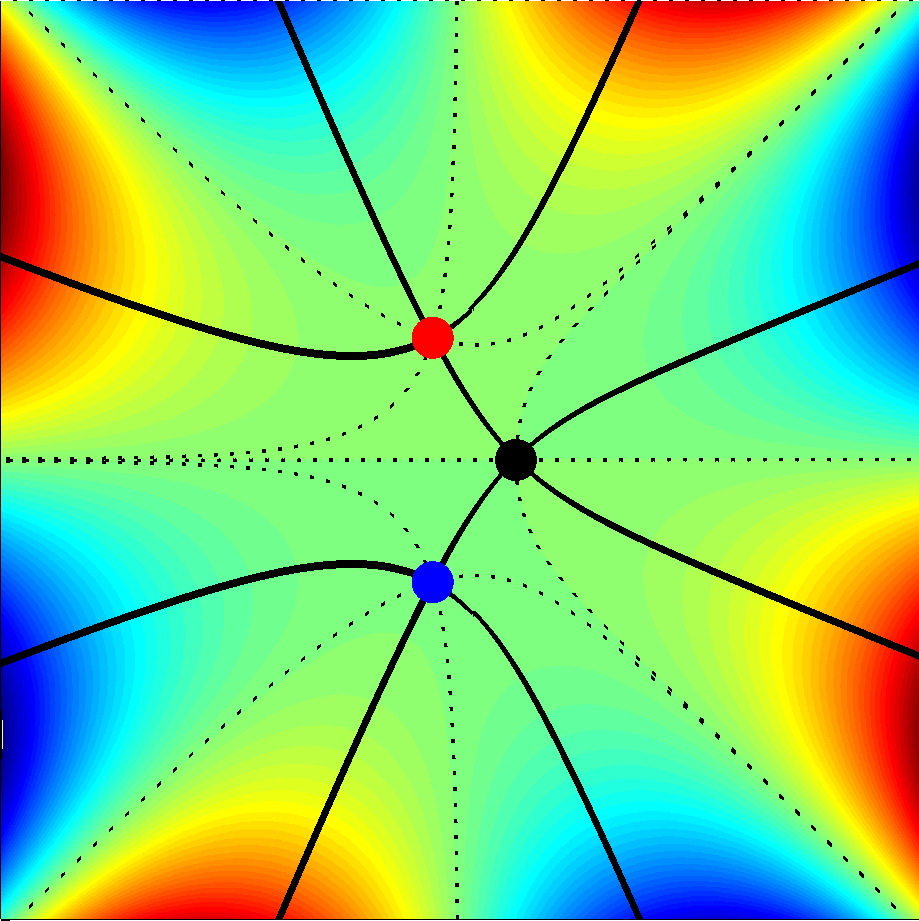}
}
\subfigure[(f) Point 5]{\includegraphics[width=\size]{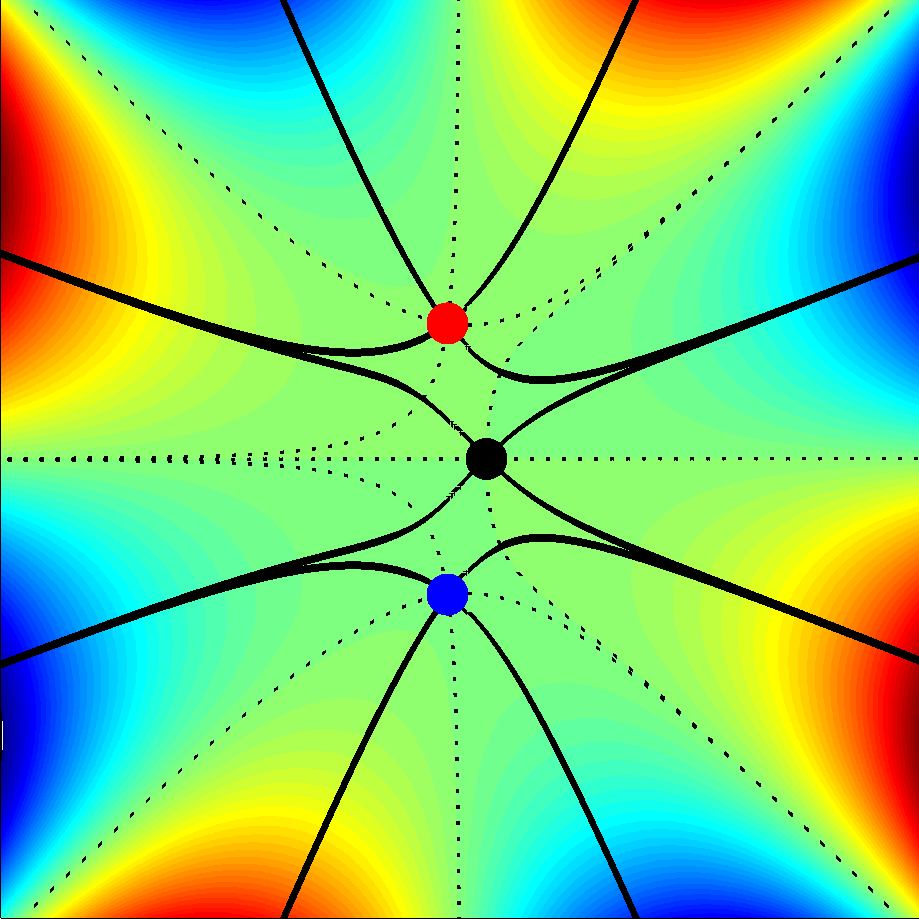}
}
\subfigure[(g) Point 6]{\includegraphics[width=\size]{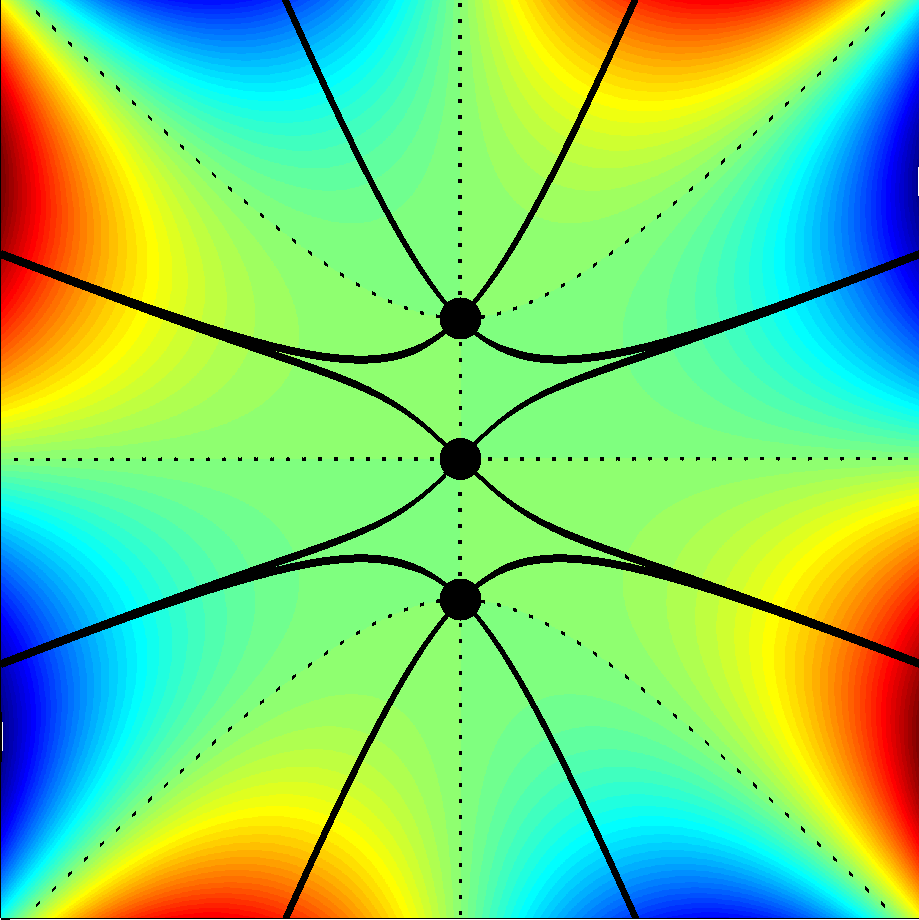}
}
\subfigure[(h) Point 7]{\includegraphics[width=\size]{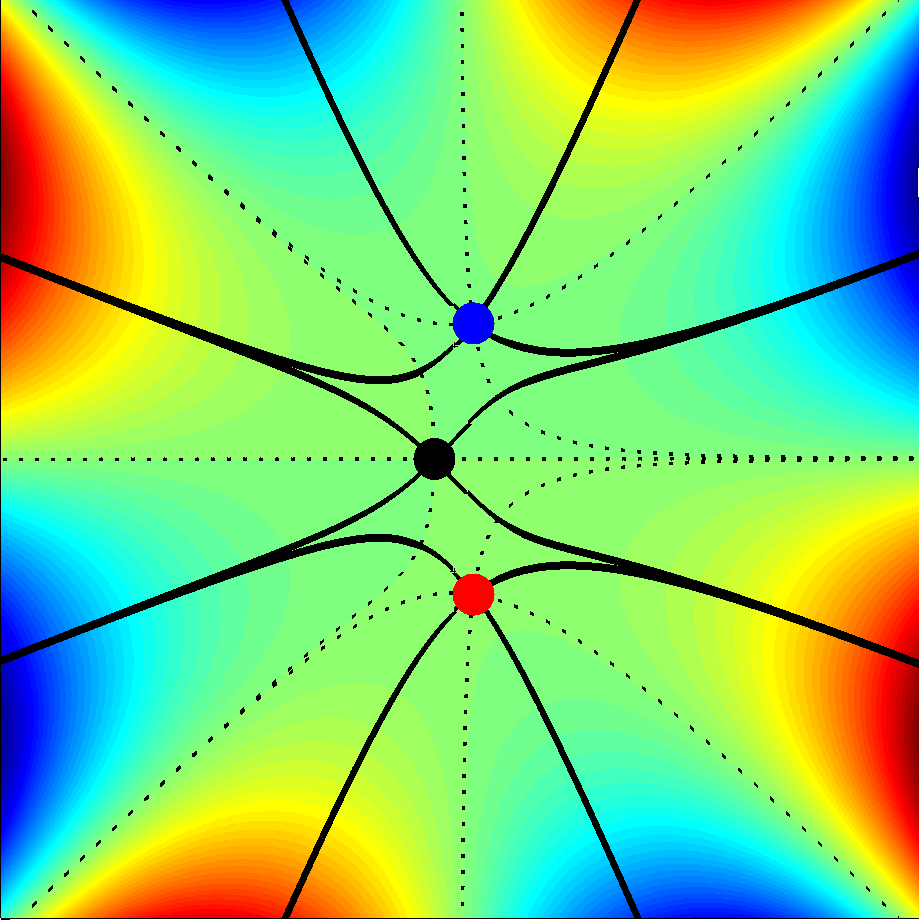}
}
\subfigure[(i) Point 8]{\includegraphics[width=\size]{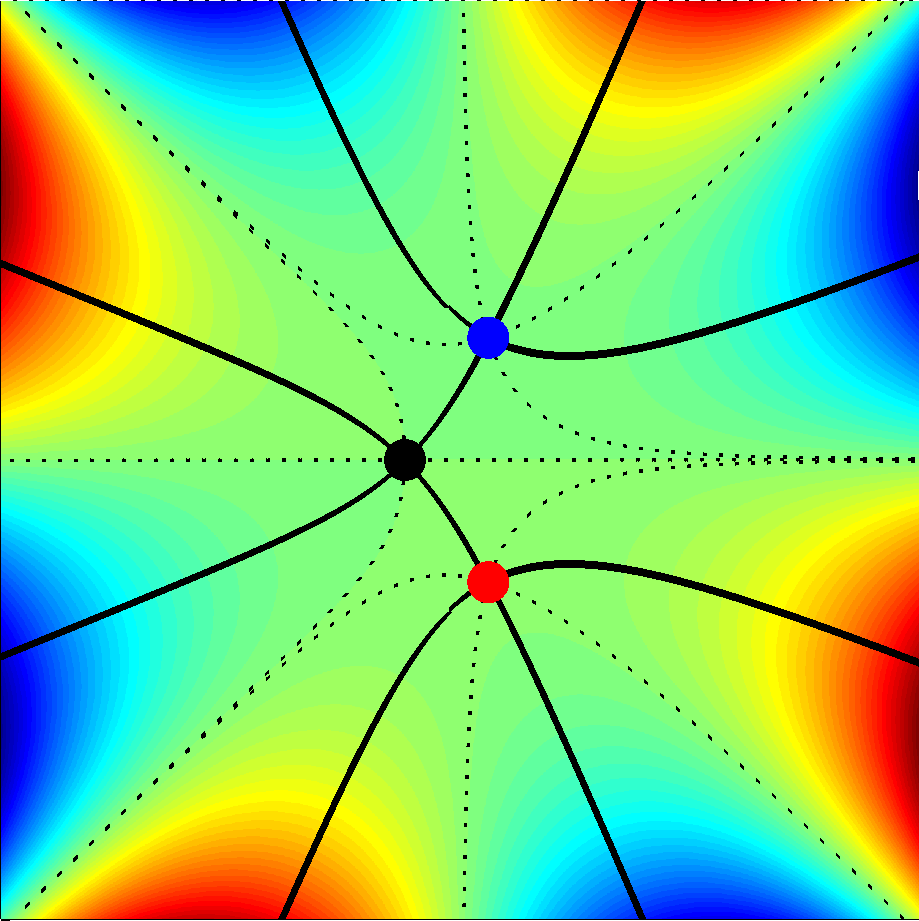}
}
\subfigure[(j) Point 9]{\includegraphics[width=\size]{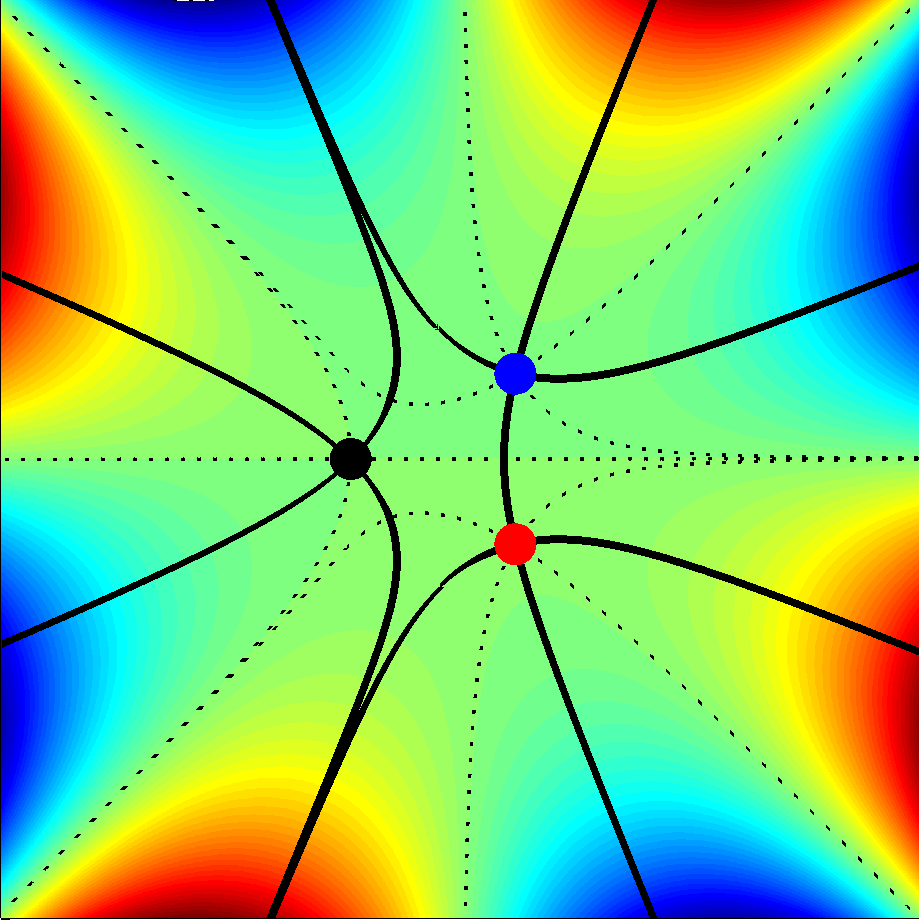}
}
\subfigure[(k) Point 10]{\includegraphics[width=\size]{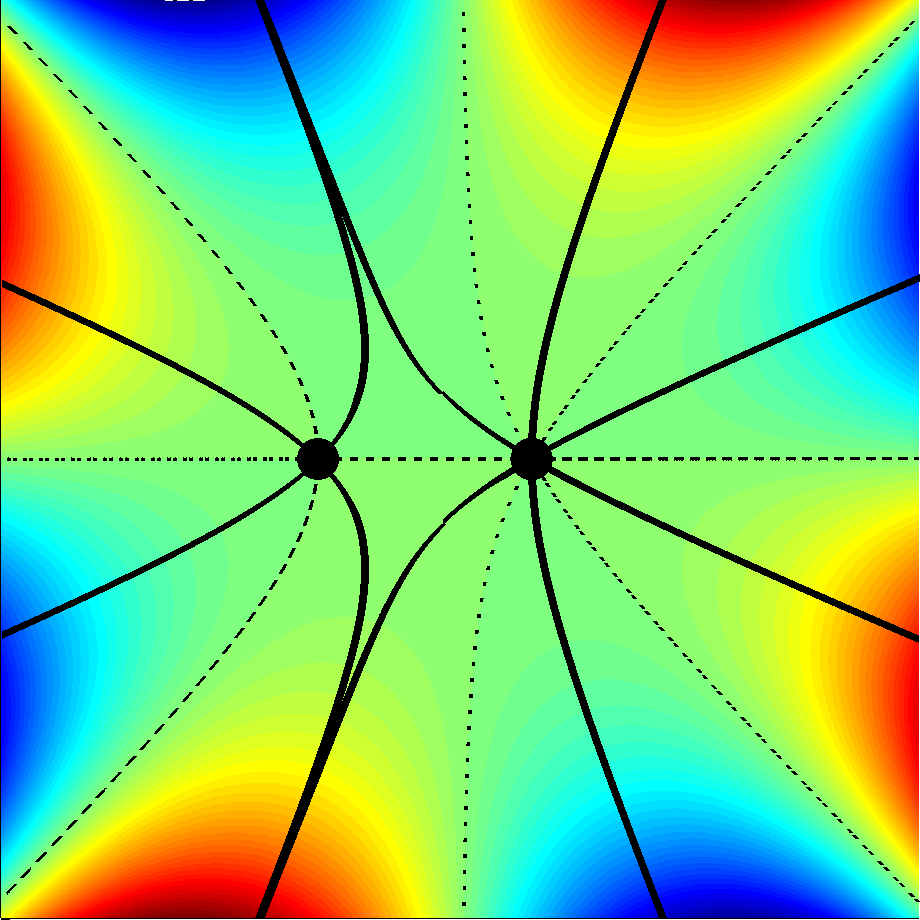}
}
\caption{Analogue of \F\ref{fig:AirySaddlesFull} 
for the case $(l,m)=(2,1)$, $\alpha=1$. (a) shows the localisation curve $\y=\pm 2\x^{3/2}/(3\sqrt{3})$, $\x\geq0$ (solid curve), the Stokes lines $\y=\pm\rho_*\x^{3/2}$, $\x\leq 0$ (dashed) and the anti-Stokes line $\x<0$, $\y=0$ (dotted). %
(b)-(k) show the saddle point configurations for points 1--10 of (a).
}
\label{fig:PearceySaddlesFull}
\end{center}
\end{figure}

Localisation is now near the curve $\Y^2=16\kappa \X^3/9$, i.e.\ $\Y=\pm(4/3)\kappa^{1/2}\X^{3/2}$, $\X\geq0$, 
and %
\begin{align}
\label{eqn:AIntCusp}
A_{ij} = \int_{\Gamma_{ij}} \re^{\ri (-\Y t-\X t^2/2 + \alpha t^4/4)} \,\rd t = \int_{\Gamma_{ij}} \re^{\ri (-\Y t-\X t^2/2 + t^4/(48\kappa))} \,\rd t,
\end{align}
where $\Gamma_{ij}$ goes from $S_i$ to $S_j$ in \F\ref{Sectors}(b). 
The scalings \rf{eqn:scaling2} are 
$\X=k^{1/2}\x$, $\Y=k^{3/4}\y$, $t=k^{1/4}\tau$, and the saddle points in the $\tau$-plane are the roots of \rf{eqn:StatPoints}, which in this case is the cubic $-\y-\x\tau+\tau^3/(12\kappa)=0$. Typical saddle point configurations are illustrated in \F\ref{fig:PearceySaddlesFull} in the case $\alpha=1$ ($\kappa=1/12$). Note that by \rf{eqn:pRelnGeneral} the saddle point configuration for a point $(\x,\y)$ can be obtained from that for $(\x,-\y)$ by the transformation $\tau\to -\tau$ (compare \Fs\ref{fig:PearceySaddlesFull}(c)-(f) with \Fs\ref{fig:PearceySaddlesFull}(h)-(k)). 
The basic configurations are as follows:%
\begin{itemize}
\item 
To the right of the localisation curve (point 1 in \F\ref{fig:PearceySaddlesFull}(a)) there are three distinct real saddle points (\F\ref{fig:PearceySaddlesFull}(b)). 
\item
On the localisation curve (points 2 and 10 in \F\ref{fig:PearceySaddlesFull}(a)) there are three real saddle points, but two of them coincide (\F\ref{fig:PearceySaddlesFull}(c) and (k)). 
\item
To the left of the localisation curve (points 3-9 in \F\ref{fig:PearceySaddlesFull}(a)) there is one real saddle point and a pair of complex conjugate saddle points (\F\ref{fig:PearceySaddlesFull}(d)-(j)). 
\end{itemize}
To the left of the localisation curve there are Stokes lines (shown as dashed lines in \F\ref{fig:PearceySaddlesFull}(a)) along which the real part of the phase is the same for all three saddle points (as for point 4 in \F\ref{fig:PearceySaddlesFull}(a); see \F\ref{fig:PearceySaddlesFull}(e)). One can show that these Stokes lines are $\y\pm\sqrt{12\kappa} \rho_*(-\x)^{3/2}=0$, $\x\leq 0$, where $\rho_* = \sqrt{5+\sqrt{27}}/\sqrt{27}$ \cite{Wr:80,StSp:83}.  
Second, there is an anti-Stokes line (shown as a dotted line in \F\ref{fig:PearceySaddlesFull}(a)) at $\x<0$, $\y=0$, 
on which the amplitudes of the two complex conjugate saddle points are equal.

As before, 
the far-field behaviour of \rf{eqn:AIntCusp} depends on the choice of integration contour. %
There are $\binom{4}{2}=6$ solutions, 
three of which are linearly independent, 
satisfying the symmetry relations (cf.\ \rf{eqn:SymmetryGeneral})
\begin{align}
A_{31}(\X,\Y) &= A_{31}(\X,-\Y), \qquad
A_{42}(\X,\Y) = A_{42}(\X,-\Y),\notag \\
A_{21}(\X,\Y) &= A_{34}(\X,-\Y), \qquad
A_{41}(\X,\Y) = A_{32}(\X,-\Y).
\label{QuarticSymmetry}
\end{align}
Furthermore, \rf{eqn:pRelnGeneral} also implies that the steepest descent contour for $A_{31}(\X,\Y)$ can be obtained from that for $A_{31}(\X,-\Y)$ by the transformation $\tau\to -\tau$. A similar relationship exists between the steepest descent contours for the other pairs in \rf{QuarticSymmetry}. As a result there are essentially four distinct far-field behaviours (modulo reflections in the $\X$-axis), which are illustrated schematically in \F\ref{fig:PearceyBehaviours}. 
We will consider $A_{31}$ in detail, and then make brief remarks on the other cases.%
\newcommand{\qaa}{\draw [dotted] (2.03,0.36)-- (5.28,4.67);}
\newcommand{\qab}{\draw [dotted] (2.03,0.36)-- (0.1,0.01);}
\newcommand{\qac}{\draw [dotted] (2.03,0.36)-- (6.82,-6.86);}
\newcommand{\qad}{\draw [->] (2.03,0.36) -- (3.01,1.66);}
\newcommand{\qae}{\draw [->] (2.03,0.36) -- (3.55,0.64);}
\newcommand{\qaf}{\draw [->] (2.03,0.36) -- (2.79,-0.78);}
\newcommand{\qag}{\fill (2.03,0.36) circle (1.5pt);}
\newcommand{\qba}{\draw [dotted] (-0.7,2.45)-- (4.14,-3.24);}
\newcommand{\qbb}{\draw [->] (-0.7,2.45) -- (0.2,1.38);}
\newcommand{\qbc}{\fill (-0.7,2.45) circle (1.5pt);}
\newcommand{\qca}{\draw [dotted] (-0.7,-2.45)-- (4.14,3.24);}
\newcommand{\qcb}{\draw [->] (-0.7,-2.45) -- (0.2,-1.38);}
\newcommand{\qcc}{\fill (-0.7,-2.45) circle (1.5pt);}
\newcommand{\qda}{\draw [dotted] (-3.91,1.39)-- (0.36,-0.08);}
\newcommand{\qdb}{\draw [->] (-3.91,1.39) -- (-2.49,0.9);}
\newcommand{\qdc}{\fill (-3.91,1.39) circle (1.5pt);}
\newcommand{\qea}{\draw [dotted] (-3.91,-1.39)-- (0.36,0.08);}
\newcommand{\qeb}{\draw [->] (-3.91,-1.39) -- (-2.49,-0.9);}
\newcommand{\qec}{\fill (-3.91,-1.39) circle (1.5pt);}

\begin{figure}[t!]
\begin{center}
\subfigure[(a) $A_{21}(\X,\Y)$ and $A_{43}(\X,-\Y)$] {
\begin{tikzpicture}[line cap=round,line join=round,>=stealth,x=0.3cm,y=0.3cm,scale=1]
\def\size{5};\def\lambdastar{0.614520361676536};
\def\Ta{R,R};\def\Tb{R,DR};\def\Tc{L,R,S};\def\Td{L,R};\def\Te{L};\def\Tf{B};\def\Tg{S};\def\Th{S};\def\Ti{S};\def\Tj{DR};
\draw (1.5*\size,0) node {\tf\Ta};
\draw[thick, smooth,domain=0:\size,variable=\t] plot({\t},{2*sqrt(\t^3)/(3*sqrt(3))}) node[right] {\tf\Tb};
\draw (0,\size) node {\tf\Tc};
\draw[thick, dashed, smooth,domain=0:(\size/\lambdastar)^(2/3),variable=\t] plot({-\t},{\lambdastar*sqrt(\t^3)}) node[left] {\tf\Td};
\draw (-\size,\size/2) node[left] {\tf\Te};
\draw[thick, dotted] (0,0) -- (-\size,0) node[left] {\tf\Tf};
\draw (-\size,-\size/2) node[left] {\tf\Tg};
\draw[thick, dashed, smooth,domain=0:(\size/\lambdastar)^(2/3),variable=\t] plot({-\t},{-\lambdastar*sqrt(\t^3)}) node[left] {\tf\Th};
\draw (0,-\size) node {\tf\Ti};
\draw[thick, smooth,domain=0:\size,variable=\t] plot({\t},{-2*sqrt(\t^3)/(3*sqrt(3))}) node[right] {\tf\Tj};
\begin{scope}\clip (-\size,-\size) rectangle(\size,\size);
\qab\qac\qae\qaf\qag
\qba\qbb\qbc
\end{scope}
\end{tikzpicture}}
\hs{5}
\subfigure[(b) $A_{31}(\X,\Y)$ {[}cusp caustic{]}] {
\begin{tikzpicture}[line cap=round,line join=round,>=stealth,x=0.3cm,y=0.3cm,scale=1]
\def\size{5};\def\lambdastar{0.614520361676536};
\def\Ta{R,R,R};\def\Tb{R,DR};\def\Tc{R,S};\def\Td{R,S};\def\Te{R};\def\Tf{R};\def\Tg{R};\def\Th{R,S};\def\Ti{R,S};\def\Tj{R,DR};
\draw (1.5*\size,0) node {\tf\Ta};
\draw[thick, smooth,domain=0:\size,variable=\t] plot({\t},{2*sqrt(\t^3)/(3*sqrt(3))}) node[right] {\tf\Tb};
\draw (0,\size) node {\tf\Tc};
\draw[thick, dashed, smooth,domain=0:(\size/\lambdastar)^(2/3),variable=\t] plot({-\t},{\lambdastar*sqrt(\t^3)}) node[left] {\tf\Td};
\draw (-\size,\size/2) node[left] {\tf\Te};
\draw[thick, dotted] (0,0) -- (-\size,0) node[left] {\tf\Tf};
\draw (-\size,-\size/2) node[left] {\tf\Tg};
\draw[thick, dashed, smooth,domain=0:(\size/\lambdastar)^(2/3),variable=\t] plot({-\t},{-\lambdastar*sqrt(\t^3)}) node[left] {\tf\Th};
\draw (0,-\size) node {\tf\Ti};
\draw[thick, smooth,domain=0:\size,variable=\t] plot({\t},{-2*sqrt(\t^3)/(3*sqrt(3))}) node[right] {\tf\Tj};
\begin{scope}\clip (-\size,-\size) rectangle(\size,\size);
\qaa\qab\qac\qad\qae\qaf\qag
\qba\qbb\qbc
\qca\qcb\qcc
\qda\qdb\qdc
\qea\qeb\qec
\end{scope}
\end{tikzpicture}}
\\
\subfigure[(c) $A_{41}(\X,\Y)$ and $A_{32}(\X,-\Y)$] {
\begin{tikzpicture}[line cap=round,line join=round,>=stealth,x=0.3cm,y=0.3cm,scale=1]
\def\size{5};\def\lambdastar{0.614520361676536};
\def\Ta{R};\def\Tb{R};\def\Tc{R};\def\Td{R,S};\def\Te{R,S};\def\Tf{R,B};\def\Tg{L,R};\def\Th{L,R,S};\def\Ti{L,S};\def\Tj{DR};
\draw (1.5*\size,0) node {\tf\Ta};
\draw[thick, smooth,domain=0:\size,variable=\t] plot({\t},{2*sqrt(\t^3)/(3*sqrt(3))}) node[right] {\tf\Tb};
\draw (0,\size) node {\tf\Tc};
\draw[thick, dashed, smooth,domain=0:(\size/\lambdastar)^(2/3),variable=\t] plot({-\t},{\lambdastar*sqrt(\t^3)}) node[left] {\tf\Td};
\draw (-\size,\size/2) node[left] {\tf\Te};
\draw[thick, dotted] (0,0) -- (-\size,0) node[left] {\tf\Tf};
\draw (-\size,-\size/2) node[left] {\tf\Tg};
\draw[thick, dashed, smooth,domain=0:(\size/\lambdastar)^(2/3),variable=\t] plot({-\t},{-\lambdastar*sqrt(\t^3)}) node[left] {\tf\Th};
\draw (0,-\size) node {\tf\Ti};
\draw[thick, smooth,domain=0:\size,variable=\t] plot({\t},{-2*sqrt(\t^3)/(3*sqrt(3))}) node[right] {\tf\Tj};
\begin{scope}\clip (-\size,-\size) rectangle(\size,\size);
\qac\qaf\qag
\qba\qbb\qbc
\qda\qdb\qdc
\qea\qeb\qec
\end{scope}
\end{tikzpicture}}
\hs{5}
\subfigure[(d) $A_{42}(\X,\Y)$] {
\begin{tikzpicture}[line cap=round,line join=round,>=stealth,x=0.3cm,y=0.3cm,scale=1]
\def\size{5};\def\lambdastar{0.614520361676536};
\def\Ta{R};\def\Tb{DR};\def\Tc{L,S};\def\Td{L,R,S};\def\Te{L,R,S};\def\Tf{R,B,B};\def\Tg{L,R,S};\def\Th{L,R,S};\def\Ti{L,S};\def\Tj{DR};
\draw (1.5*\size,0) node {\tf\Ta};
\draw[thick, smooth,domain=0:\size,variable=\t] plot({\t},{2*sqrt(\t^3)/(3*sqrt(3))}) node[right] {\tf\Tb};
\draw (0,\size) node {\tf\Tc};
\draw[thick, dashed, smooth,domain=0:(\size/\lambdastar)^(2/3),variable=\t] plot({-\t},{\lambdastar*sqrt(\t^3)}) node[left] {\tf\Td};
\draw (-\size,\size/2) node[left] {\tf\Te};
\draw[thick, dotted] (0,0) -- (-\size,0) node[left] {\tf\Tf};
\draw (-\size,-\size/2) node[left] {\tf\Tg};
\draw[thick, dashed, smooth,domain=0:(\size/\lambdastar)^(2/3),variable=\t] plot({-\t},{-\lambdastar*sqrt(\t^3)}) node[left] {\tf\Th};
\draw (0,-\size) node {\tf\Ti};
\draw[thick, smooth,domain=0:\size,variable=\t] plot({\t},{-2*sqrt(\t^3)/(3*sqrt(3))}) node[right] {\tf\Tj};
\begin{scope}\clip (-\size,-\size) rectangle(\size,\size);
\qab\qae\qag
\qda\qdb\qdc
\qea\qeb\qec
\end{scope}
\end{tikzpicture}}
\caption{Schematic showing saddle point configurations for different contour choices in the case $(l,m)=(2,1)$, as a function of position in the $(\x,\y)$-plane. The localisation curve (solid line), the Stokes lines (dashed line) and the anti-Stokes line (dotted line) are also shown. Key: R=real, DR=double real, L=exponentially large, S=exponentially small, B=complex but with $\im{p}=0$. 
}
\label{fig:PearceyBehaviours}
\end{center}
\end{figure}
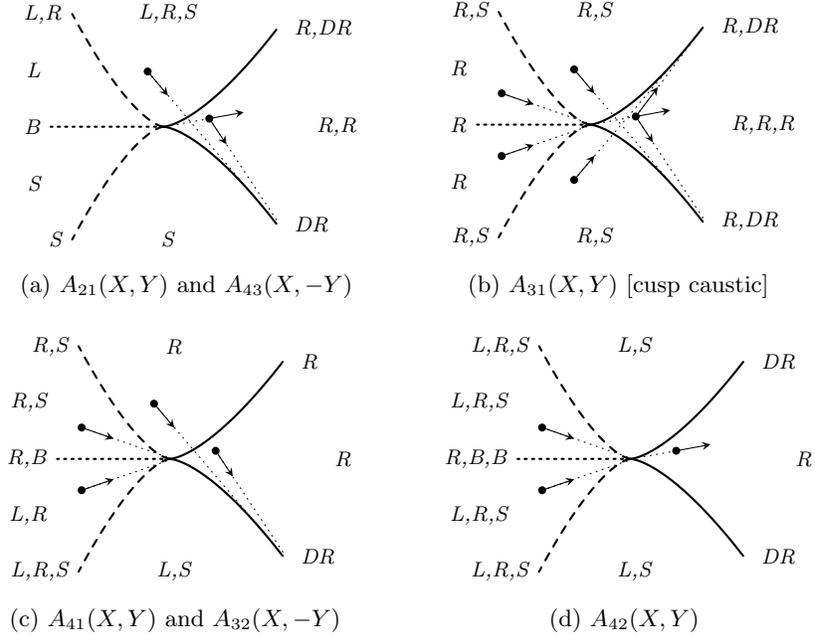
\begin{figure}[t!]
\begin{center}
\hs{-5}
\subfigure[(a) $|A_{31}|$]{\includegraphicsDave[width=40mm]{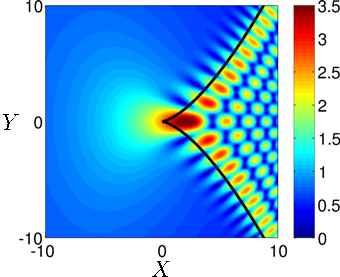}
}
\subfigure[(b) $\real{A_{31}\re^{\ri k\x}}$]{\includegraphicsDave[width=72mm]{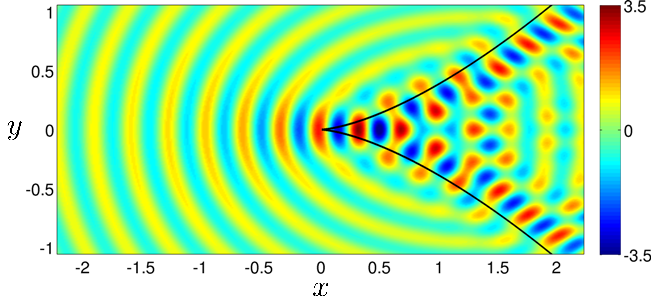}
}
\caption{Plots of $|A_{31}(\X,\Y)|$ and $\real{A_{31}\re^{\ri k \x}}$ for $(l,m)=(2,1)$, with \mbox{$\lambda= l/(l+2m)=1/2$} and $k=20$. The curve near which the solution is localised is superimposed in black.}
\label{fig:PearceyFieldPlots}
\end{center}
\end{figure}

The solution $A_{31}$ describes the field near a cusped caustic. Indeed, deforming the contour $\Gamma_{31}$ to the real axis reveals that $A_{31}$ 
can be written in terms of the cusp canonical integral 
\cite{CoCu:82} 
$C_4(a_1,a_2):=\int_{-\infty}^\infty \re^{\ri(a_1t+a_2t^2+t^4)} \,\rd t$, 
as 
$A_{31} =(\alpha/4)^{-1/4} C_4(-\sqrt{2}\alpha^{-1/4}\Y,-\alpha^{-1/2}\X)$. 
We note that 
\begin{align}
\label{eq:PearceyDefn}
P(\X,\Y):=C_4(\Y,\X) = \int_{-\infty}^\infty \re^{\ri(\Y t+\X t^2+t^4)} \,\rd t
\end{align}
is the Pearcey function \cite{CoCu:82}. %
Plots of $|A_{31}|$ in the case $\alpha=1$, and the real part of the associated approximate solution $A_{31}\re^{\ri k\x}$ of the Helmholtz equation, are given in \F\ref{fig:PearceyFieldPlots}. These plots were generated using the \texttt{cuspint} Fortran routines for cuspoid canonical integrals in \cite{KiCoHo:00}, which implement the numerical quadrature strategy from \cite{CoCu:82}. 
The qualitative asymptotic behaviour is summarised in \F\ref{fig:PearceyBehaviours}(b), and typical steepest descent contours are given in \F\ref{fig:PearceySaddlesA31}. 
\def\psize{25mm}
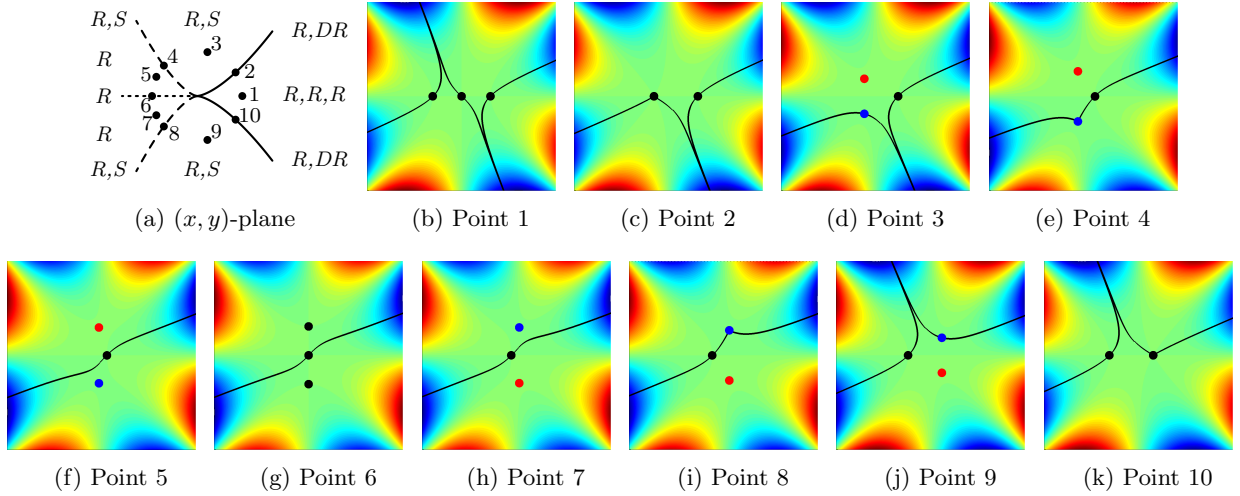
\begin{figure}[h]
\begin{center}
\subfigure[(a) $(\x,\y)$-plane] {
\begin{tikzpicture}[line cap=round,line join=round,>=triangle 45,x=0.2cm,y=0.2cm]
\def\size{5};\def\lambdastar{0.614520361676536};
\def\Ta{R,R,R};\def\Tb{R,DR};\def\Tc{R,S};\def\Td{R,S};\def\Te{R};\def\Tf{R};\def\Tg{R};\def\Th{R,S};\def\Ti{R,S};\def\Tj{R,DR};
\draw (1.5*\size,0) node {\tf\Ta};
\draw[thick, smooth,domain=0:\size,variable=\t] plot({\t},{2*sqrt(\t^3)/(3*sqrt(3))}) node[right] {\tf\Tb};
\draw (0,\size) node {\tf\Tc};
\draw[thick, dashed, smooth,domain=0:(\size/\lambdastar)^(2/3),variable=\t] plot({-\t},{\lambdastar*sqrt(\t^3)}) node[left] {\tf\Td};
\draw (-\size,\size/2) node[left] {\tf\Te};
\draw[thick, dotted] (0,0) -- (-\size,0) node[left] {\tf\Tf};
\draw (-\size,-\size/2) node[left] {\tf\Tg};
\draw[thick, dashed, smooth,domain=0:(\size/\lambdastar)^(2/3),variable=\t] plot({-\t},{-\lambdastar*sqrt(\t^3)}) node[left] {\tf\Th};
\draw (0,-\size) node {\tf\Ti};
\draw[thick, smooth,domain=0:\size,variable=\t] plot({\t},{-2*sqrt(\t^3)/(3*sqrt(3))}) node[right] {\tf\Tj};
\begin{scriptsize}
\fill (3,0) circle (1.5pt);
\draw (3.8,0.1) node {1};
\fill (2.5551,1.5721) circle (1.5pt);
\draw (3.5,1.67) node {2};
\fill (0.68167,2.9215) circle (1.5pt);
\draw (1.2817,3.5215) node {3};
\fill (-2.214,2.0244) circle (1.5pt);
\draw (-1.614,2.6244) node {4};
\fill (-2.7143,1.2778) circle (1.5pt);
\draw (-3.4,1.4778) node {5};
\fill (-3,0) circle (1.5pt);
\draw (-3.4,-0.6) node {6};
\fill (-2.7143,-1.2778) circle (1.5pt);
\draw (-3.3,-1.8) node {7};
\fill (-2.214,-2.0244) circle (1.5pt);
\draw (-1.5,-2.5) node {8};
\fill (0.68167,-2.9215) circle (1.5pt);
\draw (1.2817,-2.3215) node {9};
\fill (2.5551,-1.5721) circle (1.5pt);
\draw (3.5,-1.3) node {10};
\end{scriptsize}
\end{tikzpicture}}
\subfigure[(b) Point 1]{\includegraphics[width=\psize]{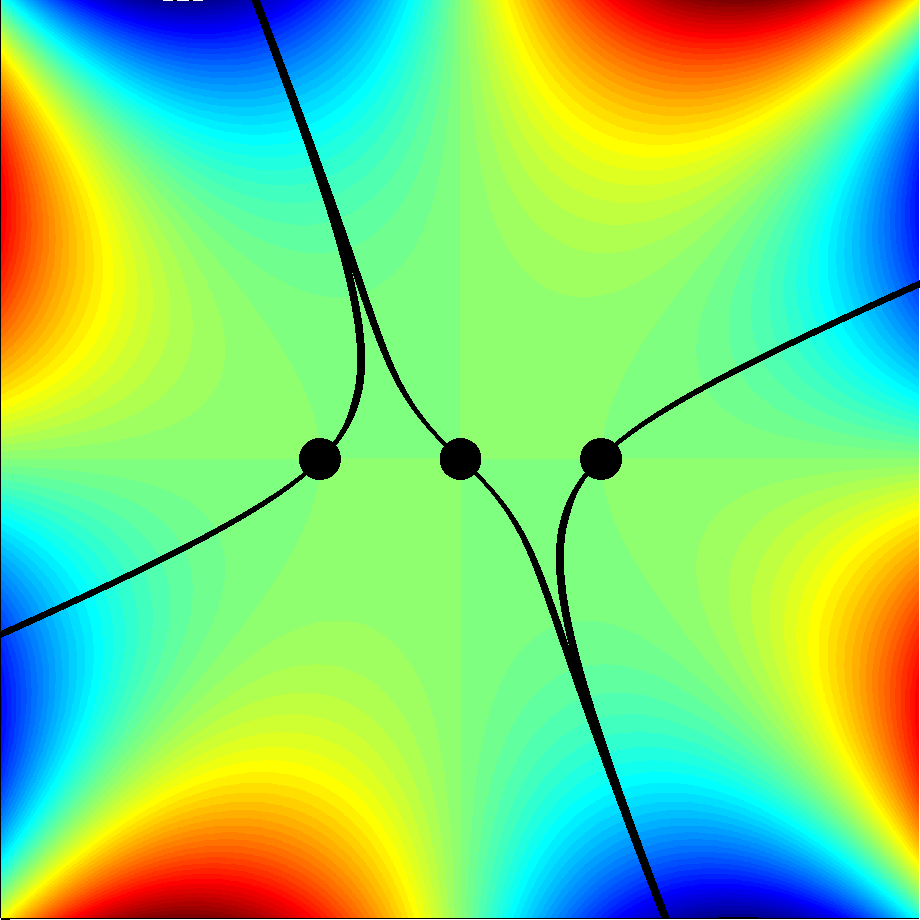}
}
\subfigure[(c) Point 2]{\includegraphics[width=\psize]{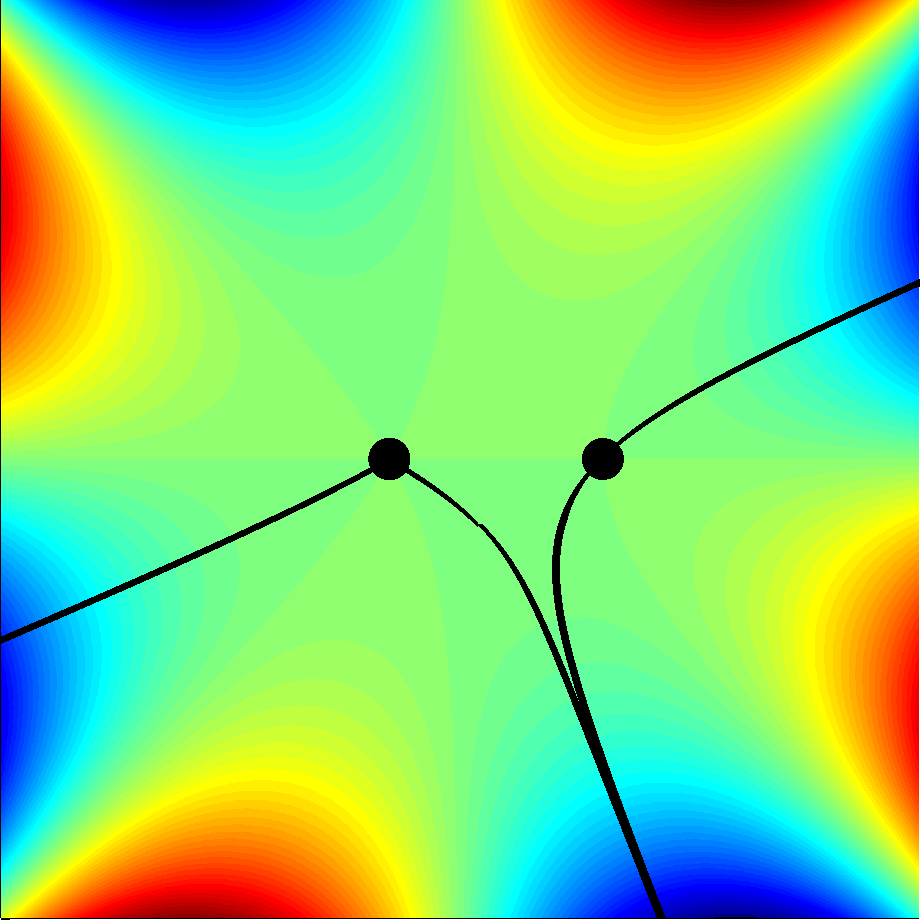}
}
\subfigure[(d) Point 3]{\includegraphics[width=\psize]{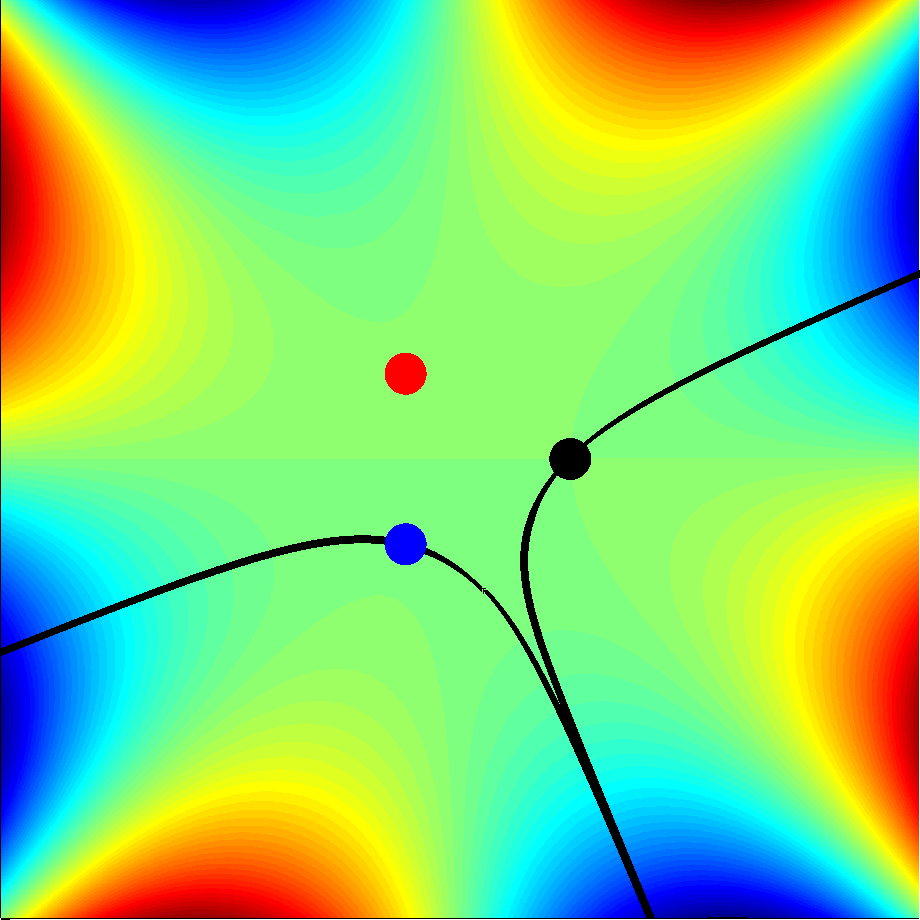}
}
\subfigure[(e) Point 4]{\includegraphics[width=\psize]{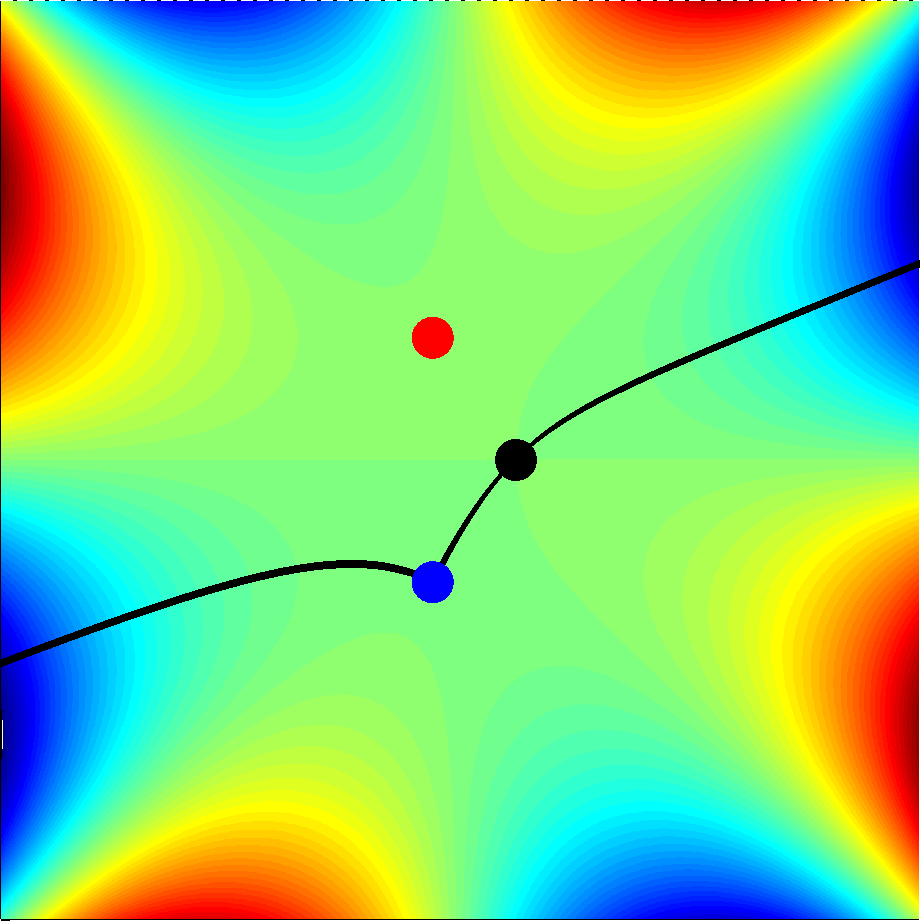}
}
\subfigure[(f) Point 5]{\includegraphics[width=\psize]{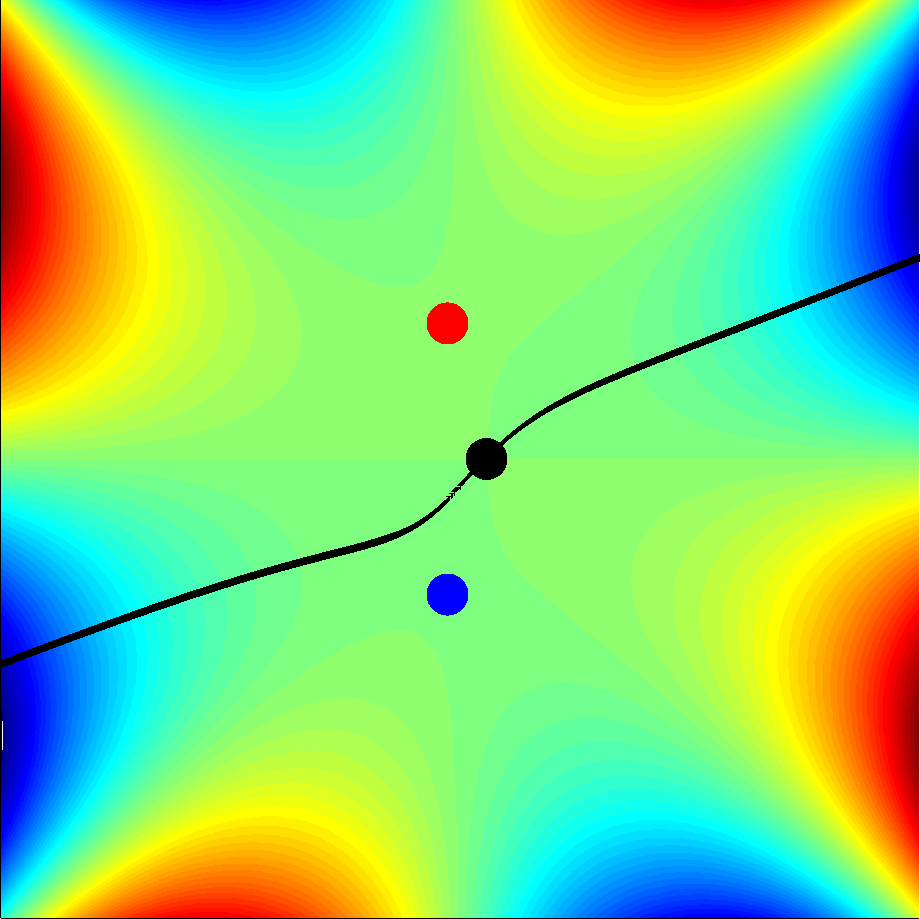}
}
\subfigure[(g) Point 6]{\includegraphics[width=\psize]{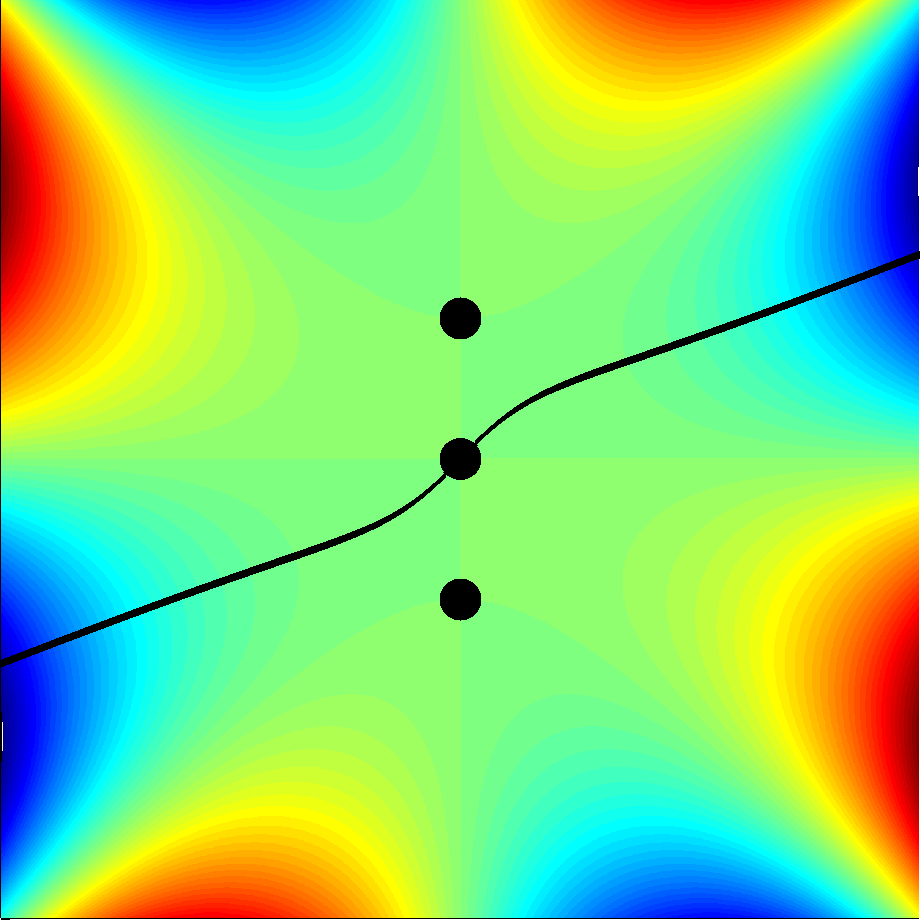}
}
\subfigure[(h) Point 7]{\includegraphics[width=\psize]{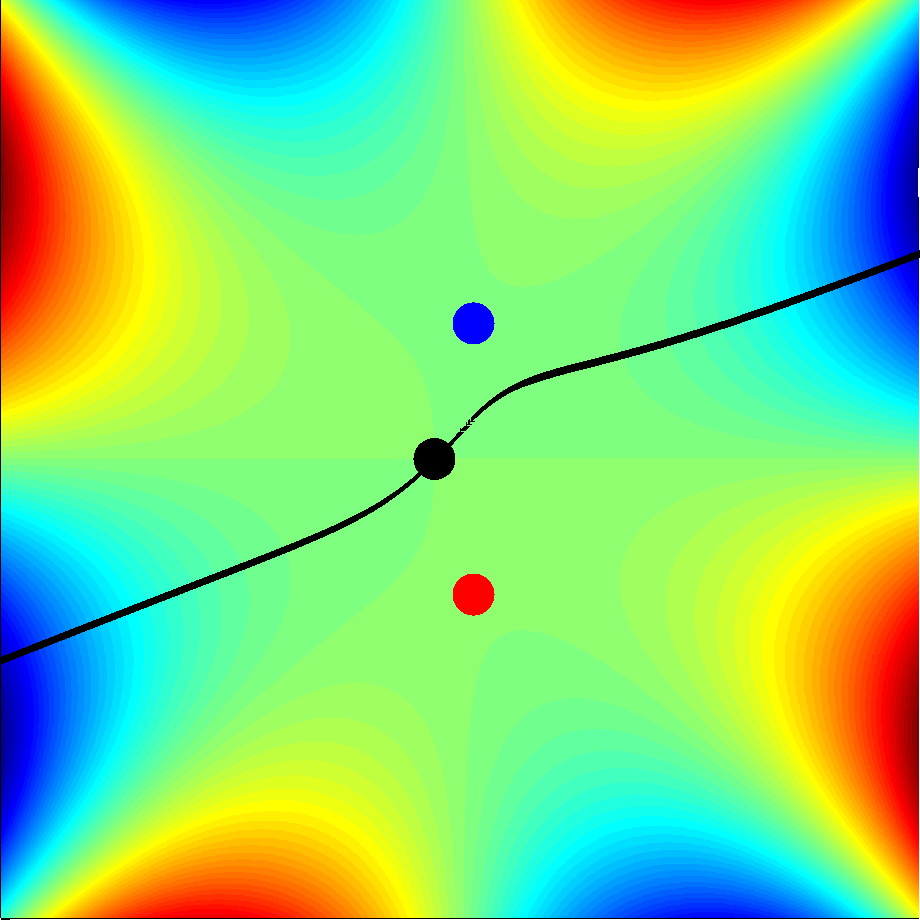}
}
\subfigure[(i) Point 8]{\includegraphics[width=\psize]{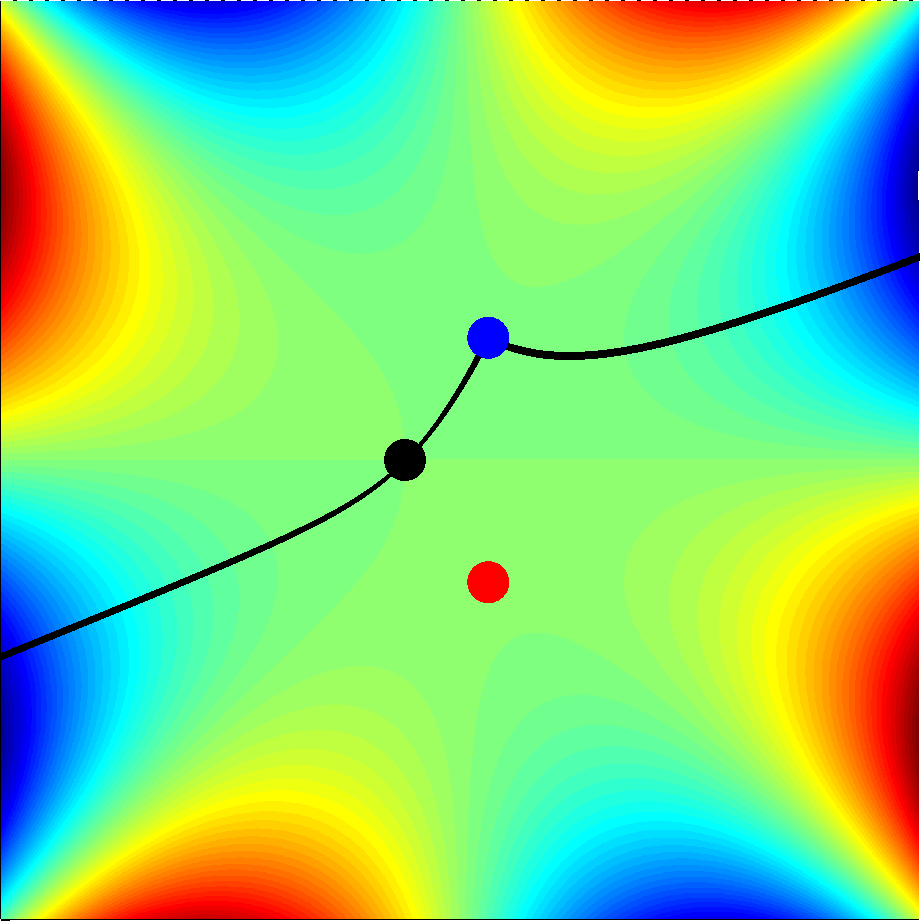}
}
\subfigure[(j) Point 9]{\includegraphics[width=\psize]{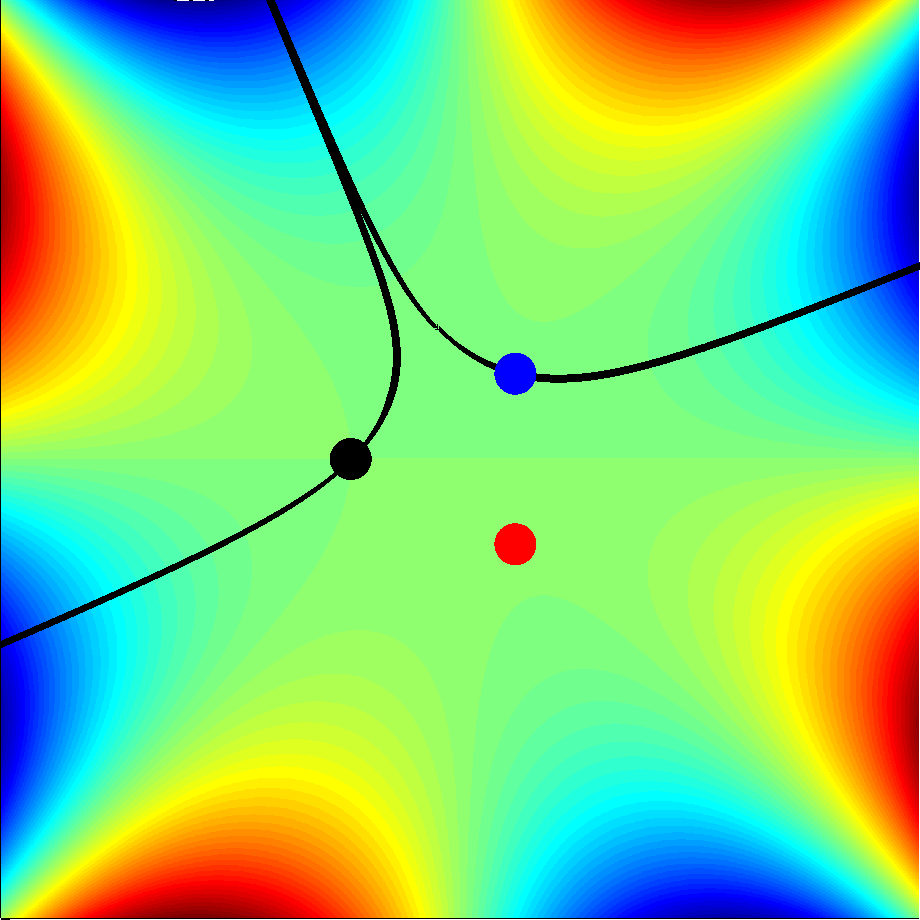}
}
\subfigure[(k) Point 10]{\includegraphics[width=\psize]{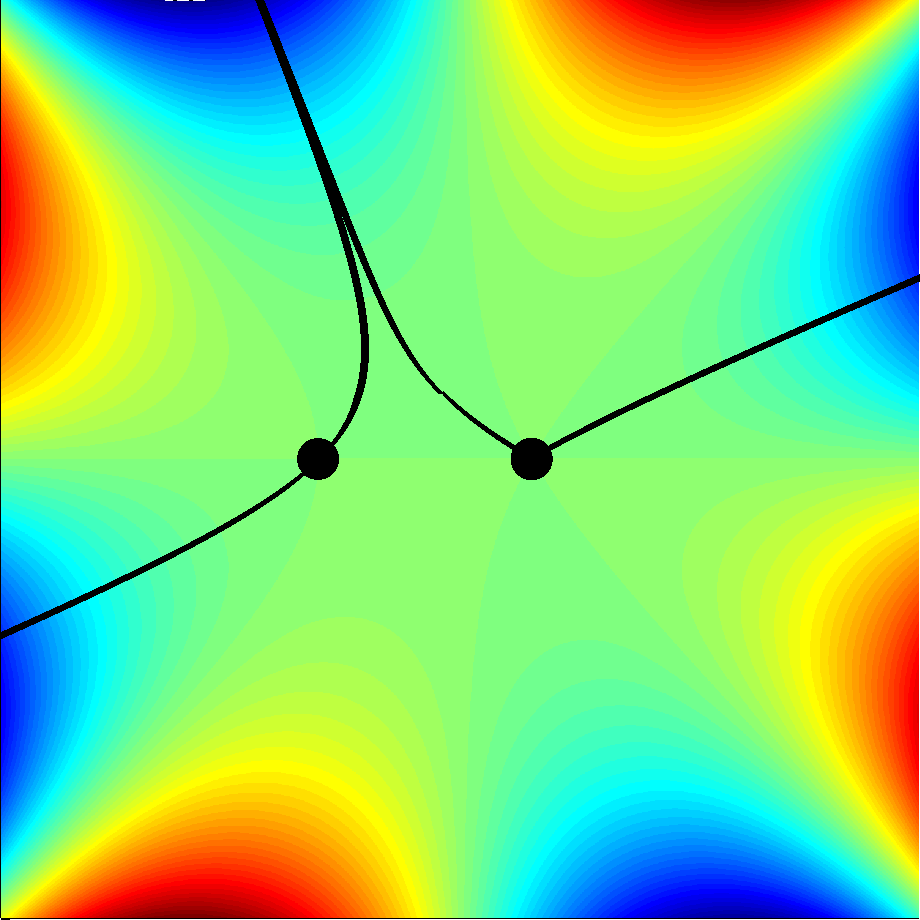}
}
\caption{Steepest descent contours for $A_{31}(\X,\Y)$ in the case $(l,m)=(2,1)$, $\alpha=1$, evaluated at points 1-10 from \F\ref{fig:PearceySaddlesFull}(a).
}
\label{fig:PearceySaddlesA31}
\end{center}
\end{figure}
\begin{itemize}
\item 
To the right of the localisation curve (cf.\ \F\ref{fig:PearceySaddlesA31}(b)) the steepest descent contour passes through the three real saddle points corresponding to the three families of real rays that exist everywhere in this region. The same arguments leading to 
\Fs\ref{fig:AirySaddlesA21}, \ref{fig:AirySaddlesA32} and \ref{fig:AirySaddlesA31} 
show that the ray picture is as in \F\ref{fig:PearceyBehaviours}(b), the second outgoing ray from the upper branch of the cusped caustic being necessary to generate the lower branch.
\item On the localisation curve (cf.\ \F\ref{fig:PearceySaddlesA31}(c) and (k)), the contour passes through the single real saddle point and the double real saddle point which generates the localisation.
\item Between the localisation curve and the Stokes lines (cf.\ \F\ref{fig:PearceySaddlesA31}(d) and (j)) the contour passes through the real saddle point and the complex saddle point which gives an exponentially small contribution in the far field. 
\item On the Stokes lines (cf.\ \F\ref{fig:PearceySaddlesA31}(e) and (i)) there is an abrupt change to the configuration of the contour, associated with the switching off of the exponentially small wave mentioned above.
\item To the left of the Stokes lines (cf.\ \F\ref{fig:PearceySaddlesA31}(f)-(h)) the contour passes only through the single real saddle point.
\end{itemize}
We now investigate the far-field localisation, as we did in \rf{eqn:AiryAsympt}. Without loss of generality, we focus on the upper branch of the localisation curve. With $\Gamma_{31}$ deformed to the real axis, we let
\begin{align*}
\label{}
\x = \x_0+\delta \x^*, \quad \y = (4/3)\kappa^{1/2} \x_0^{3/2}+\delta \y^*, \quad \tau = 2(\kappa \x_0)^{1/2}  + \tau',
\end{align*} 
for some fixed $\x_0>0$. The analogue of \rf{eqn:AiryPhase} is then
\begin{align}
\label{eqn:PearceyPhase}
-\frac{13\kappa \x_0^2}{3} - \delta (2 \y^* (\kappa \x_0)^{1/2} &+ 2\kappa \x_0 \x^*) - \tau' (\delta \y^* + 2(\kappa \x_0)^{1/2}\delta \x^*)- (\tau')^2\frac{\delta \x^*}{2} + \left(\frac{\x_0}{\kappa}\right)^{1/2}\frac{(\tau')^3}{6} + \frac{(\tau')^4}{48\kappa}.
\end{align}
Hence, with $\delta=k^{-2/3}$ and $\tau'=(4\kappa/\x_0)^{1/6} k^{-1/3}\zeta$, the analogue of \rf{eqn:AiryAsympt} is 
\begin{align}
k^{-1/12}\left(\frac{4\kappa}{\x_0}\right)^{1/6}&\exp  \left[-\ri\left(k\frac{13\kappa \x_0^2}{3} + k^{1/3}\left( 2(\kappa \x_0)^{1/2}\y^* + 2\kappa \x_0 \x^* \right)\right)\right]  \notag \\
&\times\int_{-\infty}^\infty \exp{\left[ \ri\left(-\left(\frac{4\kappa}{\x_0}\right)^{1/6}(\y^*+2(\kappa \x_0)^{1/2} \x^*)\zeta + \frac{\zeta^3}{3}\right) \right]}\,\rd \zeta. \notag \\
= 2\pi k^{-1/12}\left(\frac{4\kappa}{\x_0}\right)^{1/6}&\exp  \left[-\ri\left(k\frac{13\kappa \x_0^2}{3} + k^{1/3}\left( 2(\kappa \x_0)^{1/2}\y^* + 2\kappa \x_0 \x^* \right)\right)\right] \notag \\ 
& \times \Ai\left[
-\left(\frac{4\kappa}{\x_0}\right)^{1/6}(\y^*+2(\kappa \x_0)^{1/2} \x^*)\right].
\label{eqn:PearceyAsympt}
\end{align}
As the cusp is approached (i.e.\ as $\x_0\downarrow 0$), the local curvature of $\y=(4/3)\kappa^{1/2}\x^{3/2}$ at $\x=\x_0$ is approximately $(\kappa/\x_0)^{1/2}$, and it is this curvature that plays the role of $\kappa$ in \rf{eqn:AiryAsympt}.

Similar localisation occurs for the other five permissible contour choices, as \F\ref{fig:PearceyBehaviours} illustrates. 
But while $A_{21}$, $A_{43}$ and $A_{42}$ all exhibit localisation on both branches of the curve, $A_{41}$ and $A_{32}$ exhibit localisation only on one branch. All the solutions, except the classical solution $A_{31}$, suffer from exponential blow-up to the left of the localisation curve, which may limit their possible relevance to physical problems.
\subsection{\label{sec:Inflection} Cubic parabola ($(l,m)=(1,2)$, $2m+l=5$, $\lambda=1/5$, $\alpha = 2\sqrt{2}/(3\kappa)$)}
Localisation is now near the cubic parabola $y+\kappa^2\X^3/6=0$, 
and 
\begin{align}
\label{eqn:AIntCubic}
A_{ij} =  \int_{\Gamma_{ij}} \re^{\ri (-\Y t^2-\X t^4/2 + 2\alpha t^5/5)} \,\rd t = \int_{\Gamma_{ij}} \re^{\ri (-\Y t^2-\X t^4/2 + 4\sqrt{2}t^5/(15\kappa))} \,\rd t,
\end{align}
where $\Gamma_{ij}$ goes from $S_i$ to $S_j$ in \F\ref{Sectors}(c). 
Following the usual procedure we scale $\X=k^{1/5}\x$, $\Y=k^{3/5}\y$ and $t=k^{1/5}\tau$, giving the phase
\begin{align*}
\phi(\tau) = - \y\tau^2 -\frac{\x\tau^{4}}{2} + \frac{2\alpha \tau^{5}}{5}.
\end{align*}
There is always a saddle point at $\tau=0$, and the other three saddle points are at the roots of \mbox{$-\y-\x\tau^2+\alpha\tau^3 = 0$}. 
\begin{figure}[t!]
\def\size{25mm}
\begin{center}
\subfigure[(a)]{\includegraphics[width=27mm]{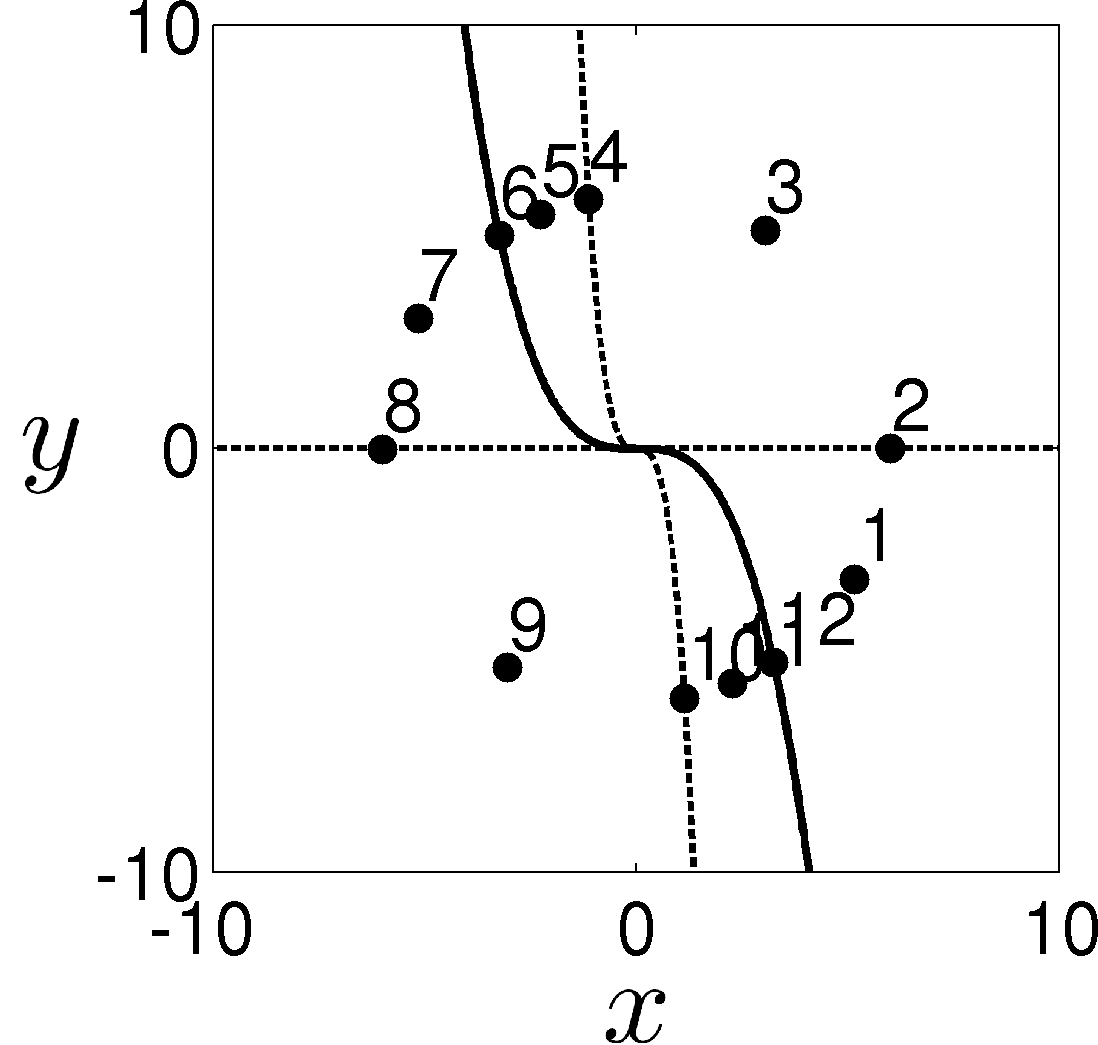}
}
\subfigure[(b) Point 1]{\includegraphics[width=\size]{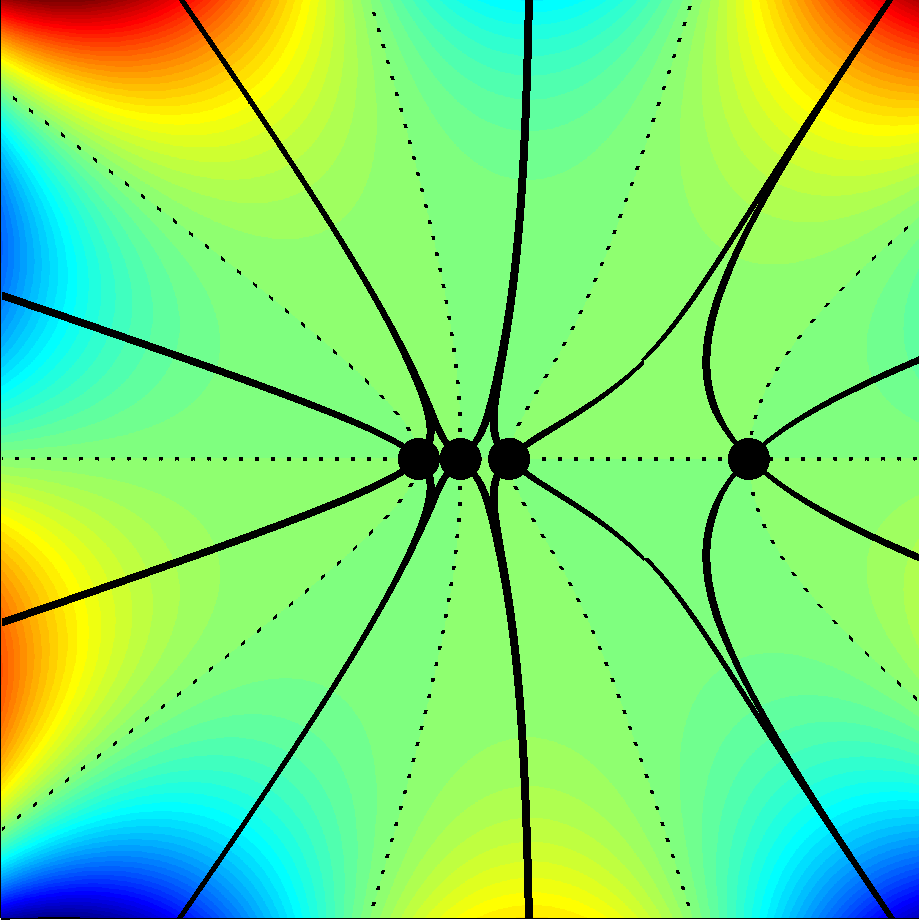}
}
\subfigure[(c) Point 2]{\includegraphics[width=\size]{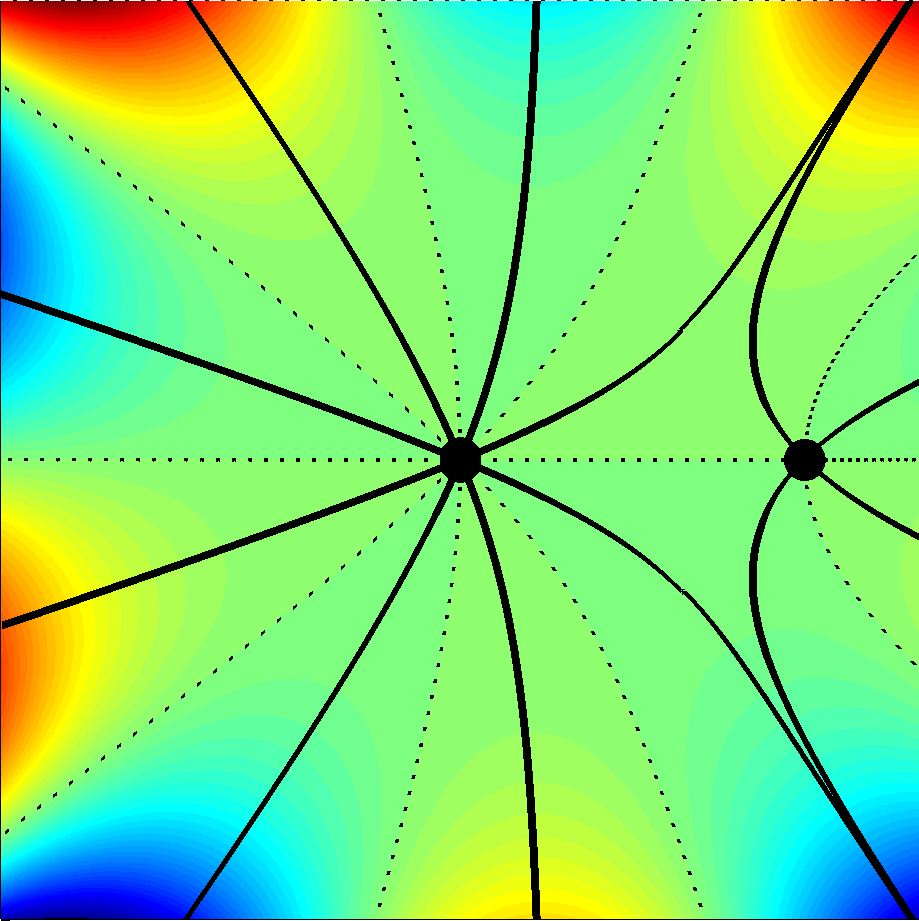}
}
\subfigure[(d) Point 3]{\includegraphics[width=\size]{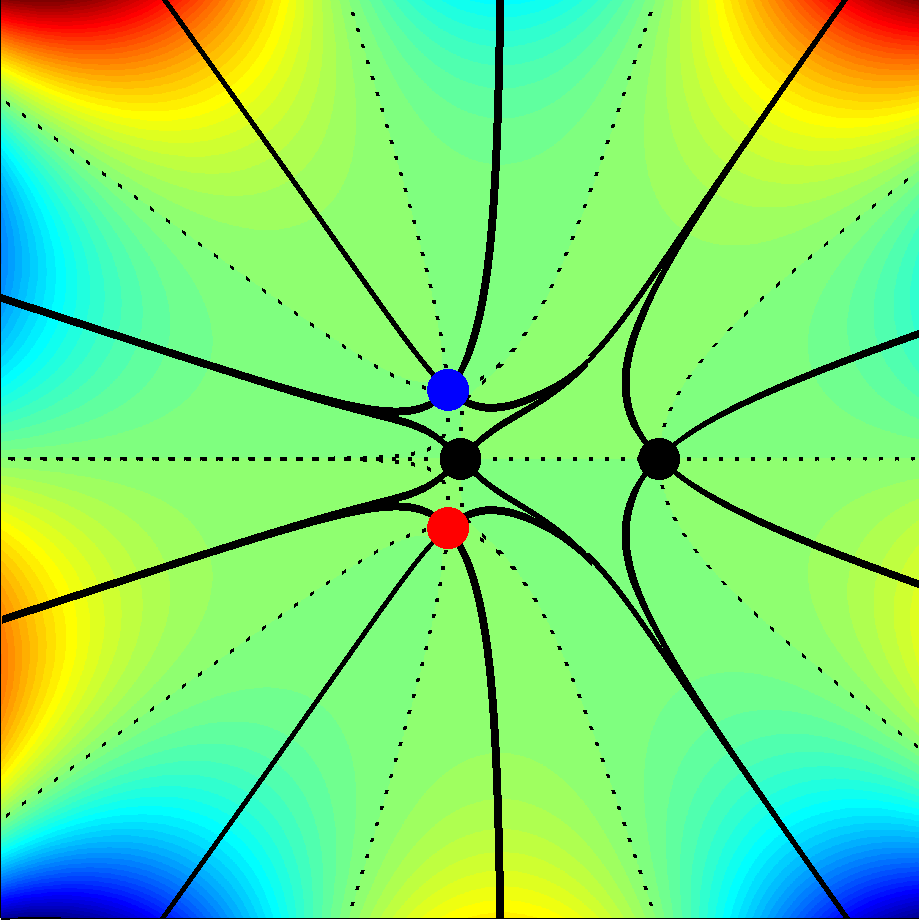}
}
\subfigure[(e) Point 4]{\includegraphics[width=\size]{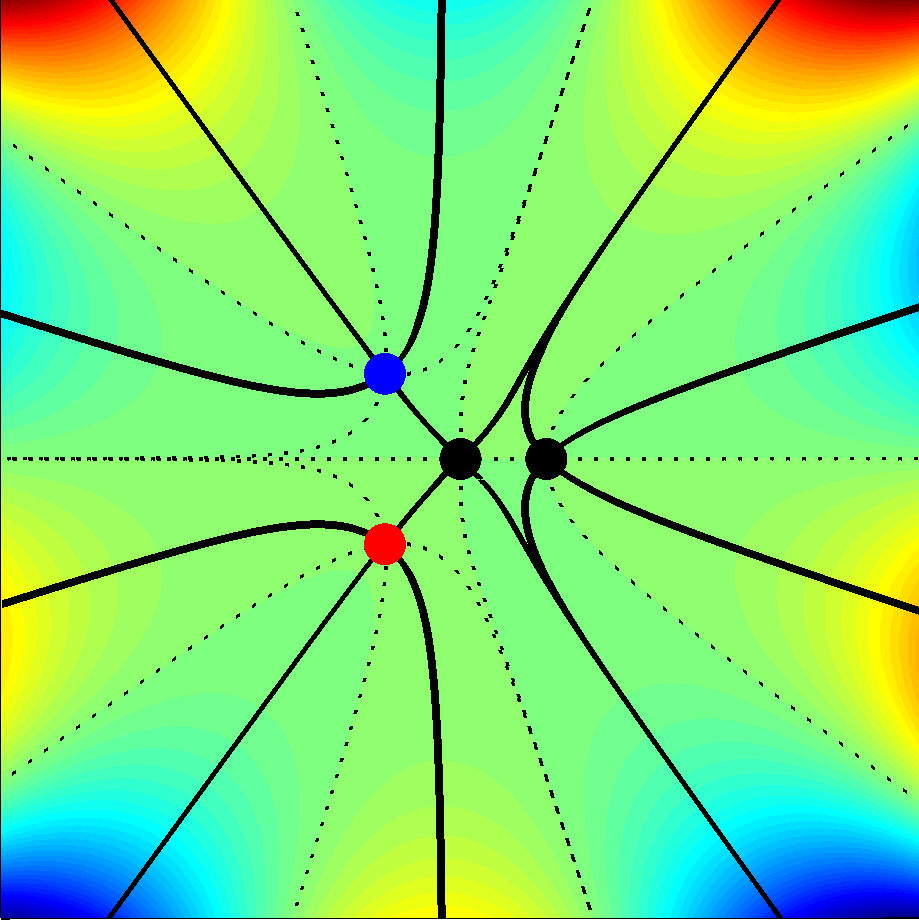}
}
\subfigure[(f) Point 5]{\includegraphics[width=\size]{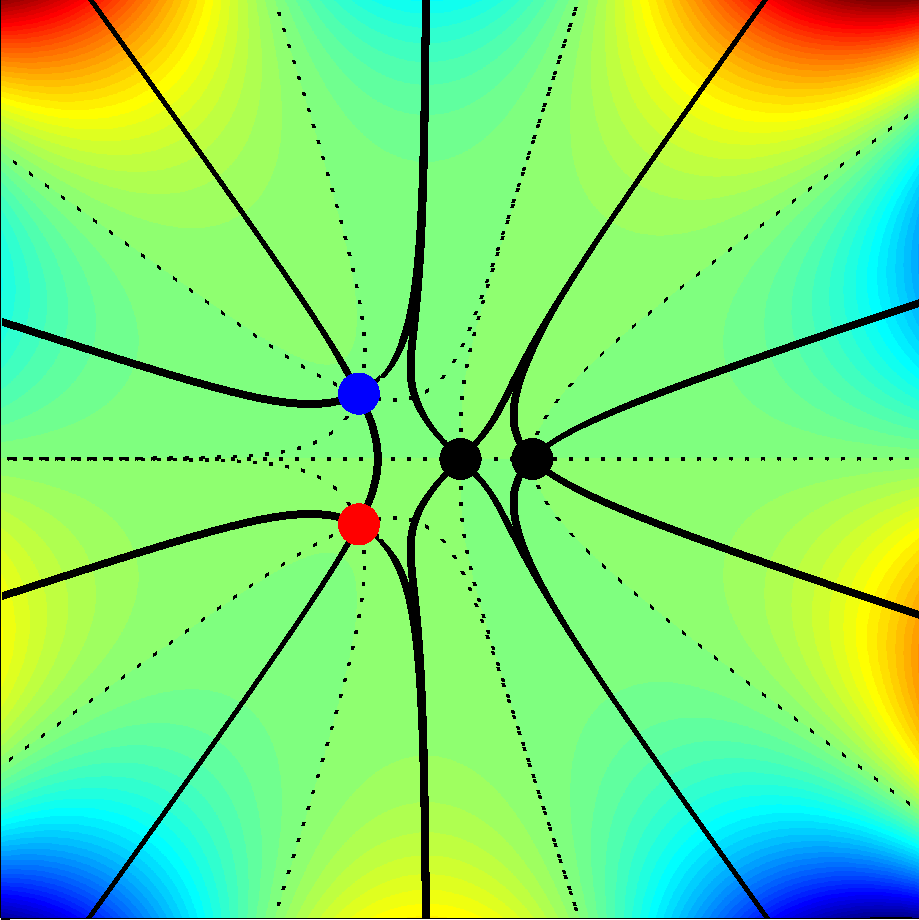}
}
\subfigure[(g) Point 6]{\includegraphics[width=\size]{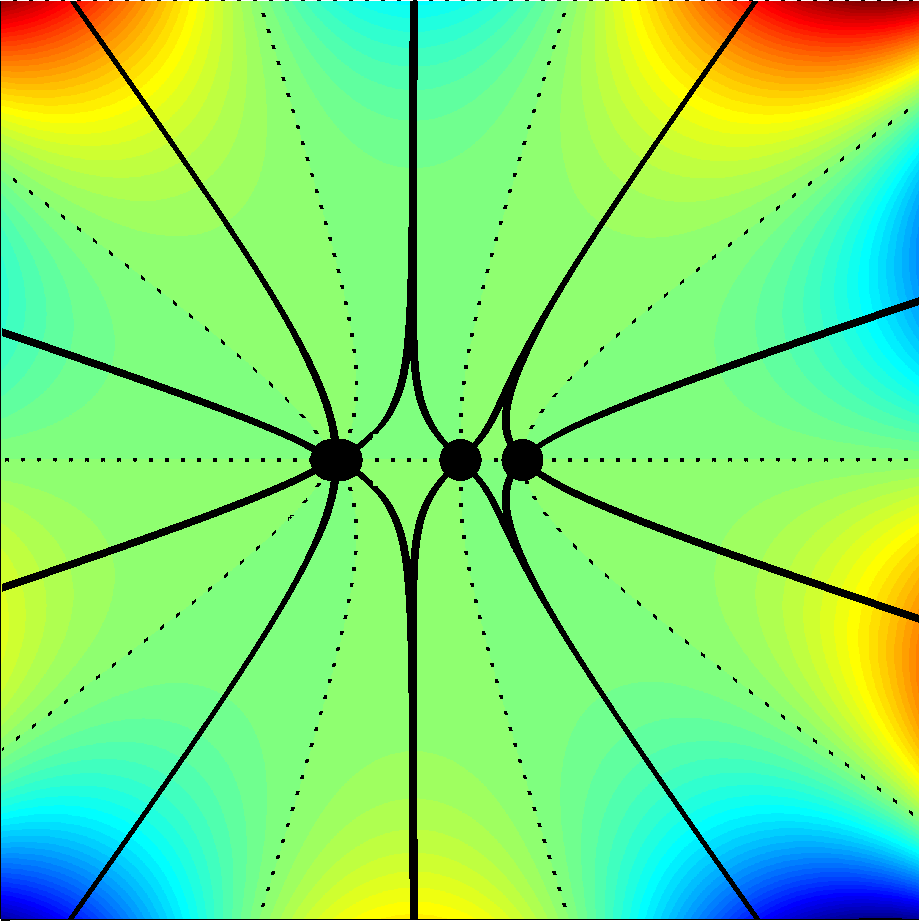}
}
\subfigure[(h) Point 7]{\includegraphics[width=\size]{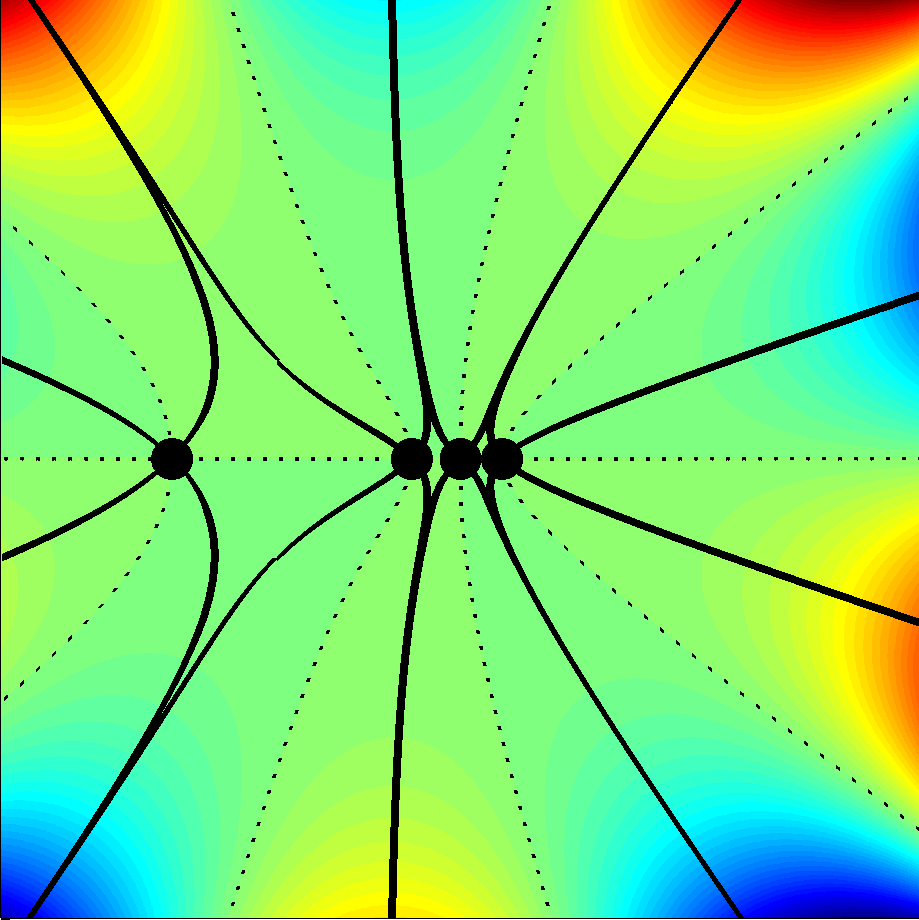}
}
\subfigure[(i) Point 8]{\includegraphics[width=\size]{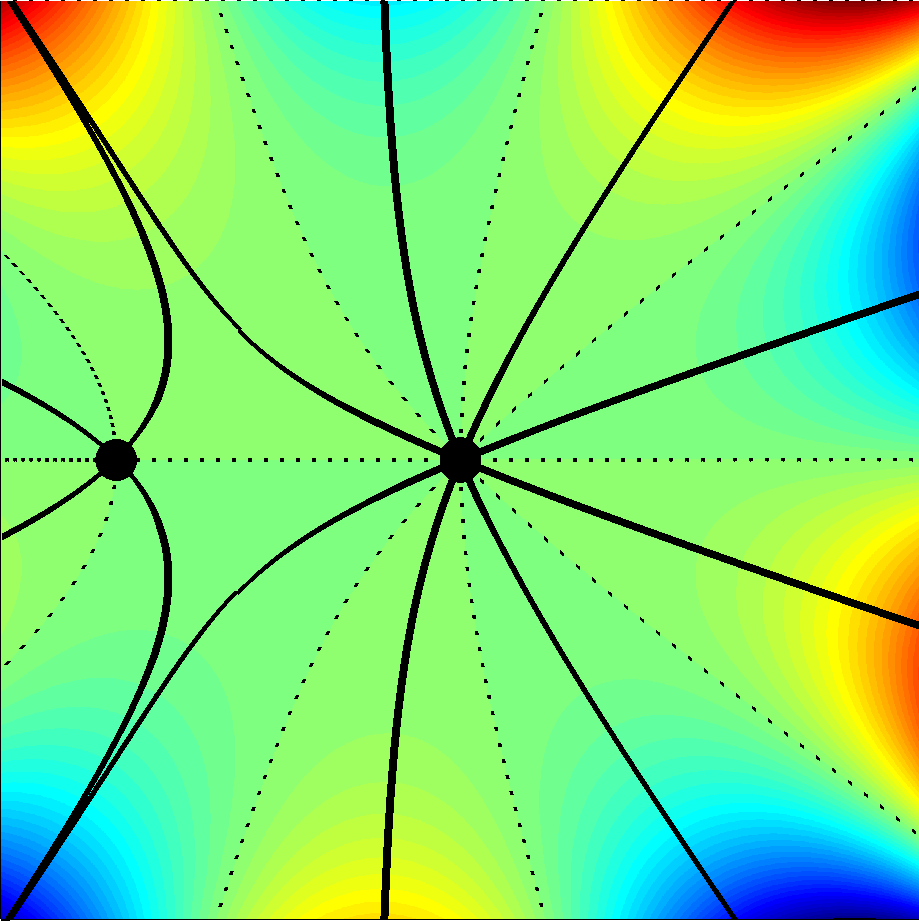}
}
\subfigure[(j) Point 9]{\includegraphics[width=\size]{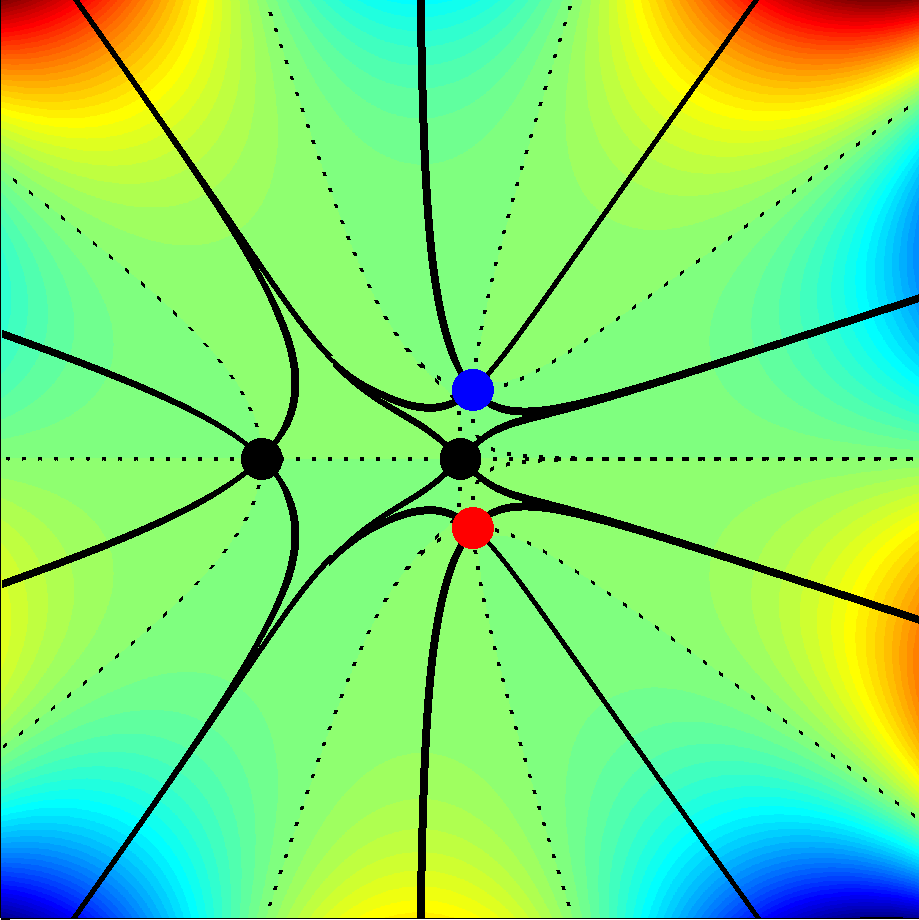}
}
\subfigure[(k) Point 10]{\includegraphics[width=\size]{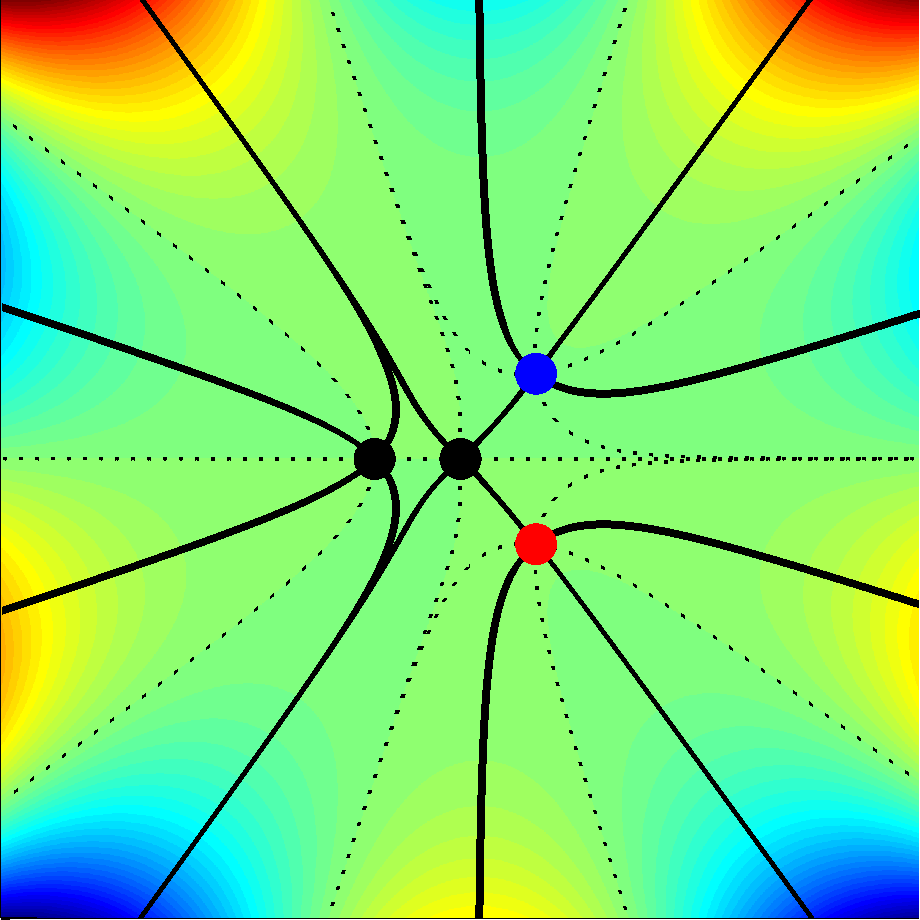}
}
\subfigure[(l) Point 11]{\includegraphics[width=\size]{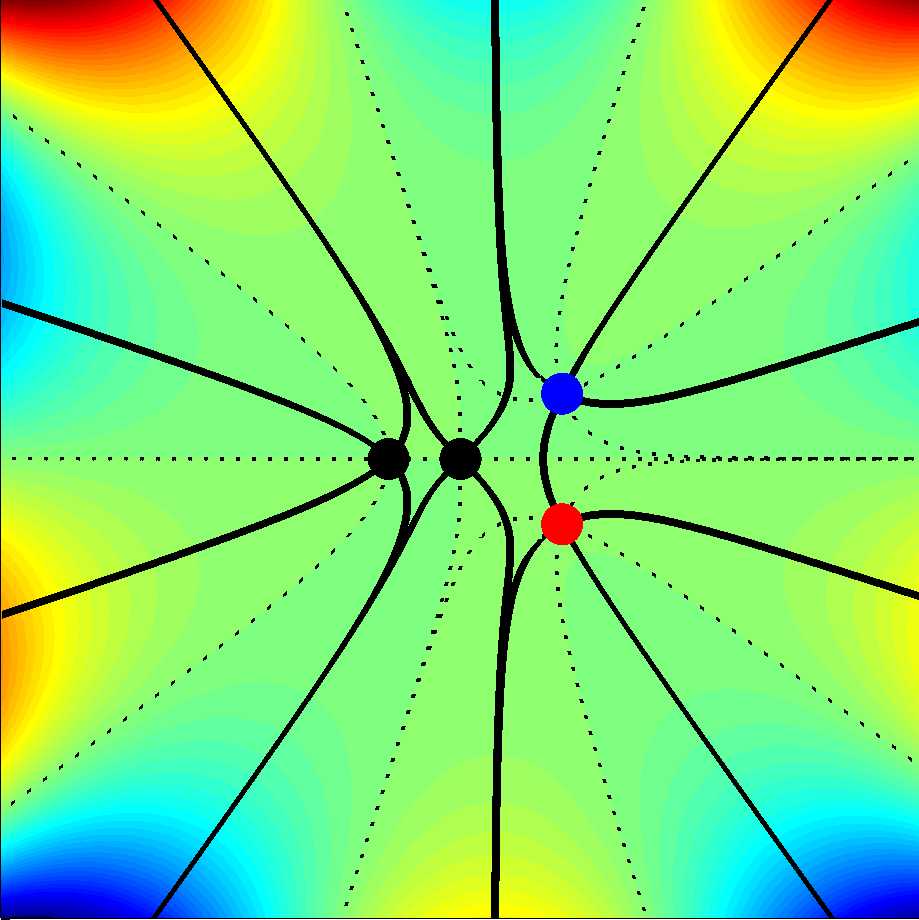}
}
\subfigure[(m) Point 12]{\includegraphics[width=\size]{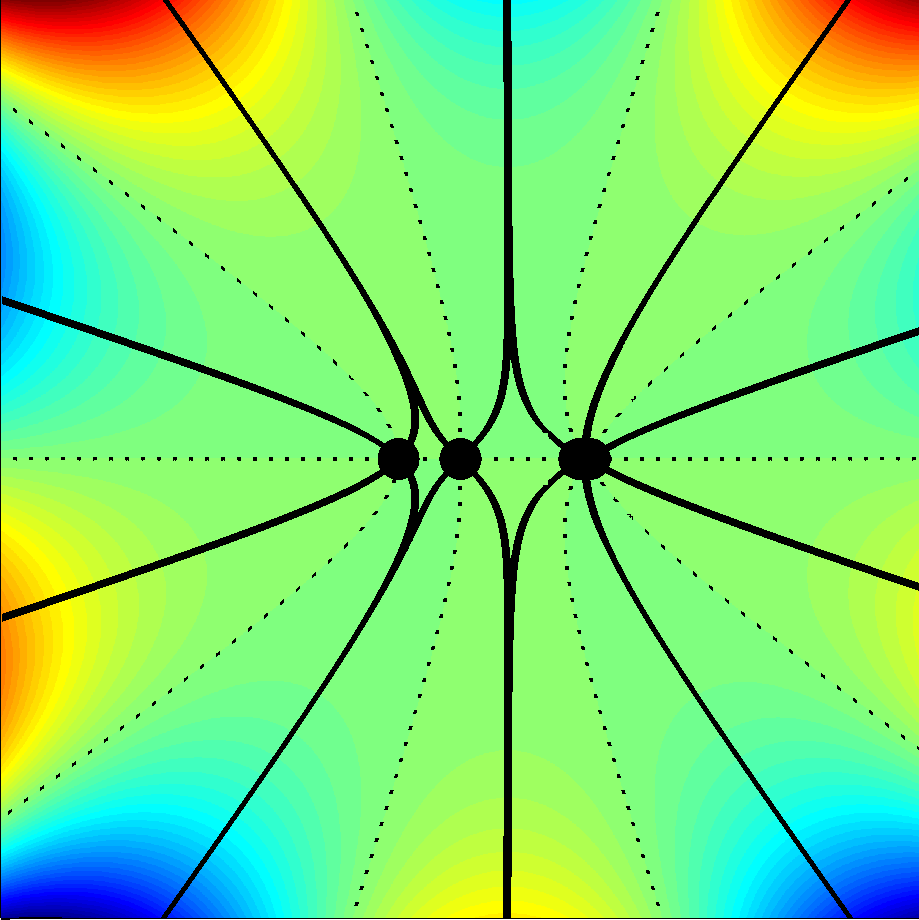}
}
\caption{Analogue of \F\ref{fig:AirySaddlesFull} 
for the case $(l,m)=(1,2)$, $\alpha=1$. In (a) we show the localisation curve $\y+4\x^3/27=0$ (solid curve), and the Stokes lines $\y+\mu_*\x^3=0$ and $\y=0$ (dashed), %
and in (b)-(m) the saddle point configurations for points 1--12 of (a).
}
\label{fig:QuinticSaddlesFull}
\end{center}
\end{figure}
\F\ref{fig:QuinticSaddlesFull} illustrates the saddle point configurations that can occur. By symmetry (cf.\ \rf{eqn:pRelnGeneral}), it is enough to consider in detail only points $(\x,\y)$ lying to the right of the localisation curve, i.e. with $\y+\kappa^2\x^3/6>0$, since the saddle point configuration for a point $(-\x,-\y)$ is simply the reflection in the imaginary $\tau$-axis of the configuration for the point $(\x,\y)$ (compare \Fs\ref{fig:QuinticSaddlesFull}(b)-(g) with \Fs\ref{fig:QuinticSaddlesFull}(h)-(m)). 
There are therefore four possibilities:
\begin{itemize}
\item 
If $\y+\kappa^2\x^3/6>0$ and $\y<0$ (point 1 in \F\ref{fig:QuinticSaddlesFull}(a)) there are four distinct real saddle points (\F\ref{fig:QuinticSaddlesFull}(b)). 
\item
If $\y+\kappa^2\x^3/6>0$ and $\y=0$ (point 2 in \F\ref{fig:QuinticSaddlesFull}(a)) there are again four real saddle points, but three of them coincide at $\tau=0$ (\F\ref{fig:QuinticSaddlesFull}(c)).%
\item
If $\y+\kappa^2\x^3/6>0$ and $\y>0$ (points 3-5 in \F\ref{fig:QuinticSaddlesFull}(a)) there are two distinct real saddle points and a pair of complex conjugate saddle points (\F\ref{fig:QuinticSaddlesFull}(d)-(f)). 
\item If $\y+\kappa^2\x^3/6=0$ and $\y>0$ (point 6 in \F\ref{fig:QuinticSaddlesFull}(a)) there are four real saddle points, but two of them coincide (\F\ref{fig:QuinticSaddlesFull}(g)).
\end{itemize}
There is a Stokes line at $y=0$ (shown as a dashed line in \F\ref{fig:QuinticSaddlesFull}(a)) associated with the coalescence of the saddle points at $\tau=0$. Furthermore, in the region $\y+\kappa^2\x^3/6>0$ and $\y>0$ there is another Stokes line (also shown as a dashed line in \F\ref{fig:QuinticSaddlesFull}(a)) along which the real part of the phase is the same for three of the saddle points (as for point 4 in \F\ref{fig:QuinticSaddlesFull}(a); see \F\ref{fig:QuinticSaddlesFull}(e)). By generalising the approach of \cite{Wr:80,StSp:83}, one can show that this Stokes line is %
$\y+(9\kappa^2/8)\mu_*\x^{3}=0$, where $\mu_*\approx 4.069$ is the unique root in the interval $(4/27,\infty)$ of the equation
\begin{align*}
\label{}
\real{g(\mu)^4-6\mu g(\mu)^2}=0,
\end{align*}
where 
\begin{align*}
\label{}
g(\mu)=\re^{-\ri\pi/3}d(\mu) + \re^{\ri\pi/3}/(9d(\mu)) + 1/3,
\qquad
d(\mu)=\left(\mu/2-1/27-\sqrt{\mu^2/4-\mu/27}\right)^{1/3}.
\end{align*}
\newcommand{\paa}{\draw [->] (2.83,-1.21) -- (3.71,-1.51);}
\newcommand{\pab}{\draw [->] (2.83,-1.21) -- (3.68,-1.73);}
\newcommand{\pac}{\draw [->] (2.83,-1.21) -- (3.01,-2.39);}
\newcommand{\pad}{\draw [dotted] (2.83,-1.21)-- (-0.89,0.1);}
\newcommand{\pae}{\draw [dotted] (2.83,-1.21)-- (1.16,-0.23);}
\newcommand{\paf}{\draw [dotted] (3.01,-2.39)-- (4.01,-9.52);}
\newcommand{\pag}{\draw [->] (2.83,-1.21) -- (3.7,-1.21);}
\newcommand{\pah}{\fill (2.83,-1.21) circle (1.5pt);}
\newcommand{\pba}{\draw [->] (-2.83,1.21) -- (-2.03,0.92);}
\newcommand{\pbb}{\draw [->] (-2.83,1.21) -- (-2.03,0.75);}
\newcommand{\pbc}{\draw [->] (-2.83,1.21) -- (-2.68,0.04);}
\newcommand{\pbd}{\draw [dotted] (-2.83,1.21)-- (0.89,-0.1);}
\newcommand{\pbe}{\draw [dotted] (-2.83,1.21)-- (-1.16,0.23);}
\newcommand{\pbf}{\draw [dotted] (-2.68,0.04)-- (-4.01,9.55);}
\newcommand{\pbg}{\draw [->] (-2.83,1.21) -- (-2.13,1.21);}
\newcommand{\pbh}{\fill (-2.83,1.21) circle (1.5pt);}
\newcommand{\pca}{\draw [->] (-1.7,2.5) -- (-1.07,1.92);}
\newcommand{\pcb}{\draw [dotted] (-1.7,2.5)-- (1.46,-0.46);}
\newcommand{\pcc}{\draw [->] (-1.7,2.5) -- (-0.97,2.5);}
\newcommand{\pcd}{\fill (-1.7,2.5) circle (1.5pt);}
\newcommand{\pda}{\draw [->] (1.67,-2.53) -- (2.34,-3.14);}
\newcommand{\pdb}{\draw [dotted] (1.67,-2.53)-- (-1.46,0.46);}
\newcommand{\pdc}{\draw [->] (1.67,-2.53) -- (2.43,-2.53);}
\newcommand{\pdd}{\fill (1.7,-2.5) circle (1.5pt);}
\newcommand{\pea}{\draw [->] (1.5,2.5) -- (1.81,1.15);}
\newcommand{\peb}{\draw [dotted] (1.5,2.5)-- (3.12,-4.48);}
\newcommand{\pec}{\draw [->] (1.5,2.5) -- (2.3,2.5);}
\newcommand{\ped}{\fill (1.5,2.5) circle (1.5pt);}
\newcommand{\pfa}{\draw [->] (-1.5,-2.5) -- (-1.18,-3.89);}
\newcommand{\pfb}{\draw [dotted] (-1.5,-2.5)-- (-3.12,4.5);}
\newcommand{\pfc}{\draw [->] (-1.5,-2.5) -- (-0.71,-2.5);}
\newcommand{\pfd}{\fill (-1.5,-2.5) circle (1.5pt);}

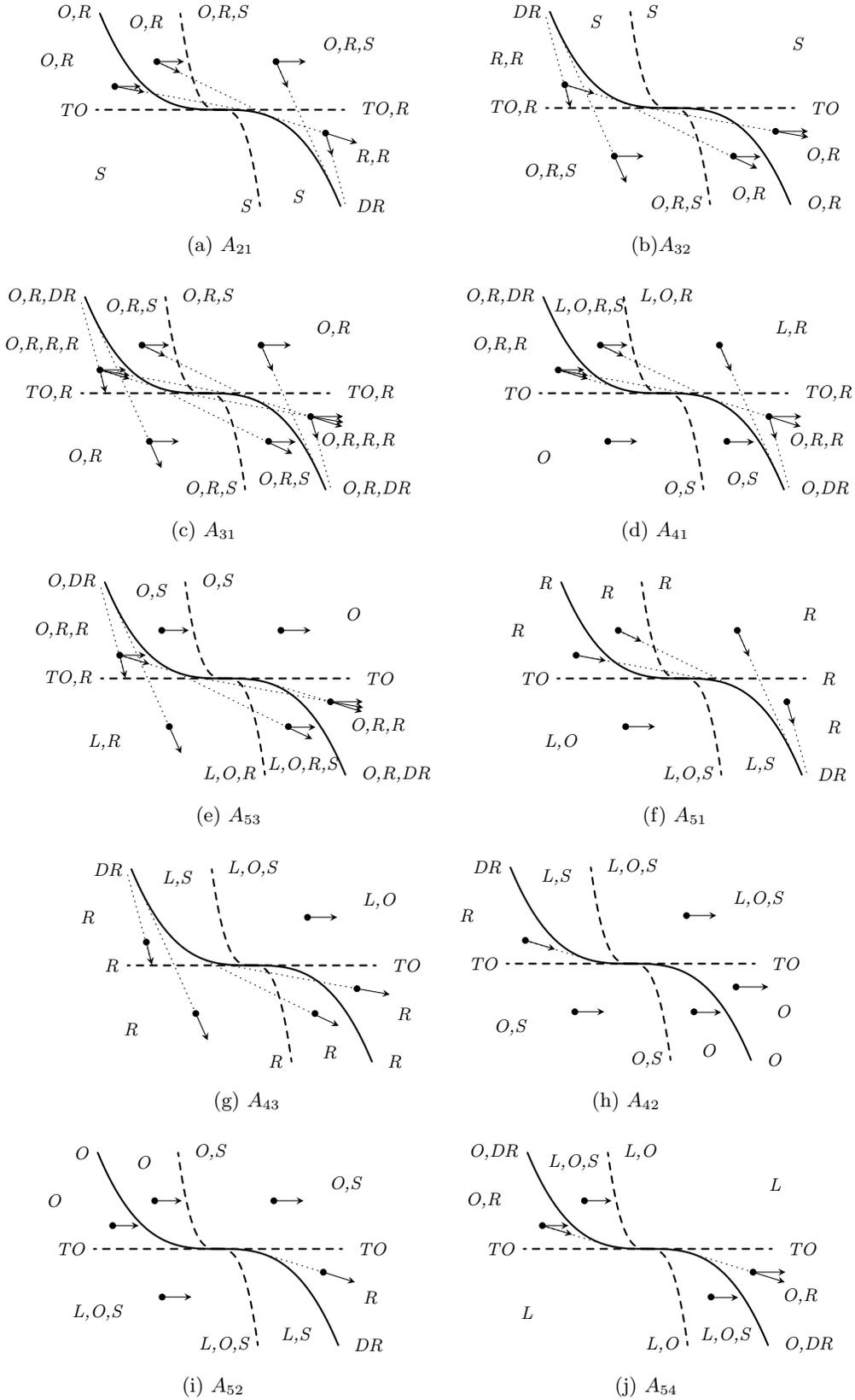
\begin{figure}[tbp!]
\begin{center}
\subfigure[(a) $A_{21} $] {
\begin{tikzpicture}[line cap=round,line join=round,>=stealth,x=0.58cm,y=0.3cm,scale=1]
\def\size{5};\def\lambdastar{4.069026673648571};
\def\Ta{R,R};\def\Tb{TO,R};\def\Tc{O,R,S};\def\Td{O,R,S};\def\Te{O,R};\def\Tf{O,R};\def\Tg{O,R};\def\Th{TO};\def\Ti{S};\def\Tj{S};\def\Tk{S};\def\Tl{DR};
\draw (4*\size/5,-\size/2) node {\tf\Ta};
\draw[thick, dashed] (0,0) -- (2*\size/3,0) node[right] {\tf\Tb};
\draw (2*\size/3,2*\size/3) node {\tf\Tc};
\draw[thick, dashed, smooth,domain=0:(\size/\lambdastar)^(1/3),variable=\t] plot({-\t},{\lambdastar*\t^3}) node[right] {\tf\Td};
\draw (-0.42*\size,0.9*\size) node {\tf\Te};
\draw[thick, smooth,domain=0:(27*\size/4)^(1/3),variable=\t] plot({-\t},{4*\t^3/27}) node[left] {\tf\Tf};
\draw (-0.9*\size,\size/2) node {\tf\Tg};
\draw[thick, dashed] (0,0) -- (-2*\size/3,0) node[left] {\tf\Th};
\draw (-2*\size/3,-2*\size/3) node {\tf\Ti};
\draw[thick, dashed, smooth,domain=0:(\size/\lambdastar)^(1/3),variable=\t] plot({\t},{-\lambdastar*\t^3}) node[left] {\tf\Tj};
\draw (0.4*\size,-0.9*\size) node {\tf\Tk};
\draw[thick, smooth,domain=0:(27*\size/4)^(1/3),variable=\t] plot({\t},{-4*\t^3/27}) node[right] {\tf\Tl};
\begin{scope}\clip (-2*\size/3,-\size) rectangle(1.2*2*\size/3,\size);
\pab\pac\pae\paf\pah
\pba\pbd\pbg\pbh
\pca\pcb\pcc\pcd
\pea\peb\pec\ped
\end{scope}
\end{tikzpicture}}
\hs{5}
\subfigure[(b)$A_{32} $] {
\begin{tikzpicture}[line cap=round,line join=round,>=stealth,x=0.58cm,y=0.3cm,scale=1]
\def\size{5};\def\lambdastar{4.069026673648571};
\def\Tg{R,R};\def\Th{TO,R};\def\Ti{O,R,S};\def\Tj{O,R,S};\def\Tk{O,R};\def\Tl{O,R};\def\Ta{O,R};\def\Tb{TO};\def\Tc{S};\def\Td{S};\def\Te{S};\def\Tf{DR};
\draw (4*\size/5,-\size/2) node {\tf\Ta};
\draw[thick, dashed] (0,0) -- (2*\size/3,0) node[right] {\tf\Tb};
\draw (2*\size/3,2*\size/3) node {\tf\Tc};
\draw[thick, dashed, smooth,domain=0:(\size/\lambdastar)^(1/3),variable=\t] plot({-\t},{\lambdastar*\t^3}) node[right] {\tf\Td};
\draw (-0.42*\size,0.9*\size) node {\tf\Te};
\draw[thick, smooth,domain=0:(27*\size/4)^(1/3),variable=\t] plot({-\t},{4*\t^3/27}) node[left] {\tf\Tf};
\draw (-0.9*\size,\size/2) node {\tf\Tg};
\draw[thick, dashed] (0,0) -- (-2*\size/3,0) node[left] {\tf\Th};
\draw (-2*\size/3,-2*\size/3) node {\tf\Ti};
\draw[thick, dashed, smooth,domain=0:(\size/\lambdastar)^(1/3),variable=\t] plot({\t},{-\lambdastar*\t^3}) node[left] {\tf\Tj};
\draw (0.4*\size,-0.9*\size) node {\tf\Tk};
\draw[thick, smooth,domain=0:(27*\size/4)^(1/3),variable=\t] plot({\t},{-4*\t^3/27}) node[right] {\tf\Tl};
\begin{scope}\clip (-2*\size/3,-\size) rectangle(1.2*2*\size/3,\size);
\paa\pad\pag\pah
\pbb\pbc\pbe\pbf\pbh
\pda\pdb\pdc\pdd
\pfa\pfb\pfc\pfd
\end{scope}
\end{tikzpicture}}\\
\hs{-5}
\subfigure[(c) $A_{31} $] {
\begin{tikzpicture}[line cap=round,line join=round,>=stealth,x=0.58cm,y=0.3cm]
\def\size{5};\def\lambdastar{4.069026673648571};
\def\Ta{O,R,R,R};\def\Tb{TO,R};\def\Tc{O,R};\def\Td{O,R,S};\def\Te{O,R,S};\def\Tf{O,R,DR};\def\Tg{O,R,R,R};\def\Th{TO,R};\def\Ti{O,R};\def\Tj{O,R,S};\def\Tk{O,R,S};\def\Tl{O,R,DR};
\draw (4*\size/5,-\size/2) node {\tf\Ta};
\draw[thick, dashed] (0,0) -- (2*\size/3,0) node[right] {\tf\Tb};
\draw (2*\size/3,2*\size/3) node {\tf\Tc};
\draw[thick, dashed, smooth,domain=0:(\size/\lambdastar)^(1/3),variable=\t] plot({-\t},{\lambdastar*\t^3}) node[right] {\tf\Td};
\draw (-0.42*\size,0.9*\size) node {\tf\Te};
\draw[thick, smooth,domain=0:(27*\size/4)^(1/3),variable=\t] plot({-\t},{4*\t^3/27}) node[left] {\tf\Tf};
\draw (-0.9*\size,\size/2) node {\tf\Tg};
\draw[thick, dashed] (0,0) -- (-2*\size/3,0) node[left] {\tf\Th};
\draw (-2*\size/3,-2*\size/3) node {\tf\Ti};
\draw[thick, dashed, smooth,domain=0:(\size/\lambdastar)^(1/3),variable=\t] plot({\t},{-\lambdastar*\t^3}) node[left] {\tf\Tj};
\draw (0.4*\size,-0.9*\size) node {\tf\Tk};
\draw[thick, smooth,domain=0:(27*\size/4)^(1/3),variable=\t] plot({\t},{-4*\t^3/27}) node[right] {\tf\Tl};
\begin{scope}\clip (-2*\size/3,-\size) rectangle(1.2*2*\size/3,\size);
\paa\pab\pac\pad\pae\paf\pag\pah
\pba\pbb\pbc\pbd\pbe\pbf\pbg\pbh
\pca\pcb\pcc\pcd
\pda\pdb\pdc\pdd
\pea\peb\pec\ped
\pfa\pfb\pfc\pfd
\end{scope}
\end{tikzpicture}}
\hs{1}
\subfigure[(d) $A_{41} $] {
\begin{tikzpicture}[line cap=round,line join=round,>=stealth,x=0.58cm,y=0.3cm]
\def\size{5};\def\lambdastar{4.069026673648571};
\def\Ta{O,R,R};\def\Tb{TO,R};\def\Tc{L,R};\def\Td{L,O,R};\def\Te{L,O,R,S};\def\Tf{O,R,DR};\def\Tg{O,R,R};\def\Th{TO};\def\Ti{O};\def\Tj{O,S};\def\Tk{O,S};\def\Tl{O,DR};
\draw (4*\size/5,-\size/2) node {\tf\Ta};
\draw[thick, dashed] (0,0) -- (2*\size/3,0) node[right] {\tf\Tb};
\draw (2*\size/3,2*\size/3) node {\tf\Tc};
\draw[thick, dashed, smooth,domain=0:(\size/\lambdastar)^(1/3),variable=\t] plot({-\t},{\lambdastar*\t^3}) node[right] {\tf\Td};
\draw (-0.42*\size,0.9*\size) node {\tf\Te};
\draw[thick, smooth,domain=0:(27*\size/4)^(1/3),variable=\t] plot({-\t},{4*\t^3/27}) node[left] {\tf\Tf};
\draw (-0.9*\size,\size/2) node {\tf\Tg};
\draw[thick, dashed] (0,0) -- (-2*\size/3,0) node[left] {\tf\Th};
\draw (-2*\size/3,-2*\size/3) node {\tf\Ti};
\draw[thick, dashed, smooth,domain=0:(\size/\lambdastar)^(1/3),variable=\t] plot({\t},{-\lambdastar*\t^3}) node[left] {\tf\Tj};
\draw (0.4*\size,-0.9*\size) node {\tf\Tk};
\draw[thick, smooth,domain=0:(27*\size/4)^(1/3),variable=\t] plot({\t},{-4*\t^3/27}) node[right] {\tf\Tl};
\begin{scope}\clip (-2*\size/3,-\size) rectangle(1.2*2*\size/3,\size);
\pab\pac\pae\paf\pag\pah
\pba\pbb\pbd\pbe\pbg\pbh
\pca\pcb\pcc\pcd
\pdc\pdd
\pea\peb\ped
\pfc\pfd
\end{scope}
\end{tikzpicture}}\\
\subfigure[(e) $A_{53} $] {
\begin{tikzpicture}[line cap=round,line join=round,>=stealth,x=0.58cm,y=0.3cm]
\def\size{5};\def\lambdastar{4.069026673648571};
\def\Tg{O,R,R};\def\Th{TO,R};\def\Ti{L,R};\def\Tj{L,O,R};\def\Tk{L,O,R,S};\def\Tl{O,R,DR};\def\Ta{O,R,R};\def\Tb{TO};\def\Tc{O};\def\Td{O,S};\def\Te{O,S};\def\Tf{O,DR};
\draw (4*\size/5,-\size/2) node {\tf\Ta};
\draw[thick, dashed] (0,0) -- (2*\size/3,0) node[right] {\tf\Tb};
\draw (2*\size/3,2*\size/3) node {\tf\Tc};
\draw[thick, dashed, smooth,domain=0:(\size/\lambdastar)^(1/3),variable=\t] plot({-\t},{\lambdastar*\t^3}) node[right] {\tf\Td};
\draw (-0.42*\size,0.9*\size) node {\tf\Te};
\draw[thick, smooth,domain=0:(27*\size/4)^(1/3),variable=\t] plot({-\t},{4*\t^3/27}) node[left] {\tf\Tf};
\draw (-0.9*\size,\size/2) node {\tf\Tg};
\draw[thick, dashed] (0,0) -- (-2*\size/3,0) node[left] {\tf\Th};
\draw (-2*\size/3,-2*\size/3) node {\tf\Ti};
\draw[thick, dashed, smooth,domain=0:(\size/\lambdastar)^(1/3),variable=\t] plot({\t},{-\lambdastar*\t^3}) node[left] {\tf\Tj};
\draw (0.4*\size,-0.9*\size) node {\tf\Tk};
\draw[thick, smooth,domain=0:(27*\size/4)^(1/3),variable=\t] plot({\t},{-4*\t^3/27}) node[right] {\tf\Tl};
\begin{scope}\clip (-2*\size/3,-\size) rectangle(1.2*2*\size/3,\size);
\paa\pab\pad\pae\pag\pah
\pbb\pbc\pbe\pbf\pbg\pbh
\pcc\pcd
\pda\pdb\pdc\pdd
\pec\ped
\pfa\pfb\pfd
\end{scope}
\end{tikzpicture}}
\hs{5}
\subfigure[(f) $A_{51} $] {
\begin{tikzpicture}[line cap=round,line join=round,>=stealth,x=0.58cm,y=0.3cm]
\def\size{5};\def\lambdastar{4.069026673648571};
\def\Ta{R};\def\Tb{R};\def\Tc{R};\def\Td{R};\def\Te{R};\def\Tf{R};\def\Tg{R};\def\Th{TO};\def\Ti{L,O};\def\Tj{L,O,S};\def\Tk{L,S};\def\Tl{DR};
\draw (4*\size/5,-\size/2) node {\tf\Ta};
\draw[thick, dashed] (0,0) -- (2*\size/3,0) node[right] {\tf\Tb};
\draw (2*\size/3,2*\size/3) node {\tf\Tc};
\draw[thick, dashed, smooth,domain=0:(\size/\lambdastar)^(1/3),variable=\t] plot({-\t},{\lambdastar*\t^3}) node[right] {\tf\Td};
\draw (-0.42*\size,0.9*\size) node {\tf\Te};
\draw[thick, smooth,domain=0:(27*\size/4)^(1/3),variable=\t] plot({-\t},{4*\t^3/27}) node[left] {\tf\Tf};
\draw (-0.9*\size,\size/2) node {\tf\Tg};
\draw[thick, dashed] (0,0) -- (-2*\size/3,0) node[left] {\tf\Th};
\draw (-2*\size/3,-2*\size/3) node {\tf\Ti};
\draw[thick, dashed, smooth,domain=0:(\size/\lambdastar)^(1/3),variable=\t] plot({\t},{-\lambdastar*\t^3}) node[left] {\tf\Tj};
\draw (0.4*\size,-0.9*\size) node {\tf\Tk};
\draw[thick, smooth,domain=0:(27*\size/4)^(1/3),variable=\t] plot({\t},{-4*\t^3/27}) node[right] {\tf\Tl};
\begin{scope}\clip (-2*\size/3,-\size) rectangle(1.2*2*\size/3,\size);
\pac\paf\pah
\pba\pbd\pbh
\pca\pcb\pcd
\pea\peb\ped
\pfc\pfd
\end{scope}
\end{tikzpicture}}\\
\subfigure[(g) $A_{43} $] {
\begin{tikzpicture}[line cap=round,line join=round,>=stealth,x=0.58cm,y=0.3cm]
\def\size{5};\def\lambdastar{4.069026673648571};
\def\Tg{R};\def\Th{R};\def\Ti{R};\def\Tj{R};\def\Tk{R};\def\Tl{R};\def\Ta{R};\def\Tb{TO};\def\Tc{L,O};\def\Td{L,O,S};\def\Te{L,S};\def\Tf{DR};
\draw (4*\size/5,-\size/2) node {\tf\Ta};
\draw[thick, dashed] (0,0) -- (2*\size/3,0) node[right] {\tf\Tb};
\draw (2*\size/3,2*\size/3) node {\tf\Tc};
\draw[thick, dashed, smooth,domain=0:(\size/\lambdastar)^(1/3),variable=\t] plot({-\t},{\lambdastar*\t^3}) node[right] {\tf\Td};
\draw (-0.42*\size,0.9*\size) node {\tf\Te};
\draw[thick, smooth,domain=0:(27*\size/4)^(1/3),variable=\t] plot({-\t},{4*\t^3/27}) node[left] {\tf\Tf};
\draw (-0.9*\size,\size/2) node {\tf\Tg};
\draw[thick, dashed] (0,0) -- (-2*\size/3,0) node[left] {\tf\Th};
\draw (-2*\size/3,-2*\size/3) node {\tf\Ti};
\draw[thick, dashed, smooth,domain=0:(\size/\lambdastar)^(1/3),variable=\t] plot({\t},{-\lambdastar*\t^3}) node[left] {\tf\Tj};
\draw (0.4*\size,-0.9*\size) node {\tf\Tk};
\draw[thick, smooth,domain=0:(27*\size/4)^(1/3),variable=\t] plot({\t},{-4*\t^3/27}) node[right] {\tf\Tl};
\begin{scope}\clip (-2*\size/3,-\size) rectangle(1.2*2*\size/3,\size);
\paa\pad\pah
\pbc\pbf\pbh
\pda\pdb\pdd
\pec\ped
\pfa\pfb\pfd
\end{scope}
\end{tikzpicture}}
\subfigure[(h) $A_{42} $] {
\begin{tikzpicture}[line cap=round,line join=round,>=stealth,x=0.58cm,y=0.3cm]
\def\size{5};\def\lambdastar{4.069026673648571};
\def\Ta{O};\def\Tb{TO};\def\Tc{L,O,S};\def\Td{L,O,S};\def\Te{L,S};\def\Tf{DR};\def\Tg{R};\def\Th{TO};\def\Ti{O,S};\def\Tj{O,S};\def\Tk{O};\def\Tl{O};
\draw (4*\size/5,-\size/2) node {\tf\Ta};
\draw[thick, dashed] (0,0) -- (2*\size/3,0) node[right] {\tf\Tb};
\draw (2*\size/3,2*\size/3) node {\tf\Tc};
\draw[thick, dashed, smooth,domain=0:(\size/\lambdastar)^(1/3),variable=\t] plot({-\t},{\lambdastar*\t^3}) node[right] {\tf\Td};
\draw (-0.42*\size,0.9*\size) node {\tf\Te};
\draw[thick, smooth,domain=0:(27*\size/4)^(1/3),variable=\t] plot({-\t},{4*\t^3/27}) node[left] {\tf\Tf};
\draw (-0.9*\size,\size/2) node {\tf\Tg};
\draw[thick, dashed] (0,0) -- (-2*\size/3,0) node[left] {\tf\Th};
\draw (-2*\size/3,-2*\size/3) node {\tf\Ti};
\draw[thick, dashed, smooth,domain=0:(\size/\lambdastar)^(1/3),variable=\t] plot({\t},{-\lambdastar*\t^3}) node[left] {\tf\Tj};
\draw (0.4*\size,-0.9*\size) node {\tf\Tk};
\draw[thick, smooth,domain=0:(27*\size/4)^(1/3),variable=\t] plot({\t},{-4*\t^3/27}) node[right] {\tf\Tl};
\begin{scope}\clip (-2*\size/3,-\size) rectangle(1.2*2*\size/3,\size);
\pag\pah
\pbb\pbe\pbh
\pdc\pdd
\pec\ped
\pfc\pfd
\end{scope}
\end{tikzpicture}}\\
\subfigure[(i) $A_{52} $] {
\begin{tikzpicture}[line cap=round,line join=round,>=stealth,x=0.58cm,y=0.3cm]
\def\size{5};\def\lambdastar{4.069026673648571};
\def\Tg{O};\def\Th{TO};\def\Ti{L,O,S};\def\Tj{L,O,S};\def\Tk{L,S};\def\Tl{DR};\def\Ta{R};\def\Tb{TO};\def\Tc{O,S};\def\Td{O,S};\def\Te{O};\def\Tf{O};
\draw (4*\size/5,-\size/2) node {\tf\Ta};
\draw[thick, dashed] (0,0) -- (2*\size/3,0) node[right] {\tf\Tb};
\draw (2*\size/3,2*\size/3) node {\tf\Tc};
\draw[thick, dashed, smooth,domain=0:(\size/\lambdastar)^(1/3),variable=\t] plot({-\t},{\lambdastar*\t^3}) node[right] {\tf\Td};
\draw (-0.42*\size,0.9*\size) node {\tf\Te};
\draw[thick, smooth,domain=0:(27*\size/4)^(1/3),variable=\t] plot({-\t},{4*\t^3/27}) node[left] {\tf\Tf};
\draw (-0.9*\size,\size/2) node {\tf\Tg};
\draw[thick, dashed] (0,0) -- (-2*\size/3,0) node[left] {\tf\Th};
\draw (-2*\size/3,-2*\size/3) node {\tf\Ti};
\draw[thick, dashed, smooth,domain=0:(\size/\lambdastar)^(1/3),variable=\t] plot({\t},{-\lambdastar*\t^3}) node[left] {\tf\Tj};
\draw (0.4*\size,-0.9*\size) node {\tf\Tk};
\draw[thick, smooth,domain=0:(27*\size/4)^(1/3),variable=\t] plot({\t},{-4*\t^3/27}) node[right] {\tf\Tl};
\begin{scope}\clip (-2*\size/3,-\size) rectangle(1.2*2*\size/3,\size);
\pab\pae\pah
\pbg\pbh
\pcc\pcd
\pec\ped
\pfc\pfd
\end{scope}
\end{tikzpicture}}
\hs{5}
\subfigure[(j) $A_{54} $] {
\begin{tikzpicture}[line cap=round,line join=round,>=stealth,x=0.58cm,y=0.3cm]
\def\size{5};\def\lambdastar{4.069026673648571};
\def\Ta{O,R};\def\Tb{TO};\def\Tc{L};\def\Td{L,O};\def\Te{L,O,S};\def\Tf{O,DR};\def\Tg{O,R};\def\Th{TO};\def\Ti{L};\def\Tj{L,O};\def\Tk{L,O,S};\def\Tl{O,DR};
\draw (4*\size/5,-\size/2) node {\tf\Ta};
\draw[thick, dashed] (0,0) -- (2*\size/3,0) node[right] {\tf\Tb};
\draw (2*\size/3,2*\size/3) node {\tf\Tc};
\draw[thick, dashed, smooth,domain=0:(\size/\lambdastar)^(1/3),variable=\t] plot({-\t},{\lambdastar*\t^3}) node[right] {\tf\Td};
\draw (-0.42*\size,0.9*\size) node {\tf\Te};
\draw[thick, smooth,domain=0:(27*\size/4)^(1/3),variable=\t] plot({-\t},{4*\t^3/27}) node[left] {\tf\Tf};
\draw (-0.9*\size,\size/2) node {\tf\Tg};
\draw[thick, dashed] (0,0) -- (-2*\size/3,0) node[left] {\tf\Th};
\draw (-2*\size/3,-2*\size/3) node {\tf\Ti};
\draw[thick, dashed, smooth,domain=0:(\size/\lambdastar)^(1/3),variable=\t] plot({\t},{-\lambdastar*\t^3}) node[left] {\tf\Tj};
\draw (0.4*\size,-0.9*\size) node {\tf\Tk};
\draw[thick, smooth,domain=0:(27*\size/4)^(1/3),variable=\t] plot({\t},{-4*\t^3/27}) node[right] {\tf\Tl};
\begin{scope}\clip (-2*\size/3,-\size) rectangle(1.2*2*\size/3,\size);
\pab\pae\pag\pah
\pbb\pbe\pbg\pbh
\pcc\pcd
\pdc\pdd
\end{scope}
\end{tikzpicture}}
\caption{Schematic showing saddle point configurations for different contour choices in the case $(l,m)=(1,2)$ as a function of position in the $(\x,\y)$-plane. The localisation curve (solid line), and the Stokes lines (dashed line) are also shown. Key: R=real, DR=double real, L=exponentially large, S=exponentially small, O=origin, TO=triple at origin. 
}
\label{fig:QuinticBehaviours2}
\end{center}
\end{figure}

In this case \rf{eqn:AIntCubic} provides $\binom{5}{2}=10$ solutions, four of which are linearly independent, satisfying %
\begin{align}
A_{31}(\X,\Y) &= \overline{A_{31}(-\X,-\Y)}, \qquad
A_{54}(\X,\Y) = \overline{A_{54}(-\X,-\Y)},\notag \\
A_{21}(\X,\Y) &= \overline{A_{32}(-\X,-\Y)},\qquad
A_{42}(\X,\Y) = \overline{A_{25}(-\X,-\Y)},\notag\\
A_{41}(\X,\Y) &= \overline{A_{35}(-\X,-\Y)},\qquad
A_{15}(\X,\Y) = \overline{A_{43}(-\X,-\Y)}.
\label{QuinticSymmetry}
\end{align}
Furthermore, the steepest descent contour for $A_{31}(\X,\Y)$ can be obtained from that for $A_{31}(-\X,-\Y)$ by reflection in the imaginary $\tau$-axis. A similar relationship exists between the steepest descent contours for the other pairs in \rf{QuinticSymmetry}. As a result, there are essentially six distinct far-field behaviours. However, for completeness and ease of reference we illustrate all ten schematically in \F\ref{fig:QuinticBehaviours2}. 

\begin{figure}[t!]
\begin{center}
\hs{-5}
\subfigure[(a) $|A_{31}|$]{\includegraphicsDave[width=40mm]{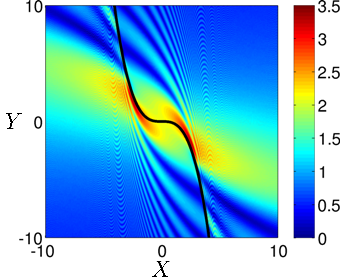}
}
\subfigure[(b) $\real{A_{31}\re^{\ri k\x}}$]{\includegraphicsDave[width=90mm]{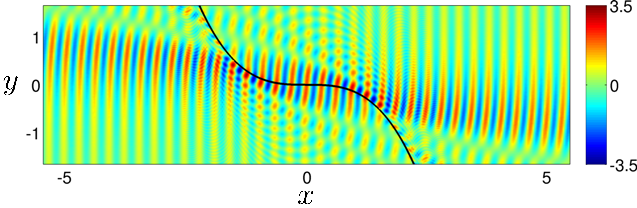}
}
\caption{Plots of $|A_{31}(\X,\Y)|$ and $\real{A_{31}\re^{\ri k \x}}$ for $(l,m)=(1,2)$, with $\lambda=l/(l+2m)=1/5$ and $k=20$. The curve near which the solution is localised is superimposed in black.}
\label{fig:QuinticFieldPlots}
\end{center}
\end{figure}

\begin{figure}[h]
\begin{center}
\subfigure[(a) $(\x,\y)$-plane] {
\begin{tikzpicture}[line cap=round,line join=round,>=triangle 45,x=0.39cm,y=0.2cm]
\def\size{5};\def\lambdastar{4.069026673648571};
\def\Ta{O,R,R,R};\def\Tb{TO,R};\def\Tc{O,R};\def\Td{O,R,S};\def\Te{O,R,S};\def\Tf{O,R,DR};\def\Tg{O,R,R,R};\def\Th{TO,R};\def\Ti{O,R};\def\Tj{O,R,S};\def\Tk{O,R,S};\def\Tl{O,R,DR};
\draw (4*\size/5,-\size/2) node {\tf\Ta};
\draw[thick, dashed] (0,0) -- (2*\size/3,0) node[right] {\tf\Tb};
\draw (2*\size/3,2*\size/3) node {\tf\Tc};
\draw[thick, dashed, smooth,domain=0:(\size/\lambdastar)^(1/3),variable=\t] plot({-\t},{\lambdastar*\t^3}) node[right] {\tf\Td};
\draw (-0.42*\size,0.9*\size) node {\tf\Te};
\draw[thick, smooth,domain=0:(27*\size/4)^(1/3),variable=\t] plot({-\t},{4*\t^3/27}) node[left] {\tf\Tf};
\draw (-0.9*\size,\size/2) node {\tf\Tg};
\draw[thick, dashed] (0,0) -- (-2*\size/3,0) node[left] {\tf\Th};
\draw (-2*\size/3,-2*\size/3) node {\tf\Ti};
\draw[thick, dashed, smooth,domain=0:(\size/\lambdastar)^(1/3),variable=\t] plot({\t},{-\lambdastar*\t^3}) node[left] {\tf\Tj};
\draw (0.4*\size,-0.9*\size) node {\tf\Tk};
\draw[thick, smooth,domain=0:(27*\size/4)^(1/3),variable=\t] plot({\t},{-4*\t^3/27}) node[right] {\tf\Tl};
\begin{scriptsize}
\fill (2.5747,-1.5398) circle (1.5pt);
\draw (2.8,-1.0) node {1};
\fill (3,0) circle (1.5pt);
\draw (3.4,0.6) node {2};
\fill (1.2231,2.7394) circle (1.5pt);
\draw (1.6231,3.3394) node {3};
\fill (-0.88677,2.8659) circle (1.5pt);
\draw (-0.5,3.4659) node {4};
\fill (-1.957,2.2738) circle (1.5pt);
\draw (-1.7,2.8738) node {5};
\fill (-2.3324,1.8867) circle (1.5pt);
\draw (-2.3,2.6) node {6};
\fill (-2.5747,1.5398) circle (1.5pt);
\draw (-2.9,1.5) node {7};
\fill (-3,0) circle (1.5pt);
\draw (-2.8,-0.7) node {8};
\fill (-1.2231,-2.7394) circle (1.5pt);
\draw (-1.5,-2.1394) node {9};
\fill (0.88677,-2.8659) circle (1.5pt);
\draw (0.35,-3.1) node {10};
\fill (1.957,-2.2738) circle (1.5pt);
\draw (1.5,-1.8) node {11};
\fill (2.3324,-1.8867) circle (1.5pt);
\draw (2.3,-3) node {12};
\end{scriptsize}
\end{tikzpicture}}
\subfigure[(b) Point 1]{\includegraphics[width=\psize]{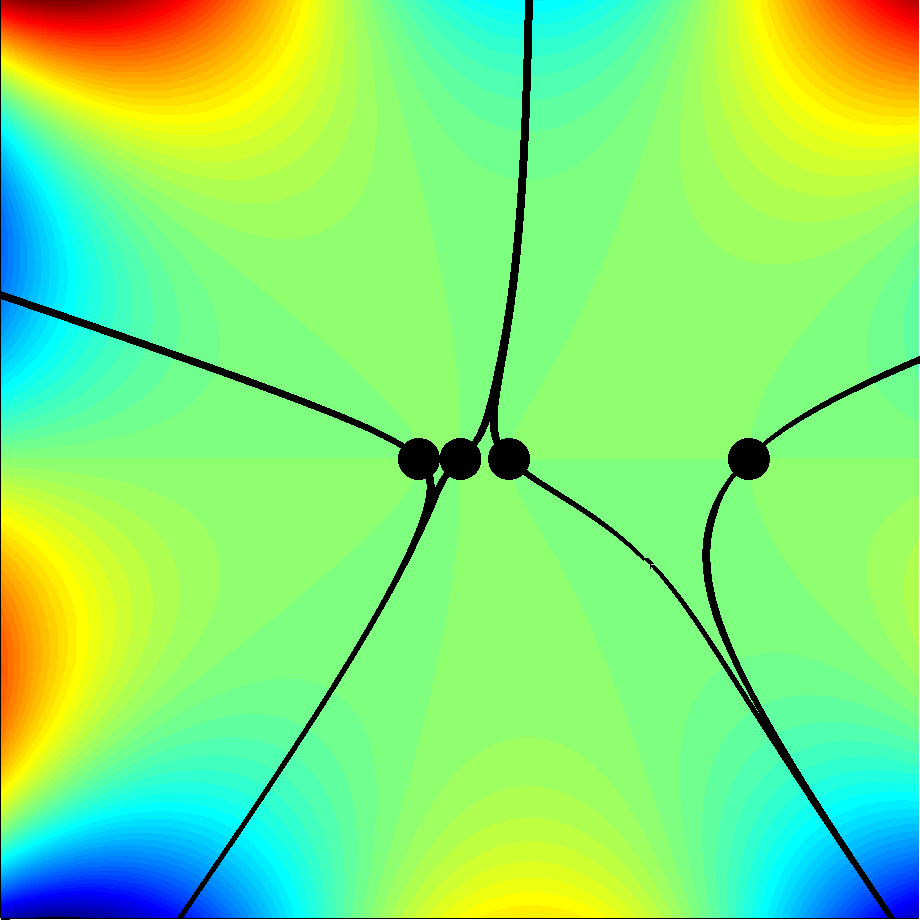}
}
\subfigure[(c) Point 2]{\includegraphics[width=\psize]{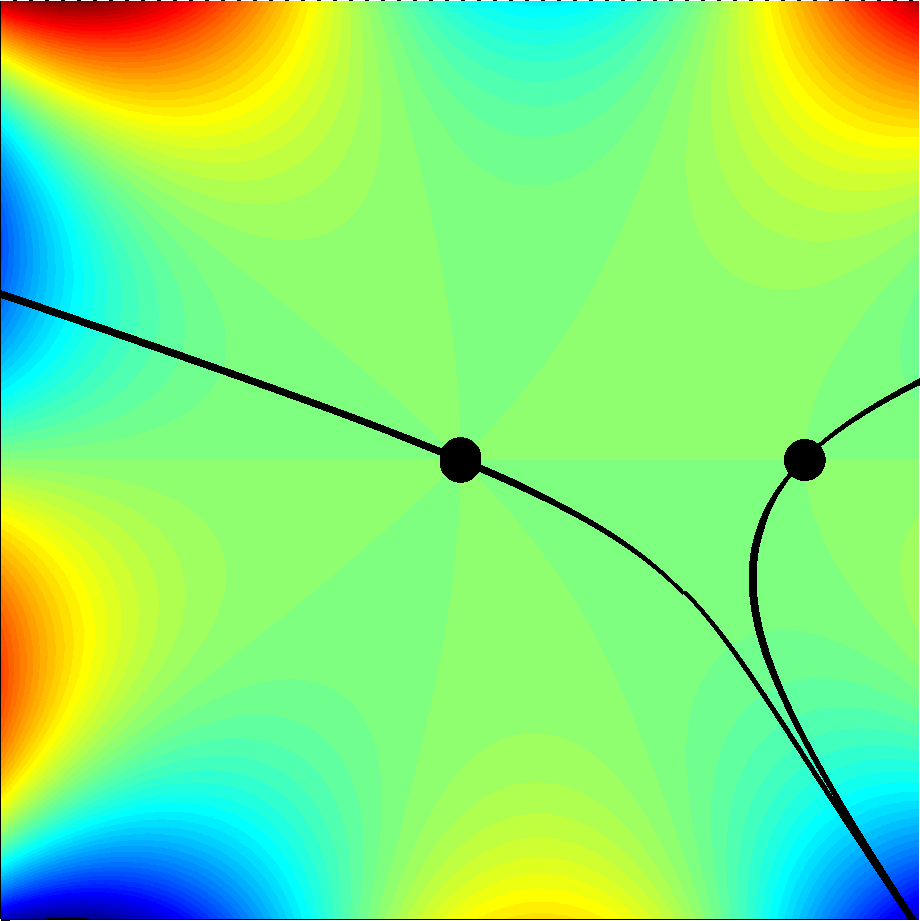}
}\\
\vs{-2}
\subfigure[(d) Point 3]{\includegraphics[width=\psize]{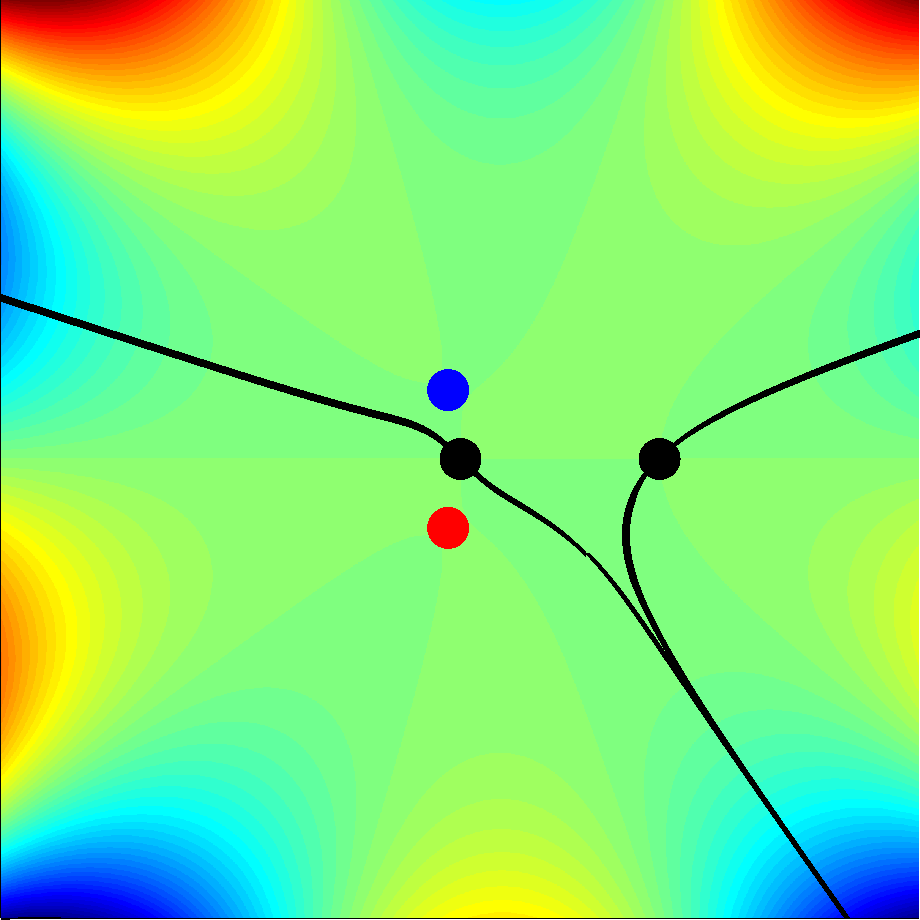}
}
\subfigure[(e) Point 4]{\includegraphics[width=\psize]{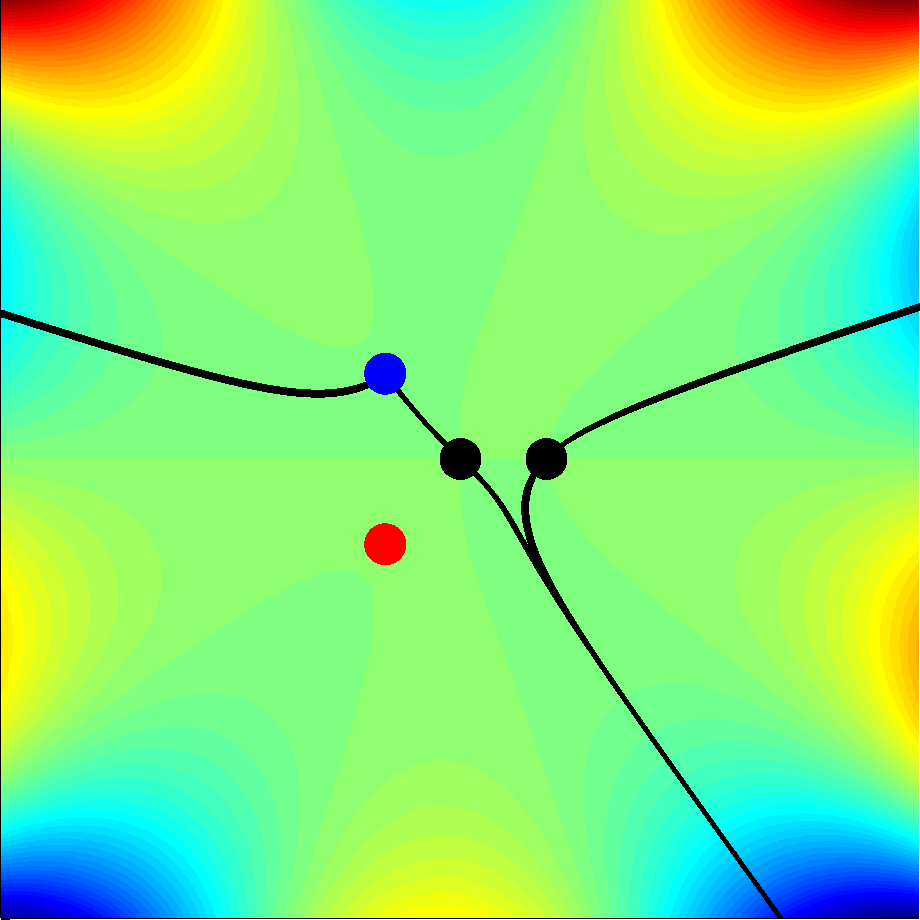}
}
\subfigure[(f) Point 5]{\includegraphics[width=\psize]{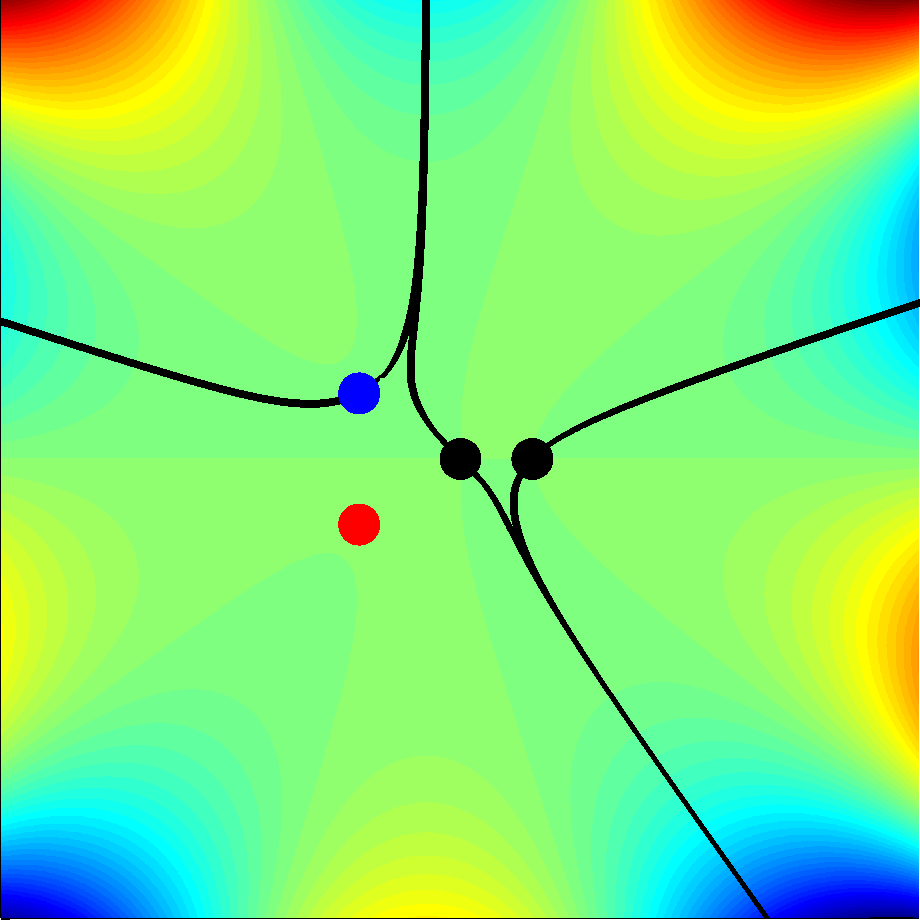}
}
\subfigure[(g) Point 6]{\includegraphics[width=\psize]{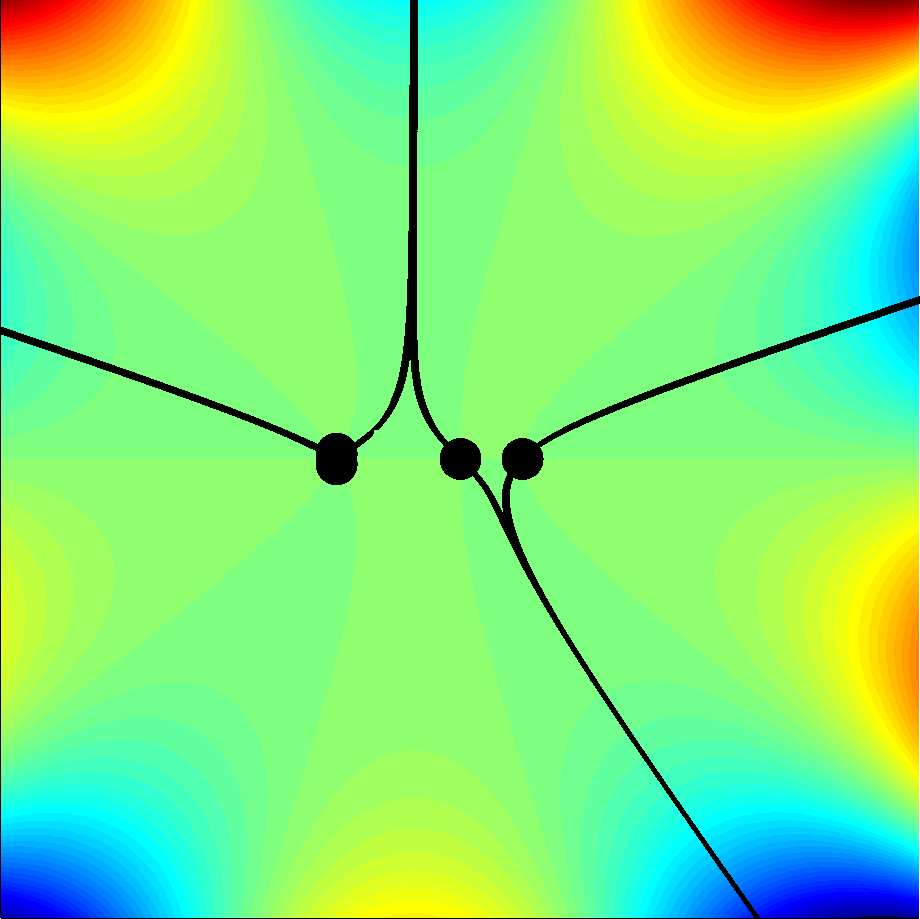}
}
\subfigure[(h) Point 7]{\includegraphics[width=\psize]{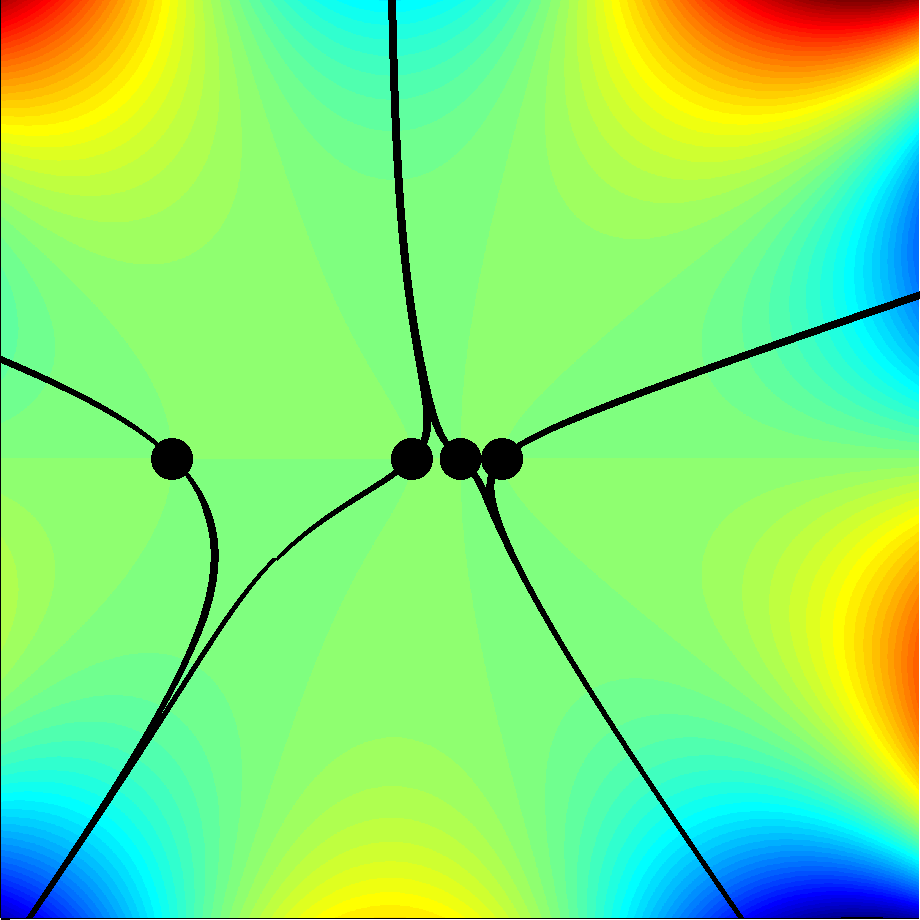}
}\\
\vs{-2}
\subfigure[(i) Point 8]{\includegraphics[width=\psize]{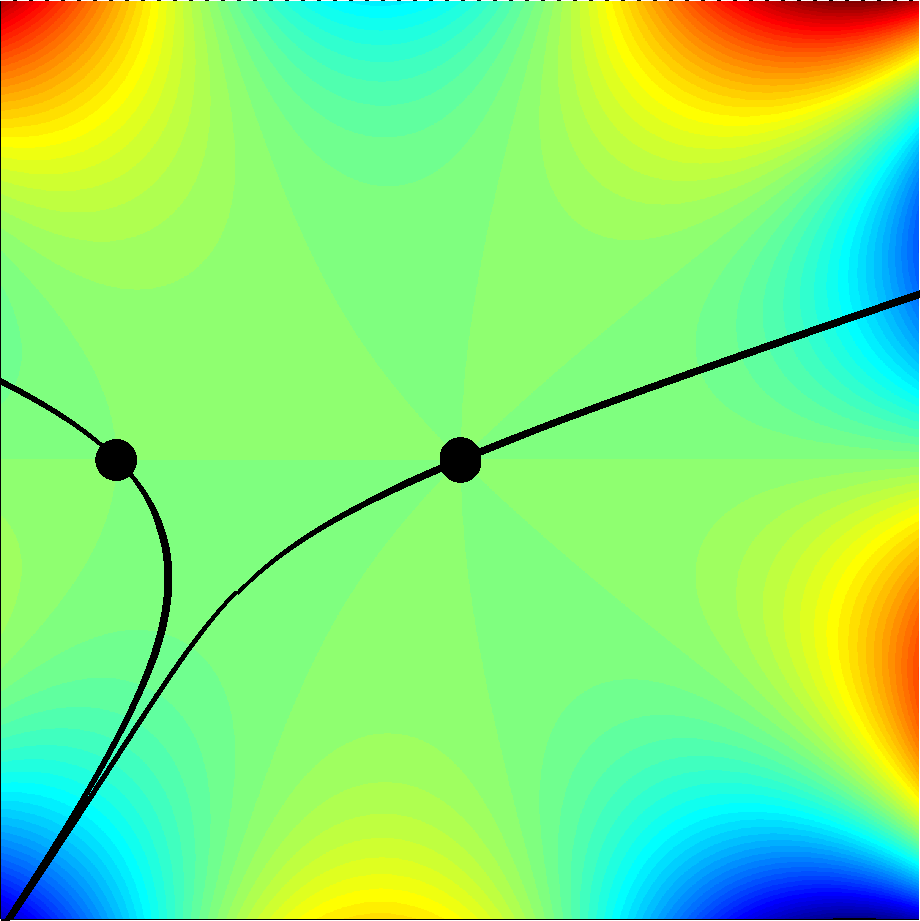}
}
\subfigure[(j) Point 9]{\includegraphics[width=\psize]{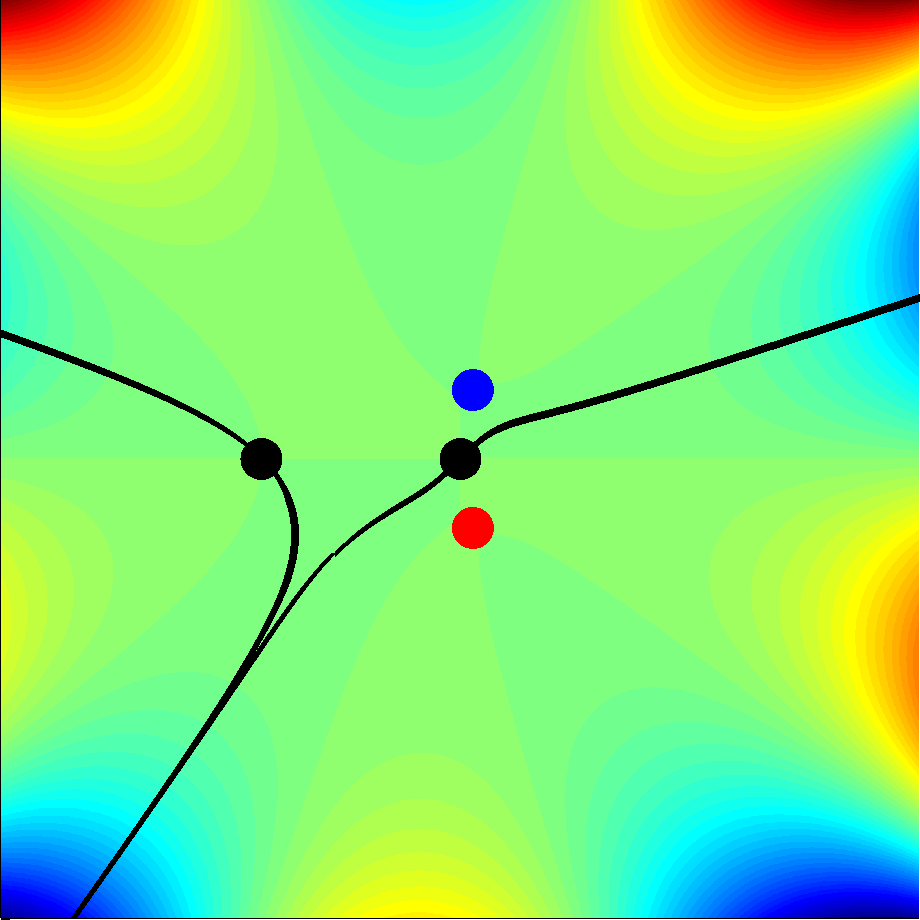}
}
\subfigure[(k) Point 10]{\includegraphics[width=\psize]{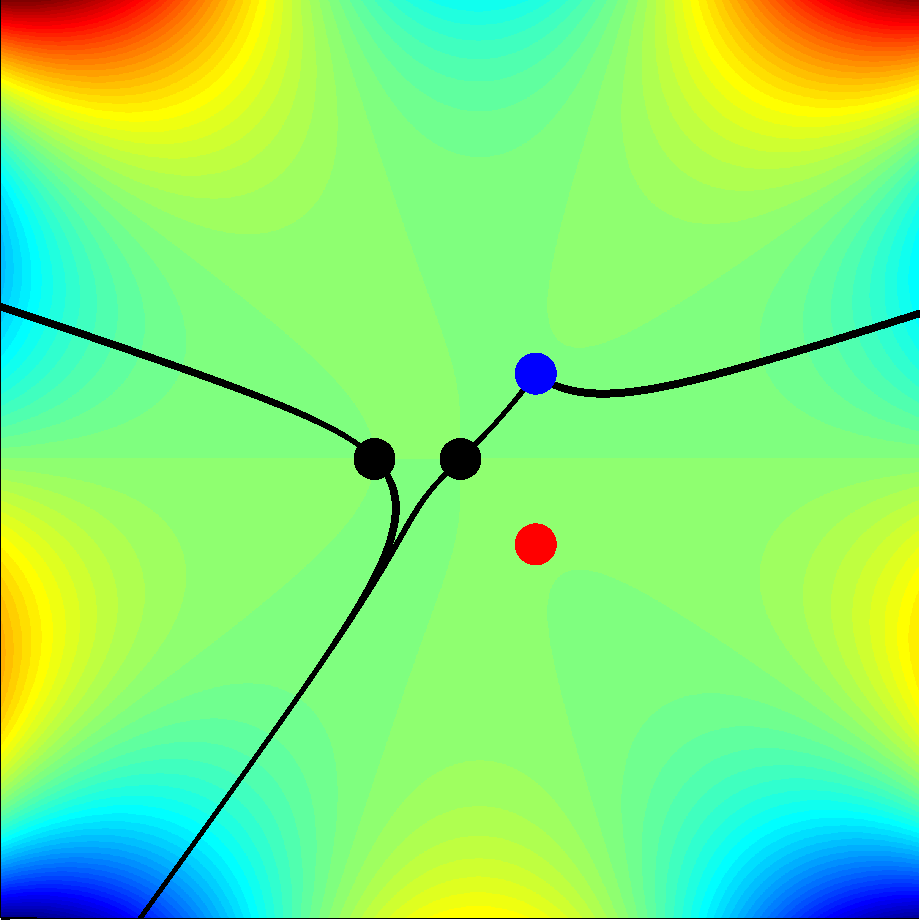}
}
\subfigure[(k) Point 11]{\includegraphics[width=\psize]{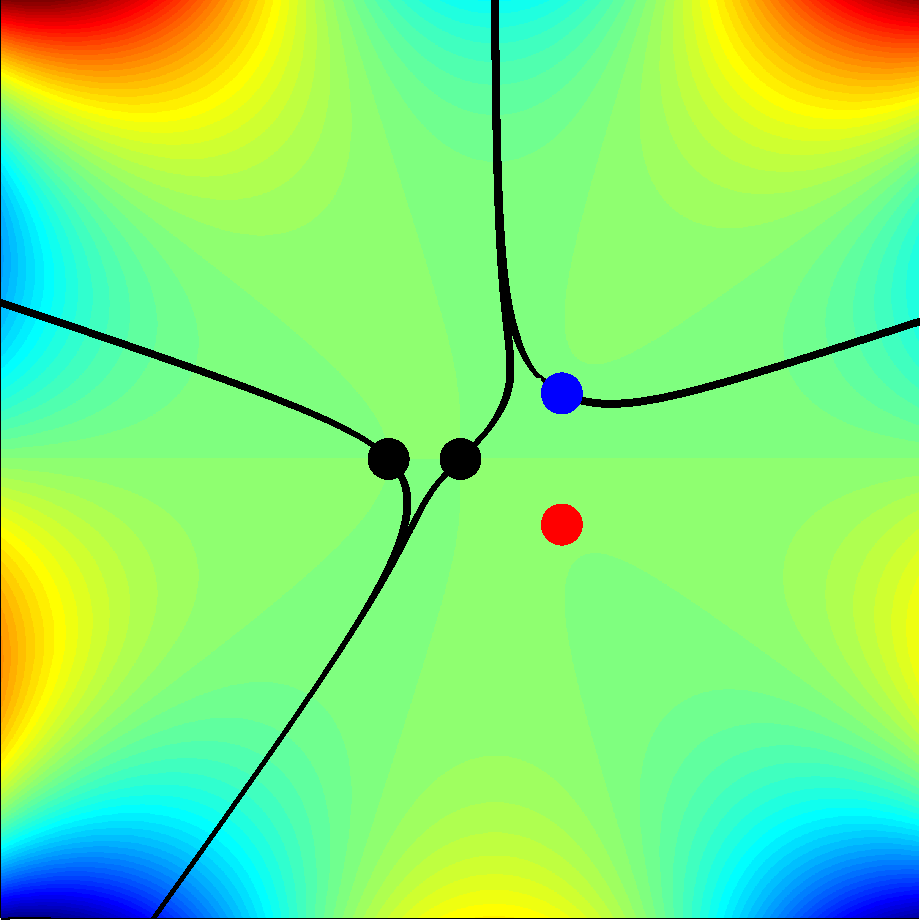}
}
\subfigure[(k) Point 12]{\includegraphics[width=\psize]{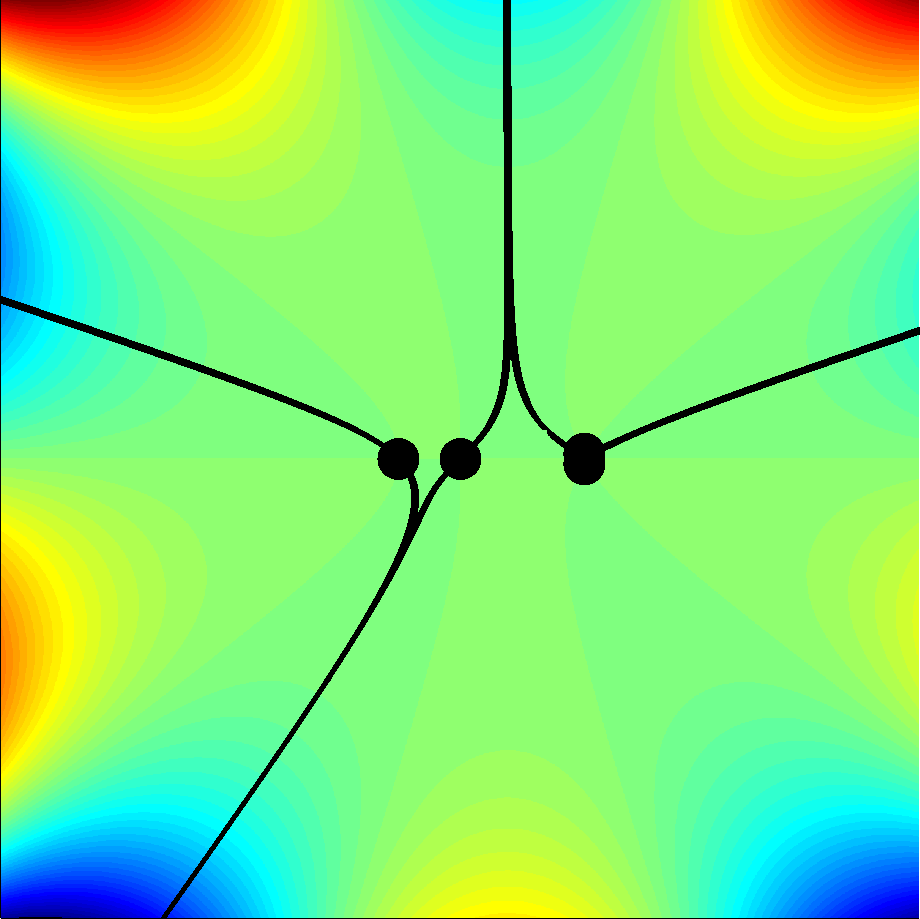}
}
\vs{-4}
\caption{Steepest descent contours for $A_{31}(\X,\Y)$ in the case $(l,m)=(1,2)$, $\alpha=1$, evaluated at points 1-12 from \F\ref{fig:QuinticSaddlesFull}(a).
}
\label{fig:QuinticSaddlesA31}
\end{center}
\end{figure}
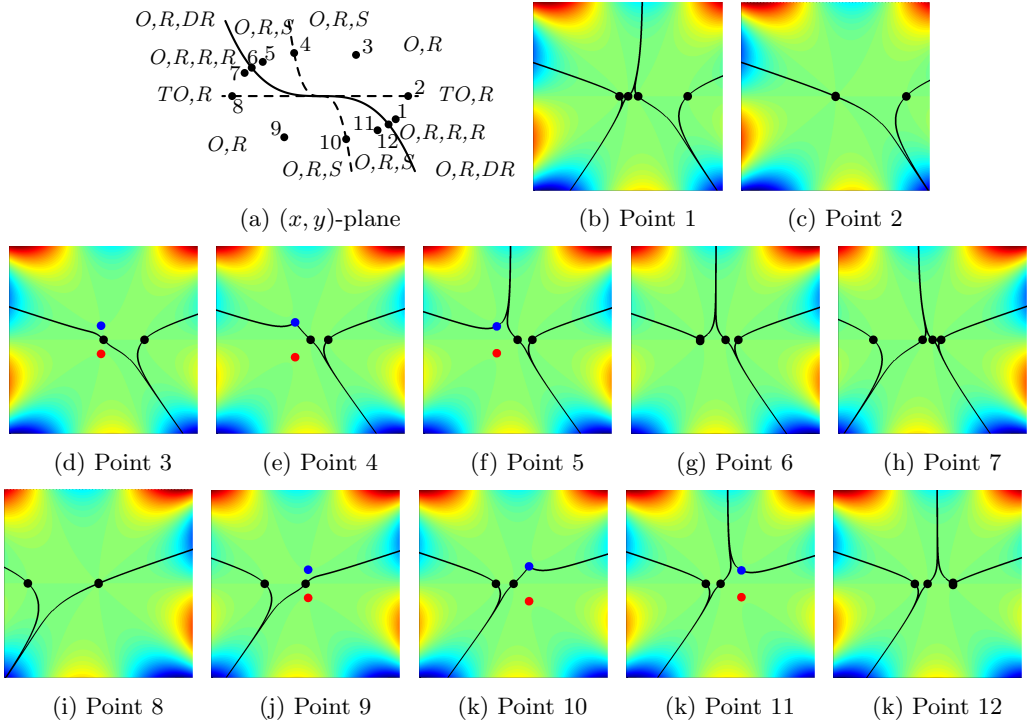

For $A_{31}$, %
one can deform the contour $\Gamma_{31}$ to the real axis, and $A_{31}$ can then be written in terms of the swallowtail canonical integral \cite{KiCoHo:00,Kaz:03}
\begin{align*}
\label{}
C_5(a_1,a_2,a_3)=\int_{-\infty}^\infty \re^{\ri(a_1t+a_2t^2+a_3t^3+t^5)} \,\rd t,
\end{align*}
as
\begin{align}
\label{}
A_{31} = \left(\frac{2\alpha}{5} \right)^{-1/5} \re^{\ri a_0} C_5(a_1,a_2,a_3),
\end{align}
where 
\begin{align*}
\label{}
a_0 = -\frac{\X^5+40\alpha^2\X^2\Y}{640\alpha^4}, \qquad 
a_1 = -\left(\frac{5}{2\alpha} \right)^{1/5}\frac{3\X^4+64\alpha^2\X \Y }{128\alpha^3},\\
a_2 = -\left(\frac{5}{2\alpha} \right)^{2/5}\frac{\X^3+8\alpha^2\Y}{8\alpha^2}, \qquad
a_3 = -\left(\frac{5}{2\alpha} \right)^{3/5}\frac{\X^2}{4\alpha}.
\end{align*}
Plots of the absolute value of $A_{31}$ in the case $\alpha=1$, and the real part of the associated approximate solution $A_{31}\re^{\ri k\x}$ of the Helmholtz equation, are given in \F\ref{fig:QuinticFieldPlots}; again, these plots were generated using the \texttt{cuspint} package \cite{KiCoHo:00}. 
The qualitative asymptotic behaviour is illustrated in \F\ref{fig:QuinticBehaviours2}(c), and some typical steepest descent contours are given in \F\ref{fig:QuinticSaddlesA31}. 

\begin{itemize}
\item 
For $\y+\kappa^2\x^3/6>0$ and $\y<0$ (\F\ref{fig:QuinticSaddlesA31}(b)) the steepest descent contour passes through all four real saddle points, corresponding to the three families of real rays that exist everywhere in this region. The same arguments leading to 
\Fs\ref{fig:AirySaddlesA21}, \ref{fig:AirySaddlesA32} and \ref{fig:AirySaddlesA31}
show that the ray picture is as in \F\ref{fig:QuinticBehaviours2}(c). 
The three non-zero saddle points correspond to rays ingoing or outgoing from the cubic (as suggested by the dotted lines in  \F\ref{fig:QuinticBehaviours2}(c)), and the saddle point at zero corresponds to the horizontal ray. 
\item For $\y+\kappa^2\x^3/6>0$ and $\y=0$ (\F\ref{fig:QuinticSaddlesA31}(c)), the contour passes through the single real saddle point and the triple real saddle point at zero.
\item For $\y+(9\kappa^2/8)\mu_*\x^{3}>0$ and $\y>0$ (\F\ref{fig:QuinticSaddlesA31}(d)) the contour passes through the two real saddle points. 
\item For $\y+(9\kappa^2/8)\mu_*\x^{3}=0$ and $\y>0$ (\F\ref{fig:QuinticSaddlesA31}(e)) the contour picks up the complex saddle point associated with an exponentially small wave.
\item For $\y+(9\kappa^2/8)\mu_*\x^{3}<0$ and $\y+\kappa^2\x^3/6>0$ (\F\ref{fig:QuinticSaddlesA31}(f)) the contour passes through the two real saddle points and the complex saddle point associated with the exponentially small wave.
\item For $\y+\kappa^2\x^3/6=0$ and $\y>0$ (\F\ref{fig:QuinticSaddlesA31}(g)) the contour passes through the two single real saddle points and double real saddle point corresponding to the localisation.
\end{itemize}

To investigate the far-field localisation near \rf{eqn:Bdy} we deform $\Gamma_{31}$ to the real axis, and set
\begin{align*}
\label{}
\x = \x_0+\delta \x^*, \quad \y = -\frac{\kappa^2\x_0^3}{6}+\delta \y^*, \quad \tau = \frac{\kappa \x_0}{\sqrt{2}} + \tau'.
\end{align*}
The expanded phase is then
\begin{align}
\label{eqn:InflectionPhase}
\frac{\kappa^4 \x_0^5}{40} &- \frac{\kappa^2\x_0^2}{2}\delta \y^*-\frac{\kappa^4\x_0^4}{8}\delta \x^* + \tau'\left(-\sqrt{2}\kappa \x_0 \left( \delta \y^* + \frac{\kappa^2\x_0^2}{2}\delta \x^*\right) \right)\notag\\
&- (\tau')^2\left(\delta \y^* + \frac{3\kappa^2 \x_0^2\delta \x^*}{2}\right) + (\tau')^3\sqrt{2}\kappa \x_0\left(\frac{ \x_0}{3} -\delta \x^*\right)+ \ord{(\tau')^4},
\end{align}
so that, when we finally set $\delta=k^{-2/3}$ and $\tau'=k^{-1/3}(\sqrt{2}\kappa \x_0^2)^{-1/3} \zeta$, we obtain a prefactor proportional to $\exp(\ri k\kappa^4 \x_0^5/40)$ multiplied by the Airy function
\begin{align}
\int_{-\infty}^\infty \exp{\left[ \ri\left(-(\sign{\x_0})\left(2\kappa^2 |\x_0|\right)^{1/3}\left(\y^*+\frac{\kappa^2 \x_0^2 \x^*}{2}\right)\zeta + \frac{\zeta^3}{3}\right) \right]}\,\rd \zeta\notag \\
\hspace{20mm} =2\pi \Ai\left(-(\sign{\x_0})\left(2\kappa^2 |\x_0|\right)^{1/3}\left(\y^*+\frac{\kappa^2 \x_0^2 \x^*}{2}\right)\right).
\label{eqn:InflectionAsympt}
\end{align}

To investigate the quite different localisation near $\y=0$, which is associated with a triple saddle point at $\tau=0$, we again deform $\Gamma_{31}$ to the real axis, and now set
\begin{align*}
\label{}
\x = \x_0, \quad \y = \delta \y^*, \quad \tau = \tau'.
\end{align*}
The phase becomes
\begin{align*}
\label{}
-\delta \y^* (\tau')^2 - \frac{\x_0}{2}(\tau')^4 + \frac{4\sqrt{2}}{15 \kappa}(\tau')^5,
\end{align*}
and, assuming for simplicity that $\x_0>0$, setting $\delta=k^{-1/2}$ and $\tau' = k^{-1/4}(\x_0/2)^{-1/4}\zeta$, we obtain a contribution
\begin{align}
\label{eqn:QuinticTripleZero}
k^{-1/20}\left(\frac{\x_0}{2}\right)^{-1/4}\int_{-\infty}^\infty \exp{\left[\ri \left( -\sqrt{2/\x_0}\y^*\zeta^2 -\zeta^4 \right)\right]}\,\rd \zeta 
= k^{-1/20}\left(\frac{\x_0}{2}\right)^{-1/4} \overline{P\left(\sqrt{\frac{2}{\x_0}}\y^*,0\right)},
\end{align}
where $P$ is the Pearcey function defined in \rf{eq:PearceyDefn}. (The function \rf{eqn:QuinticTripleZero} is also expressible in terms of the parabolic cylinder function of order $-1/2$; cf.\ \cite[\S4.3]{Kaz:03}.)
Under these scalings, we have $\X=k^{1/5}\x_0$, $\Y= k^{1/10}\y^*$, so that for fixed $\x_0$ and $\y^*=\ord{1}$, we have $\Y/\sqrt{\X}=\y^*/\sqrt{\x_0}=\ord{1}$. 

When $\Y/\sqrt{\X}=\y^*/\sqrt{\x_0}\ll 1$, \rf{eqn:QuinticTripleZero} can be approximated further, giving
\begin{align*}
k^{-1/20}\left(\frac{\x_0}{2}\right)^{-1/4}\int_{-\infty}^\infty \re^{-\ri\zeta^4}\,\rd \zeta  = k^{-1/20}\left(\frac{\x_0}{2}\right)^{-1/4}\re^{-\ri\pi/8}\Gamma[5/4],
\end{align*}
where $\Gamma[z]$ is the usual Gamma function. 
On the other hand, the limiting behaviour of \rf{eqn:QuinticTripleZero} as $|\Y|/\sqrt{\X}=|\y^*|/\sqrt{\x_0}\to \infty$ can be found using the method of stationary phase, with the stationary phase point at $\zeta=0$ producing a term
\begin{align*}
\label{}
\frac{k^{-3/10}\sqrt{\pi}\re^{-\sign{\y}\ri\pi/4}}{\sqrt{|\y|}} = \frac{\sqrt{\pi}\re^{-\sign{\Y}\ri\pi/4}}{\sqrt{|\Y|}},
\end{align*}
which matches with the leading order contribution of the $\tau=0$ saddle point in the outer region $\Y=\ord{\X^3}$. 

As can be observed from \F \ref{fig:QuinticBehaviours2}, all of the other nine solutions (other than $A_{31}$) also exhibit some sort of localisation near the cubic curve, with $A_{21}$, $A_{32}$, $A_{51}$ $A_{43}$, $A_{42}$ and $A_{52}$ being localised only near one of the two branches of the cubic (i.e.\ only in $\y>0$ or $\y<0$ but not both). The only solutions not to suffer from exponential growth in any region of the $(\x,\y)$ plane are $A_{21}$, $A_{32}$ and $A_{31}$ (note that any of these three is a linear combination of the other two). However, as we shall discuss in the next section in the context of boundary value problems, the fact that a solution grows exponentially in some region does not preclude it from being physically relevant (i.e.\ matching to a solution of the Helmholtz equation satisfying an appropriate radiation condition at infinity), provided that the regions of exponential growth lie in the appropriate domain (as was the case for $A_{32}$ and $A_{13}$ in the parabolic case $(l,m)=(1,1)$). 

The function $A_{32}$ coincides (up to a prefactor) with the ``\textit{concave-convex special function}'' (CCSF) $\Psi$ investigated by Kazakov in \cite{Kaz:03}. 
Explicitly, setting $\omega_{\rm K}=k$, $t_{\rm K}=k^{-1/5}2^{-3/5}\kappa^{4/5} \X$, $z_{\rm K}=k^{-3/5}2^{1/5}\kappa^{2/5}(\Y + \kappa^2 \X^3/6)$, $v_{\rm K} = k^{-1/5}2^{-1/10}\kappa^{-1/5}t$, where the subscript $_{\rm K}$ indicates variables appearing in \cite{Kaz:03}, we have
\begin{align}
\label{eqn:Kaz}
\Psi(z_{\rm K},t_{\rm K}) = k^{-1/5}2^{-1/10}\kappa^{-1/5}\re^{\ri(\kappa^2 \X^2 \Y /2 + 7 \kappa^4\X^5/120)}A_{32}(\X,\Y).
\end{align} 
The $(X,Y)$-dependent prefactor arises from the fact that Kazakov works in curvilinear rather than Cartesian coordinates. As Kazakov notes in \cite[\S4.4]{Kaz:03}, and as our qualitative analysis in \F\ref{fig:QuinticBehaviours2}(b) confirms, this solution is exponentially small at infinity in the region $y+\kappa^2 x^3/6>0$, $y>0$, and has asymptotic localisation near the left-hand branch of the cubic parabola (i.e.\ $y+\kappa^2 x^3/6=0$, $x<0$), where it assumes the character of the field near a smooth caustic (or equivalently, whispering gallery waves - see \S\ref{sec:BVPs}). There is also localisation on the line $y=0$, the tangent line to the curve at the inflection point, which Kazakov refers to as the ``\textit{searchlight}'' line \cite{Kaz:03}, in reference to the fact that one expects some enhancement of the field here in the inflection point boundary value problem. We consider this and other boundary value problems in the following sections.%
\section{Boundary value problems}
\label{sec:BVPs}
An interesting and important question is whether PWE solutions of the form \rf{eqn:AInt} can possibly be `modal', i.e.\ satisfy homogeneous Dirichlet, Neumann or Robin conditions for the complex function $A$ on the real curve \rf{eqn:Bdy}. For the case $(l,m)=(1,1)$ (parabolic caustic) it is clear from \rf{eqn:AAiry} and \rf{eqn:AAiryA32} that this can be arranged by translating $\Y$ by an appropriate (possibly complex) constant, and, as discussed in \cite{OckTew:12}, this leads to models for whispering gallery and creeping waves. In terms of the original integral \rf{eqn:AInt}, this corresponds to choosing the prefactor $F(t)$ in the integrand to be a certain exponential. For instance, for the Dirichlet case, let $\eta_n<0$, $n=0,1,2,\ldots$, denote the zeros of $\Ai$. Then, to model whispering gallery waves one modifies \rf{eqn:AAiry} by replacing $\Y$ with $\Y-(2\kappa)^{-1/3}\eta_n$ for some $n$, which corresponds to taking $F(t)=\exp{[\ri(2\kappa)^{-1/3}\eta_n t]}$ and $\Gamma=\Gamma_{21}$ in \rf{eqn:AInt}. To model creeping waves one modifies \rf{eqn:AAiryA32} by replacing $\Y$ with $\Y+\re^{\ri \pi/3}(2\kappa)^{-1/3}\eta_n$, which corresponds to taking $F(t)=\exp{[\re^{-\ri\pi/6}(2\kappa)^{-1/3}\eta_n t]}$ and $\Gamma=\Gamma_{32}$ in \rf{eqn:AInt}. In both cases the exponential behaviour of $F(t)$ shifts the saddle points in the $\tau$-plane by an $O(k^{-2/3})$ perturbation (as in \rf{eqn:ShiftedSaddles1} below), and accordingly shifts the localisation curve on which the saddles coalesce away from \rf{eqn:Bdy} (in the creeping wave case, into complex space). 

Less well known is the fact that, as shown in \cite{He:14}, a contour integral solution of the form \rf{eqn:AInt} (again with $(l,m)=(1,1)$) also exists for the Fock-Leontovich tangent ray diffraction problem, which describes the field near a point of grazing incidence for a plane wave $\re^{\ri kx}$ incident on a smooth boundary. Without loss of generality one can adopt Cartesian coordinates such that the boundary near the point of grazing incidence is locally approximated by the parabola $y+x^2/4 = 0$. For the Dirichlet problem, the corresponding solution of PWE should vanish on the parabola $Y+X^2/4 = 0$, and in the far-field close to the parabola match to the geometrical optics approximation (i.e.\ the incident plus specularly reflected waves) in the illuminated region (i.e.\ as $X\to-\infty$) and a creeping wave solution in the shadow region (i.e.\ as $X\to+\infty$). Classically the solution is usually represented as a generalised Fourier-type integral involving ratios of Airy functions, sometimes regularised with a `forked contour' approach (see the discussion and references in \cite[\S2.1]{He:14}). However, in \cite{He:14} it was demonstrated that, by applying suitable transformations, the classical solution can be rewritten in the form \rf{eqn:AInt}, specifically as
\begin{align}
\label{eqn:FLPsoln}
A(X,Y) = \int_{\Gamma_{32}} \hat{p}(t) \re^{\ri(-\Y t - \X t^2/2 + t^3/3)}\,\rd t,
\end{align}
where $\hat{p}(t)$ is the `Pekeris caret function' (or `Fock integral'), a meromorphic function with a single simple pole at $t=0$, defined by\footnote{An alternative integral representation, valid on the whole of $\C\setminus\{0\}$, is \cite[\S2.3]{He:14}
\begin{align*}
\label{}
\hat{p}(t) = \frac{1}{2\pi \ri t} - \frac{1}{2\pi}\left( \re^{2\pi \ri/3} \int_0^{\re^{2\pi \ri/3}\infty}\re^{\ri t \sigma}  \frac{\Ai[\re^{-2\pi \ri/3}\sigma]}{\Ai[\re^{2\pi \ri/3}\sigma]}\,\rd\sigma + \re^{-2\pi \ri/3} \int_0^{\infty} \re^{\ri t \sigma}  \frac{\Ai[\sigma]}{\Ai[\re^{2\pi \ri/3}\sigma]} \,\rd\sigma \right), \quad t\in\C\setminus\{0\},
\end{align*}
where the integration contours pass below all the poles of the integrands.} 
\begin{align}
\label{eqn:PekerisFourierRep}
\hat{p}(t) = \frac{\re^{\pi \ri/3}}{2\pi}\int_{-\infty}^\infty \re^{\ri t \sigma}\frac{\Ai[\sigma]}{\Ai[\re^{2\pi \ri/3}\sigma]}\,\rd \sigma, \qquad \im{t}<0.
\end{align} 
The expression \rf{eqn:FLPsoln} describes both the total field (provided $\Gamma_{32}=\Gamma^r_{32}$ is taken to pass to the \textit{right} of the pole at $t=0$) and also the scattered field (provided $\Gamma_{32}=\Gamma^l_{32}$ is taken to pass to the \textit{left} of the pole at $t=0$), with the incident field corresponding to the pole contribution. 
Some basic properties of $\hat{p}(t)$ are given in \cite[\S2.3]{He:14}. We highlight especially 
the large argument behaviour 
\begin{subnumcases}{\hat{p}(t) \sim}
\dfrac{\re^{-2\pi\ri/3}}{2\pi}\dfrac{\re^{\re^{-\ri \pi/6}t\eta_0}}{\Ai'[\eta_0]^2}, & $|t|\to\infty$, $\arg{t}\in (-\pi/3,2\pi/3)$,\label{eqn:PekAsympt1}\\[2mm]
\dfrac{\sqrt{-t}}{2\sqrt{\pi}}\re^{\ri \pi/4}\re^{-\ri t^3/12}, & $|t|\to\infty$, $\arg{t}\in (2\pi/3,5\pi/3)$, \label{eqn:PekAsympt2}
\end{subnumcases}
since it determines the far-field behaviour of \rf{eqn:FLPsoln} near the parabolic boundary. Details of the asymptotic matching can be found in \cite[\S3]{He:14}; here we briefly review a few of the key points, to inform our subsequent discussions of other boundary value problems. 

We start by rescaling $\X=k^{1/3} \x$, $\Y=k^{2/3}\y$, $t = k^{1/3} \tau$ as usual. Then, for the matching as $\X\to \infty$ we work with the total field, taking $\Gamma^r_{32}$ in \rf{eqn:FLPsoln} and approximating $\hat{p}(t) = \hat{p}(k^{1/3}\tau)$ by \rf{eqn:PekAsympt1}. The exponential factor shifts the saddle points from 
\rf{eqn:ParabSaddles} to 
\begin{align}
\label{eqn:ShiftedSaddles1}
\tau_\pm = \frac{1}{2}\left(\x \pm \sqrt{(\x^2+4\y)+k^{-2/3}\re^{\ri\pi/3}\eta_0}\right),
\end{align}
just as in the discussion of creeping waves above. Close to the parabola $\x+4\y^2=0$ and with $\x>0$, the steepest descent contour is still roughly as in \F\ref{fig:AirySaddlesA32} (modulo the effects of the $O(k^{-2/3})$ perturbation), with the shed creeping rays in the outer Helmholtz solution arising from $\tau_-$. It is important here that $\tau_-$, as given by \rf{eqn:ShiftedSaddles1}, is only relevant when it lies in the half-plane in which \rf{eqn:PekAsympt1} is valid. One can check that (neglecting the $O(k^{-2/3})$ perturbation) this occurs if and only if $\x>0$ and $-\x^2<\y<0$.
In particular, for $\x>0$ with $\y$ close to or greater than $0$, the above analysis has to be modified, in order to demonstrate a correct match with the outer solution behaviour in the penumbra, i.e.\ the shadow boundary zone. As shown in detail in \cite[\S3]{He:14} (and see also \cite{TeChKiOcSmZa:00}), there are two regimes to consider. Matching to the `outer' layer of the penumbra requires scaling 
$\X=k^{1/3} \x$ as before, but now scaling $\Y=k^{1/3}\hat{y}$, and leaving $t$ unscaled. This produces a steepest descent integral with large parameter $k^{1/3}$ and a single saddle at $t=-\tilde{y}/x$, meaning that the Pekeris caret function $\hat{p}$ appears in the far-field behaviour in unapproximated form. In the `inner' layer of the penumbra where $\tilde{y}$ is small, the Fresnel integral emerges due to the coalescence of this saddle point with the pole in $\hat{p}(t)$ at $t=0$.

For the matching as $\X\to -\infty$ we work with the scattered field, taking $\Gamma^l_{32}$ in \rf{eqn:FLPsoln} and approximating $\hat{p}(t) = \hat{p}(k^{1/3}\tau)$ by \rf{eqn:PekAsympt2}. The $\re^{-\ri t^3/12}$ factor now shifts the saddle points by an $O(1)$ distance from \rf{eqn:ParabSaddles} to 
\begin{align}
\label{eqn:ShiftedSaddles2}
\tau_\pm = \frac{2}{3}(\x \pm \sqrt{\x^2+3\y}),
\end{align}
and so when $x<0$ the qualitative behaviour of the steepest descent contour is still as in \F\ref{fig:AirySaddlesA32}, but with the localisation curve shifted from $\y+\x^2/4=0$ downwards to $\y+\x^2/3=0$. The specularly reflected wave corresponds to the contribution from $\tau_-$, and, interestingly, can be identified with `outgoing' rays shed tangentially from the shifted parabola $\y+\x^2/3=0$. 
Again $\tau_-$, as given by \rf{eqn:ShiftedSaddles2}, is only relevant when it lies in the half-plane in which \rf{eqn:PekAsympt2} is valid, which holds if and only if $(\x,\y)\not\in \{(\x,\y):\x>0 \text{ and }-4\x^2/3<\y<0\}$.
It is natural to ask whether \rf{eqn:FLPsoln}-\rf{eqn:PekerisFourierRep} could have been obtained \textit{without} knowledge of the classical solution representation from which it was derived. Suppose we were to seek a solution for the scattered field $A^s$ in the tangent ray diffraction problem in the form \rf{eqn:AInt} with $(l,m)=(1,1)$ and $\alpha=1$ ($\kappa=1/2$). Based on \S\ref{sec:i} (and the discussion at the start of this section) we might expect that, in order to capture the specularly reflected wave and the creeping waves, one should choose the contour $\Gamma_{32}$, since this is the only choice for which all the ray fields propagate away from the boundary (rather than towards it). Imposing the homogeneous Dirichlet boundary condition $A^s=-1$ on the parabola then leads to the following integral equation reformulation of the tangent ray diffraction problem: find a function $F(t)$ such that 
\begin{align}
\label{eqn:FockLeontovich}
A^s=\int_{\Gamma_{32}} F(t) \re^{\ri(-\Y t-\X t^2/2+t^3/3)}\,\rd t = -1 \quad \textrm{on } \Y=-\X^2/4,\; \X\in\R.
\end{align}
Based on our discussion of creeping waves above, we should expect that 
$F(t)\sim C\exp{[\re^{-\ri\pi/6}\eta_0 t]}$ in some sector around the positive real $t$-axis, so that we recover modal creeping wave behaviour in the shadow region $\x>0$. 

The integral equation \rf{eqn:FockLeontovich} can be solved, at least formally, using Fourier transform methods. Suppose we write $F(t) = \int_{-\infty}^\infty\re^{\ri t\sigma} \cF(\sigma) \,\rd\sigma$, where $\cF$ is the Fourier transform of $F$. %
Then, formally changing the order of integration in \rf{eqn:FockLeontovich} and changing variable $t=s + \X/2$, we obtain
\begin{align}
\label{}
A^s &= \re^{-\ri (\X \Y/2 +\X^3/12)}\int_{-\infty}^\infty\re^{\ri \X\sigma/2} \cF(\sigma) 
\int_{\Gamma_{32}}\re^{\ri[-(y+\X^2/4-\sigma)s + s^3/3]}
\,\rd s\,\rd\sigma \notag\\
&= 2\pi\re^{2\ri\pi/3} \re^{-\ri (\X \Y/2 + \X^3/12)}\int_{-\infty}^\infty\re^{\ri \X\sigma/2} \cF(\sigma) 
\Ai\left[-\re^{2\pi\ri/3}(y+\X^2/4-\sigma)\right]
\,\rd\sigma.
\label{eqn:FockLeontovich2}
\end{align}
Applying the boundary condition %
and replacing $\X$ by $2\X$ throughout 
gives
\begin{align*}
\label{}
\int_{-\infty}^\infty\re^{\ri \X\sigma} \cF(\sigma) 
\Ai\left[\re^{2\pi\ri/3}\sigma\right]
\,\rd\sigma = \frac{\re^{\ri\pi/3} \re^{-\ri \X^3/3}}{2\pi},
\end{align*}
and, interpreting the LHS as the inverse Fourier transform of $\cF(\sigma) 
\Ai\left[-\re^{-\ri\pi/3}\sigma\right]$, we find that
\begin{align*}
\label{}
\cF(\sigma) 
\Ai\left[-\re^{-\ri\pi/3}\sigma\right] = 
 \frac{\re^{\ri\pi/3}}{4\pi^2}\int_{-\infty}^\infty\re^{\ri (-\X\sigma- \X^3/3)}\,\rd \X
=  
\frac{\re^{\ri\pi/3}}
{2\pi}\Ai[\sigma],
\end{align*}
so that
\begin{align}
\label{eqn:FLSolution}
\cF(\sigma)  = 
\frac{\re^{\ri\pi/3}\Ai[\sigma]}
{2\pi \Ai\left[\re^{2\pi\ri/3}\sigma\right]},
\end{align}
and hence (recalling \rf{eqn:PekerisFourierRep})
\begin{align}
\label{eqn:FPek}
F(t) = \frac{\re^{\ri\pi/3}}
{2\pi} \int_{-\infty}^\infty \re^{\ri t\sigma}\frac{\Ai[\sigma]}
{\Ai\left[\re^{2\pi\ri/3}\sigma\right]}\,\rd \sigma = \hat{p}(t).
\end{align}

For the case $(l,m)=(1,2)$ (cubic parabola) the construction of modal solutions seems to be significantly more difficult, and we have not been able to find an explicit choice of $F$ which makes \rf{eqn:AInt} or its normal derivative vanish on the localisation curve $\Y+\kappa^2\X^3/6=0$ for any of the contour choices $\Gamma_{ij}$. However, for some of these contours it is possible to specify asymptotic properties of $F$ that ensure asymptotic modality at infinity. To illustrate this, consider $A_{21}$, which, as \F\ref{fig:QuinticBehaviours2}(a) and \rf{eqn:InflectionAsympt} suggest, has an asymptotic structure associated with an outgoing whispering gallery wave (below the cubic parabola); we will show how $A_{21}$ can in fact be made to vanish asymptotically on the curve $\Y+\kappa^2 \X^3/6=0$ as $\X\to+\infty$. Suppose that $F$ is such that $F(t)\sim C\re^{\ri \sigma t^\beta}$ as $|t|\to\infty$, for some constants $C\in\C$, $\sigma\in \R$ and $\beta>0$, uniformly in some sector around the positive real $t$-axis in which there is no branch cut of $t^\beta$. We expand the phase as usual, under the scalings leading to \rf{eqn:InflectionAsympt}, and note that, with $\x_0>0$,
\begin{align*}
\label{}
t^\beta = \left(k^{1/5}\tau\right)^\beta 
=k^{\beta/5}\left(\frac{\kappa \x_0}{\sqrt{2}}+\tau'\right)^\beta 
\sim k^{\beta/5}\left(\frac{\kappa \x_0}{\sqrt{2}}\right)^\beta \left(1 + \frac{2^{1/3}\beta k^{-1/3}}{\kappa^{4/3}\x_0^{5/3}}\zeta + \ord{k^{-2/3}}\right).
\end{align*}
Choosing $\beta=5/3$ makes the coefficient of the $\zeta$ term order one, so that, after neglecting higher order terms, the prefactor $F(t)$ contributes an extra factor 
\begin{align*}
\label{}
\re^{\ri k^{1/3}(\kappa \x_0/\sqrt{2})^{5/3}\sigma}
\re^{\ri 5\kappa^{1/3}\sigma \zeta/(3\sqrt{2}) }
\end{align*}
to the leading order behaviour of the integrand in \rf{eqn:InflectionAsympt}. This modifies \rf{eqn:InflectionAsympt} so that it contains as a factor
\begin{align*}
\label{}
\Ai\left[-\left(2\kappa^2 \x_0\right)^{1/3}\left(\y^*+\frac{\kappa^2 \x_0^2 \x^*}{2}\right)+ \tilde\sigma\right],
\end{align*}
where $\tilde\sigma :=5\kappa^{1/3}\sigma/(3\sqrt{2})$. By choosing $\tilde\sigma<0$ to be any zero of $\Ai$, we can make this factor vanish on the cubic curve, which corresponds to $A_{21}$ vanishing asymptotically on this curve as $\X\to+\infty$. 

By the symmetry relation \rf{QuinticSymmetry} we see immediately that a similar approach can be applied to $A_{32}$, which as \F\ref{fig:QuinticBehaviours2}(b) suggests, has an asymptotic structure associated with an incoming whispering gallery wave. (As we remarked at the end of \S\ref{sec:Inflection}, this behaviour of $A_{32}$ was noted already by Kazakov in \cite{Kaz:03}, although he did not attempt to satisfy boundary conditions.) 
The double saddle point giving rise to the localisation now occurs on the negative real axis (we take $\x_0<0$), so to obtain a solution which vanishes asymptotically on $\Y+\kappa^2 \X^3/6=0$ as $\X\to-\infty$, we can simply take as the prefactor in \rf{eqn:AInt} the function $F(-t)$, where $F$ has the asymptotic behaviour described above. 
In principle, in this way one can construct higher-order terms (to arbitrarily high order) in the asymptotic expansion of $F(t)$ as $t\to+\infty$, using the higher order terms in the expansion for the incoming whispering gallery waves, cf.\ e.g.\ \cite{Pop:79,BabSmy:86}. 
A similar approach applied to $A_{43}$ and $A_{52}$ allows the construction of solutions which, asymptotically, represent outgoing ($\X\to +\infty$) and incoming ($\X\to -\infty$) creeping waves, respectively (cf.\ \F\ref{fig:QuinticBehaviours2}(g),(i)).%

\section{\label{sec:Discussion}Discussion}
We end the paper with some remarks on 
the possibility of using a contour integral ansatz of the form \rf{eqn:AInt} to solve 
the canonical inflection point problem, in which one seeks a solution of PWE describing the transition from whispering gallery waves propagating along a concave portion of a boundary, through an inflection point, to creeping waves propagating along a convex portion of the same boundary. The first rigorous formulation of the inflection point problem, along with a proof of existence and uniqueness of the solution matching with an incoming whispering gallery mode, was presented by Popov in \cite{Pop:79}. Following this a number of results concerning the smoothness, decay properties, and asymptotic behaviour of the solution, along with some numerical results, were derived by Popov and co-workers \cite{Pop:82,Pop:82a,Pop:86,PopKra:86,Pop:79a} and Babich and Smyshlyaev \cite{BabSmy:86,BabSmy:87}. However, as yet, no closed form expression for the solution has been found.
In particular, the standard approach of transformation to curvilinear coordinates and application of a Fourier transform that delivers the solution of the Fock-Leontovich tangent ray diffraction problem fails for the inflection point problem because the transformed equation is not separable. 
A connection between the inflection point problem and contour integral solutions of PWE was made by Kazakov in \cite{Kaz:03}, but, as was alluded to in \S\ref{sec:Inflection} and \S\ref{sec:BVPs}, while Kazakov's special function \rf{eqn:Kaz} (related to our contour integral $A_{32}$) shares some properties of the solution of the inflection point problem, it fails to satisfy homogeneous boundary conditions, and, as we discuss below, does not capture creeping waves. 

We start by making two general observations. First, %
provided $F(t)$ does not grow too fast at infinity (specifically, provided the integrand decays sufficiently rapidly at infinity along $\Gamma$), the integral \rf{eqn:AInt} defines an entire function of the variables $\X$ and $\Y$ that solves the PWE globally on $\R^2$ (in fact on $\C^2$). Hence if the solution to the inflection point problem can indeed be represented in the form \rf{eqn:AInt}, then it must be analytically extendable across the boundary to an entire function on the whole plane. 
Proving this extendibility property for the inflection point problem would therefore verify a necessary condition for the existence of a contour integral solution representation. Conversely, 
showing that the analytic continuation of the solution develops singularities on the `non-physical' side of the boundary would constitute a proof of non-existence of such a representation.
Second, 
if a contour integral representation \textit{does} exist, then it will almost certainly not extend to a tempered distribution on $\R^2$. Indeed, it was already noted in \S\ref{sec:i} that the canonical creeping wave solution \rf{eqn:AAiryA32} blows up exponentially in the non-physical region of the plane. The same is true for the Fock-Leontovich solution, as is easily seen from 
a steepest descent analysis of \rf{eqn:FLPsoln}, \rf{eqn:PekerisFourierRep} as the observation point tends to infinity along the negative $\Y$-axis, for example.

Next we turn to the problem of actually deriving a solution representation of the form \rf{eqn:AInt}. %
We demonstrated above that the solutions $A_{32}$ and $A_{52}$ in \F\ref{fig:QuinticBehaviours2} can respectively model incoming whispering gallery and outgoing creeping waves which are asymptotically modal on the relevant portion of a cubic parabola (with $F(t)$ behaving as prescribed at the end of \S\ref{sec:BVPs}). However, we have been unable to find a single choice of contour $\Gamma$ and prefactor $F(t)$ that can capture both asymptotic phenomena simultaneously. In particular, consulting \F\ref{fig:QuinticBehaviours2}, we see that while $A_{32}$ captures incoming whispering gallery waves in $\x<0$, it exhibits no localisation on the portion of the cubic in $\x>0$, producing two real rays which are not associated with a creeping wave field. Similarly, while $A_{52}$ captures outgoing creeping waves in $\x>0$, it exhibits no localisation on the portion of the cubic in $\x<0$, producing a single real ray associated with the saddle point at the origin. If it is indeed possible to obtain a single integral of the form \rf{eqn:AInt} exhibiting both behaviours, it would seem that the function $F(t)$ must exhibit strongly exponential behaviour in certain sectors, in order to modify the saddle point structure in a significant way. 
We recall that strongly exponential behaviour in $F(t)$ was responsible for modifying a creeping-wave-type integral into the specularly reflected field in the tangent ray diffraction problem, with the $\re^{-\ri t^3/12}$ factor in \rf{eqn:PekAsympt2} modifying the saddle point structure by an $O(1)$ amount. For the inflection point problem the required modifications to the saddle point structure would likely have to be even more extreme. %
We point out that both $A_{32}$ and $A_{52}$ exhibit localisation near the `searchlight' line $y=0$, $x>0$ (see the discussion at the end of \S\ref{sec:Inflection}), picking up the triple saddle point contribution from the origin (see \F\ref{fig:QuinticBehaviours2}). Further investigation into the far-field structure of the solution near this searchlight line, building on \cite{Pop:86} where for Dirichlet boundary conditions the far field amplitudes near the searchlight line are expressed in terms of the (as yet unknown) Neumann data on the boundary, could be an interesting avenue to pursue, since it would provide additional information about the required far-field behaviour of any representation of the form \rf{eqn:AInt}, as was the case for the penumbra field in the tangent ray diffraction problem (see the discussion following \rf{eqn:ShiftedSaddles1}).

If a single integral representation of the form \rf{eqn:AInt} cannot be found, a more general approach might be to seek a solution in the `forked contour' form 
\begin{align}
\label{eqn:AIntInflectionCombined}
A = \int_{\Gamma_{32}}F_{i}(t)\re^{\ri p(\X,\Y,t)}\,\rd t + \int_{\Gamma_{52}}F_{s}(t)\re^{\ri p(\X,\Y,t)}\,\rd t,
\end{align}
for suitable functions $F_i(t)$ (associated with the `incident' whispering gallery wave) and $F_s(t)$ (associated with the `scattered' creeping wave).  %
However, it remains an open question whether these functions can be chosen so that \rf{eqn:AIntInflectionCombined} satisfies homogeneous boundary conditions on the cubic parabola. %

A potential avenue for progress may lie in combining the present approach with that in \cite{Pop:79,Pop:82a,Pop:82,Pop:86} and \cite{BabSmy:86,BabSmy:87}. Indeed, the detailed information on asymptotic properties of the exact solution derived in the above references would relate to the analytic and asymptotic properties of the amplitudes $F_{i}(t)$ and $F_{s}(t)$. The latter may eventually have to be found numerically, e.g.\ from an appropriate integral equation (cf.\ the use of Green's integral representations in \cite{Pop:82,Pop:86}, and a somewhat similar approach applied to a different class of non-separable problems in e.g.\ \cite{lyalinov2010scattering}). 
We leave further investigation of these and related issues to future work. 

Finally, we also mention another approach to the inflection point and related diffraction problems proposed recently in \cite{kazakov2018separation}.

\section*{Acknowledgements} The authors thank Prof. R.\ Tew for helpful discussions in relation to this work.

\bibliography{inflection_bib}
\bibliographystyle{siam}
\end{document}